\documentclass[pra,reprint,superscriptaddress,bibnotes]{revtex4-1}

\usepackage[T1]{fontenc}
\usepackage[utf8]{inputenc}
\usepackage[australian]{babel}
\usepackage[a4paper,centering,hmargin=1.75cm,vmargin=2cm]{geometry}
\usepackage{amsmath,amssymb,bm}
\usepackage{graphicx,xcolor}
\usepackage{titlesec}
\usepackage{microtype}
\usepackage{braket,siunitx}
\usepackage[version=3]{mhchem}
\usepackage{bbold}
\usepackage{lipsum}
\usepackage{mathtools}
\usepackage{hyperref} 
\usepackage{tabularx}
\usepackage[normalem]{ulem}
\usepackage{tikz}
\usetikzlibrary{shapes,math}
\usepackage{float}

\usepackage{comment}

\graphicspath{{figures/}}
\titleformat*{\paragraph}{\itshape}

\newcommand{\gridstate}{\tikz[scale=.14,baseline=-1mm]{ 
\draw (-1, -.45) -- (1, -.45);
\draw (-1, +.45) -- (1, +.45);
\draw (-.45, -1) -- (-.45, 1);
\draw (.45, -1) -- (.45, +1);}}

\newcommand{\npstate}{\tikz[scale=.14,baseline=-1mm]{ 
\draw (0,0) circle [radius=.7];
\draw (-1, 0) -- (1, 0);
\draw (0, -1) -- (0, 1);}}

\usepackage[normalem]{ulem}

\newcommand*{\bigchi}{\mbox{\Large$\chi$}}

\newcommand{\shiftparam}{\boldsymbol{\epsilon}}

\def\blue#1           {\textcolor{blue} {#1} }
\def\yellow#1           {\textcolor{orange} {#1} }

\begin{document}

\title{Quantum-Enhanced Multi-Parameter Sensing in a Single Mode}

\author{Christophe~H.~Valahu}
\email{christophe.valahu@sydney.edu.au}
\affiliation{School of Physics, University of Sydney, NSW 2006, Australia}
\affiliation{ARC Centre of Excellence for Engineered Quantum Systems, University of Sydney, NSW 2006, Australia}
\affiliation{Sydney Nano Institute, University of Sydney, NSW 2006, Australia}

\author{Matthew~P.~Stafford}
\affiliation{Quantum Engineering Technology Labs, H. H. Wills Physics Laboratory and Department of Electrical and Electronic Engineering, University of Bristol, UK.}
\affiliation{Quantum Engineering Centre for Doctoral Training, University of Bristol, UK.}

\author{Zixin~Huang}
\affiliation{School of Mathematical and
Physical Sciences, Macquarie University, NSW 2109, Australia}
\affiliation{Centre for Quantum Software and Information, Faculty of Engineering and
Information Technology, University of Technology Sydney, NSW 2007, Australia}

\author{Vassili~G.~Matsos}
\affiliation{School of Physics, University of Sydney, NSW 2006, Australia}
\affiliation{ARC Centre of Excellence for Engineered Quantum Systems, University of Sydney, NSW 2006, Australia}

\author{Maverick~J.~Millican}
\affiliation{School of Physics, University of Sydney, NSW 2006, Australia}
\affiliation{ARC Centre of Excellence for Engineered Quantum Systems, University of Sydney, NSW 2006, Australia}

\author{Teerawat~Chalermpusitarak}
\affiliation{School of Physics, University of Sydney, NSW 2006, Australia}

\author{Nicolas~C.~Menicucci}
\affiliation{Centre for Quantum Computation and Communication Technology,
School of Science, RMIT University, VIC 3000, Australia}

\author{Joshua~Combes}
\affiliation{University of Melbourne, VIC 3052, Australia}

\author{Ben~Q.~Baragiola}
\affiliation{Centre for Quantum Computation and Communication Technology,
School of Science, RMIT University, VIC 3000, Australia}

\author{Ting~Rei~Tan}
\email{tingrei.tan@sydney.edu.au}
\affiliation{School of Physics, University of Sydney, NSW 2006, Australia}
\affiliation{ARC Centre of Excellence for Engineered Quantum Systems, University of Sydney, NSW 2006, Australia}
\affiliation{Sydney Nano Institute, University of Sydney, NSW 2006, Australia}

\begin{abstract}
Precise measurements underpin scientific and technological advancements. Quantum mechanics provides an avenue to enhance precision, but it comes with a restriction: incompatible observables, such as position and momentum, cannot be simultaneously measured to arbitrary accuracy as decreed by Heisenberg’s uncertainty principle. 
This restriction can be bypassed by instead measuring commuting modular observables, which are counterparts to the naturally incompatible observables. 
Here, we measure modular observables to estimate small changes in position and momentum with a \emph{single-mode multi-parameter sensor}. We deterministically prepare grid states in the mechanical motion of a trapped ion and demonstrate uncertainties in position and momentum below the standard quantum limit (SQL). 
Further, we examine another pair of incompatible observables---number and phase. We prepare a different resource---number-phase states---and demonstrate a metrological gain over the SQL. These results introduce new measurement capabilities unavailable to classical systems and mark a significant step in quantum metrology.
\end{abstract}

\maketitle

\section{Introduction}

Advances in metrology have historically led to breakthroughs in scientific understanding. Galileo's telescope led to the rejection of the geocentric model, while Young's double-slit experiment established the wave properties of light. More recently, gravitational wave detection~\cite{Aasi2013,Abbott2016}---a feat of engineering in precision interferometry---has ushered in a new era of astrophysical discoveries.
Precision metrology also leads to transformative technologies, such as the global positioning system which harnesses the accuracy of atomic clocks.

The precision of measurements is ultimately limited by quantum mechanical noise, and this limit becomes ever more salient as capabilities improve. The standard quantum limit~(SQL) applies when measurements use only classical resources and methods. Surpassing this limit by using quantum resources or measurement strategies is known as \emph{quantum metrology}~\cite{giovannetti2004quantum}, and it promises to revolutionise precision measurements beyond the capability of conventional sensors. For example, the Laser Interferometer Gravitational-Wave Observatory~(LIGO) uses non-classical light to reduce the measurement uncertainty below the SQL~\cite{tse2019quantum}. Quantum enhancements to sensing have also been demonstrated in atomic clocks~\cite{ludlow2015optical}, biological imaging~\cite{Taylor2013,Taylor2014}, and the search for dark matter~\cite{Backes2020}.

When measuring two or more observables that are incompatible, quantum mechanics imposes trade-offs on their uncertainties. This is exemplified by Heisenberg’s uncertainty principle: one can not simultaneously reduce the measurement uncertainty of position and momentum using a probe composed of a single bosonic mode~\cite{Heisenberg1927, Ozawa2004}. In this context, a mode refers to a single, independent degree of freedom described by observables obeying the canonical commutation relations, typically expressed for bosonic modes in terms of creation and annihilation operators, $[\hat a, \hat a^\dag]=1$. The limit imposed by Heisenberg's uncertainty principle is typically circumvented by using multiple entangled modes, the simplest case being a two-mode squeezed state~\cite{Jensen2010, OckeloenKorppi2018, Barzanjeh2019, Qin2023, Leong2023, Metzner2024}, where enhancements beyond the SQL have been demonstrated at the cost of additional quantum resources~\cite{Genoni2013optimaljoint, cao2023, Li2023}.

\begin{figure*}
    \centering
    \includegraphics[]{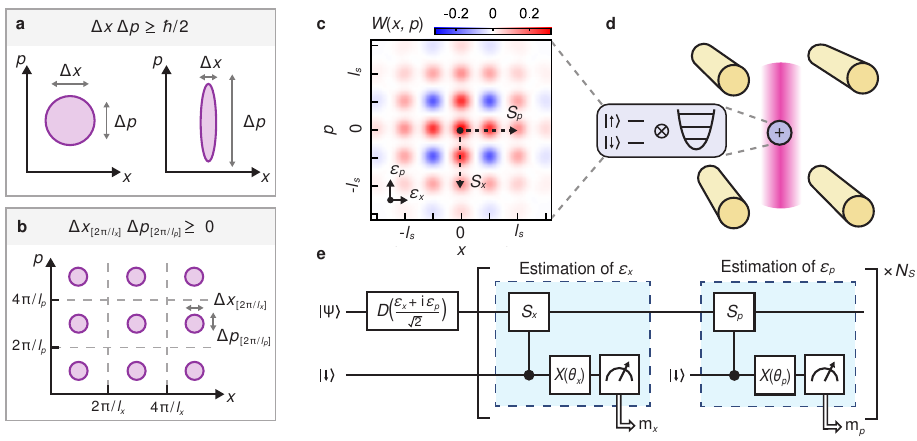}
    \caption{
    \textbf{Multi-parameter quantum enhanced sensing.}
    \textbf{(a)} The uncertainty in simultaneous position--momentum measurements is bounded by the canonical commutation relation.
    \textbf{(b)} Their modular counterparts can instead be made to commute, allowing for estimation of $\hat{x}_{[2\pi/l_x]}$ and $\hat{p}_{[2\pi/l_p]}$ with uncertainties simultaneously below the SQL, where $l_s = l_x = l_p = \sqrt{2\pi}$ are the modulus lengths.
    \textbf{(c)} The eigenstates of modular position--momentum are grid states (shown is the Wigner function), which can sense displacements in position by $\epsilon_x$ and displacements in momentum by $\epsilon_p$ by measuring commuting operators $\hat{S}_x$ and $\hat{S}_p$. 
    \textbf{(d)} Physical grid states are prepared in the bosonic mode of a trapped ion. An ancilla qubit encoded in the electronic ground state of the ion couples to the bosonic mode via a laser interaction (pink beam), allowing for measurement of position and momentum observables.
    \textbf{(e)} A Quantum Phase Estimation (QPE) circuit is used for multi-parameter estimation of $\epsilon_x$ and $\epsilon_p$. After preparing the sensing state and undergoing an unknown displacement, multiple rounds of a QPE sub-routine (blue box) are applied, where each round applies conditional $\hat{S}_x$ or $\hat{S}_p$, which gives one bit of phase estimation, $\mathrm{m}_x$ or $\mathrm{m}_p$. Since $\hat{S}_x$ and $\hat{S}_p$ commute, they can be applied alternately $N_s$ times, ideally performing many rounds of backaction-evading measurements. 
    }
    \label{fig:states_and_circuit}
\end{figure*}

Alternatively, a single mode can simultaneously reduce the uncertainties of two parameters corresponding to incompatible observables with a different trade-off: increased sensitivities at the cost of restrictions on parameter ranges. The reduction in range is unimportant when the parameters are sufficiently small. The key idea is to measure \emph{modular} variables that are made to commute, circumventing the usual constraint of naturally non-commuting observables imposed by the uncertainty principle~\cite{Aharonov1969,Gottesman2001,Ketterer2016modular,Duivenvoorden2017,Flhmann2018}. This measurement strategy has been theoretically explored for the simultaneous estimations of position and momentum using grid states---a sub-class of the Gottesman--Kitaev--Preskill~(GKP) states~\cite{Gottesman2001} that have been investigated in the context of quantum error correction~\cite{Flhmann2019, CampagneIbarcq2020, deNeeve2022, Eickbusch2022, Sivak2023, kawasaki2024broadband}. Ideal grid states allow backaction-evading measurements when sequentially measuring modular position and momentum that commute. Beyond position and momentum, there exist many pairs of non-commuting variables of metrological interest, but simultaneous estimations of them have remained relatively unexplored in the literature.

Here, we demonstrate the simultaneous reduction in uncertainty of two parameters associated with incompatible observables by measuring their modular counterparts. Our experiment uses a single probe made up of a trapped ion's vibrational mode. Our work draws on concepts and techniques developed for error-corrected quantum information processing---such as logical qubit encodings, stabiliser syndrome extractions, and optimal control---and adapts them for metrology. We first consider multi-parameter displacement sensing using grid states and demonstrate a clear metrological gain over the SQL. We then perform a quantum phase estimation algorithm with Bayesian inference, and find that an adaptive strategy performs better than a non-adaptive strategy. Furthermore, we investigate simultaneous estimations of \textit{number} and \textit{phase}, which do not commute. This is achieved by using \textit{number-phase} states~\cite{Grimsmo2020}, which are the polar counterparts of grid states that were only theoretically explored because of no previously known experimental scheme to prepare the states. We develop the necessary quantum control for operations with these states and demonstrate a metrological gain over the simultaneous SQL of number and phase. In so doing, we introduce---and experimentally realise---a novel resource for quantum sensing on which we plan further theoretical and experimental investigations.

\section{Results}

We first consider the simultaneous estimation of position, $\hat{x}$, and momentum, $\hat{p}$, whose uncertainties follow the uncertainty principle, $\Delta x \Delta p \geq \hbar / 2$~\cite{Heisenberg1927, Ozawa2004} (see Fig.~\ref{fig:states_and_circuit}a). This limit is circumvented by instead measuring their modular counterpart, $\hat{x}_{[2\pi/l_x]}$ and $\hat{p}_{[2\pi/l_p]}$, where $\hat{q}_{[m]} = \hat{q}\text{ mod }m$ (see Fig.~\ref{fig:states_and_circuit}b). Modular position and momentum are observables of $\hat{x}$ and $\hat{p}$ up to a modulus $2\pi/l_x$ and $2\pi/l_p$, respectively. By setting $l_{x,p} = l_s = \sqrt{2\pi}$, modular position and momentum commute, giving the uncertainty relation $\Delta \hat{x}_{[2\pi/l_s]} \Delta \hat{p}_{[2\pi/l_s]} \geq 0$. This commutation relation was investigated by a signaling-in-time experiment and a violation of the Leggett-Garg inequality~\cite{Flhmann2018}.

We exploit the compatibility of these observables to simultaneously sense small displacements of position and momentum below the SQL. To do so, we employ grid states, $\ket{\gridstate}$, which are simultaneous eigenstates of the shift operators $\hat{S}_x = e^{- i l_s \hat{x}_{[2\pi/l_s]}} = e^{- i l_s \hat{x}}$ and $\hat{S}_p = e^{-i l_s \hat{p}_{[2\pi/l_s]}} = e^{- i l_s \hat{p}}$. Grid states are periodic in $\hat{x}$ and $\hat{p}$ and have peaks in phase space located on points of a square lattice of size $l_{s}$ (see Fig.~\ref{fig:states_and_circuit}c). The ideal grid states are unphysical as they have infinite energy. We instead consider finite-energy approximations to these states, $\ket{\tilde{\gridstate}}$, given by a weighted superposition of squeezed states parameterised by $\Delta$, which serves as a measure of quality. Ideal grid states are recovered in the limit $\Delta \rightarrow 0$. Physical grid states are used for multi-parameter sensing in the following way: after an unknown displacement, $\ket{\tilde{\gridstate}_{\shiftparam}} = e^{i \epsilon_p \hat{x}} e^{- i  \epsilon_x \hat{p}} \ket{\tilde{\gridstate}}$ with displacement parameters $\epsilon_x, \epsilon_p \in \mathbb{R}$, one can estimate $\epsilon_x~\mathrm{mod}~\sqrt{2\pi}$ and $\epsilon_p~\mathrm{mod}~\sqrt{2\pi}$ by estimating the eigenvalues of $\hat{S}_x$ and $\hat{S}_p$. Applying these operators gives $\hat{S}_x \ket{\tilde{\gridstate}_{\shiftparam}} \approx e^{- i \sqrt{2\pi} \epsilon_x} \ket{\tilde{\gridstate}_{\shiftparam}}$ and $\hat{S}_p \ket{\tilde{\gridstate}_{\shiftparam}} \approx e^{- i \sqrt{2\pi} \epsilon_p} \ket{\tilde{\gridstate}_{\shiftparam}}$, and the displacement parameters are then retrieved by estimating the phases, $\sqrt{2\pi} \epsilon_{x}$ and $\sqrt{2\pi} \epsilon_{p}$, imprinted on the sensing state via a quantum phase estimation (QPE) algorithm. These modular measurements give unambiguous estimations of position and momentum if it is known a priori that the displacement parameters are smaller than $\sqrt{2\pi}$. In systems where this is not known, the above measurement protocol estimates the remainder of the modulus but the integer part remains unknown.

\begin{figure}
    \centering
    \includegraphics[]{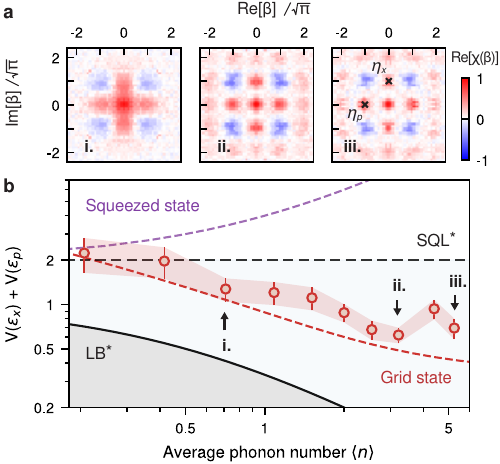}
    \caption{ 
    \textbf{Metrological gain of grid states for multi-parameter displacement sensing.}
    \textbf{(a)} Experimentally reconstructed characteristic function of prepared grid states with target parameters $\Delta$ of (i) 0.61, (ii) 0.37 and (iii) 0.30. Increasing energy results in increased squeezing along position and momentum. Crosses in (iii) show points of the characteristic function that correspond to values of the visibility parameters $\eta_x = \mathrm{Re}[\bigchi(i \sqrt{\pi})]$ and $\eta_p = \mathrm{Re}[\bigchi(- \sqrt{\pi})]$.
    \textbf{(b)} Multi-parameter variance, $\mathrm{V}(\epsilon_x) + \mathrm{V}(\epsilon_p)$, of grid states with increasing average phonon number, $\langle \hat{n} \rangle$. Variances (red circles) are calculated from the classical Fisher information of the experimentally measured probability distributions $\mathrm{P}_x$ and $\mathrm{P}_p$. The sensing signal is varied in the range $\{\epsilon_{x}, \epsilon_{p}\} \in [0, 1.4]$. Error bars correspond to one standard deviation calculated from quantum projection noise. Dashed red line is the expected multi-parameter variance of the target grid states. The simultaneous standard quantum limit (SQL*) corresponds to the multi-parameter variance of a coherent state from heterodyne measurement (see SM). The simultaneous lower bound (LB*) plots the minimum uncertainty from the quantum Fisher information,  $1/(2\langle \hat{n} \rangle + 1)$~\cite{Genoni2013optimaljoint, Duivenvoorden2017}. Dashed purple line corresponds to the multi-parameter variance from heterodyne detection, with a single-mode squeezed state squeezed in one quadrature and anti-squeezed in the other. Heterodyne detection is equivalent to double homodyne detection: the state is split by a 50:50 beam splitter, with position measured on one output and momentum on the other.
    }
    \label{fig:grid_state_metro}
\end{figure}

The circuit to perform multi-parameter sensing is depicted in Fig.~\ref{fig:states_and_circuit}e. After preparing the sensing state and undergoing an unknown displacement, a sequence of QPE sub-routines (blue boxes) is applied in an alternating fashion. Each sub-routine first applies a conditional momentum or position operator, $C\hat{S}_x = e^{- i l_s \hat{\sigma}_x \hat{x}/2}$ or $C\hat{S}_p = e^{- i l_s \hat{\sigma}_x \hat{p}/2}$, which maps information from the bosonic mode to an ancillary two-level system. A measurement of the ancilla then gives a binary measurement outcome $\mathrm{m}_{x}\in\{0, 1\}$ or $\mathrm{m}_{p}\in\{0, 1\}$, with probability distributions
\begin{align}
    & \mathrm{P}_x(\mathrm{m}_x| \epsilon_x, \theta_x) = \frac{1}{2}\left(1 + (-1)^{\mathrm{m}_x}\eta_x \cos(\epsilon_x l_s + \theta_x)\right), \label{eq:prob_distribution_a}\\
    & \mathrm{P}_p(\mathrm{m}_p | \epsilon_p, \theta_p) = \frac{1}{2}\left(1 + (-1)^{\mathrm{m}_p} \eta_p \cos(\epsilon_p l_s + \theta_p)\right). \label{eq:prob_distribution_b}
\end{align}
The visibility parameters $\eta_x = \langle \hat{S}_x + \hat{S}_x^\dagger \rangle/2 =  \mathrm{Re}[\bigchi(i \sqrt{\pi})]$ and $\eta_p = \langle \hat{S}_p + \hat{S}_p^\dagger \rangle/2 = \mathrm{Re}[\bigchi(- \sqrt{\pi})]$ correspond to values of the characteristic function on the square lattice (see Fig.~\ref{fig:grid_state_metro}a), where $\bigchi(\beta) = \langle \hat{D}^\dagger(\beta)\rangle$ and $\beta \in \mathbb{C}$ is the phase space location where the characteristic function is sampled. The visibility parameters are $\eta_x = \eta_p = 1$ for an ideal grid state with $\Delta = 0$. The phases $\theta_{x}$ and $\theta_{p}$ are controllable and are introduced by an ancilla rotation~(see Fig.~\ref{fig:states_and_circuit}e). Importantly, the probability distribution $\mathrm{P}_x$ and $\mathrm{P}_p$ are only dependent on one of the parameters, $\epsilon_x$ or $\epsilon_p$, respectively. Therefore, each measurement outcome can be used to independently retrieve $\epsilon_x$ or $\epsilon_p$.

Our experiment is performed with a single \ce{^{171}Yb^+} ion confined in a room-temperature macroscopic Paul trap. The sensing states are encoded in the vibrational bosonic mode along a radial direction with a frequency of $\omega_x  = 2\pi \times 1.33$ MHz. An ancillary qubit is encoded in the ``atomic clock'' state of the hyperfine ground state with labels $\ket{\downarrow} \equiv \ket{F=0, m_f=0}$ and $\ket{\uparrow} \equiv \ket{F=1, m_f = 0}$, and is used to assist in the preparation of sensing states and measurements; see Refs.~\cite{MacDonell2023, Valahu2023, Matsos2024} for more details on the experimental system. 

The coherent control required to prepare the grid states and measure observables is performed with a laser-driven state-dependent force (SDF) that couples the ancillary qubit and the vibrational mode (see Materials and Methods). An SDF is enacted by stimulated Raman transitions from a pair of orthogonal beams derived from a 355~nm pulsed laser. The Hamiltonian of the SDF in the interaction frame of both the qubit and the vibrational mode is 
\begin{align}
    \hat{H}_\mathrm{SDF}(t) = \frac{\Omega}{2} & (\hat{\sigma}_x \cos\phi_s(t) + \hat{\sigma}_y\sin\phi_s(t)) \nonumber \\
    &  \times (\hat{a}^\dagger e^{- i \phi_m(t)} + \hat{a} e^{ i \phi_m(t)}).
\end{align}
The interaction strength $\Omega$ is controllable by varying the laser power, while the phases $\phi_s(t)$ and $\phi_m(t)$ are tuneable by modulating an acousto-optic modulator in the path of one of the Raman beams. The controlled position and momentum operators are obtained by applying $\hat{H}_\mathrm{SDF}$ for a duration $t = \sqrt{\pi}/\Omega$. Setting $(\phi_s, \phi_m) = (0, 0)$ gives $C\hat{S}_x$, while setting $(\phi_s, \phi_m) = (\pi, \pi/2)$ gives $C\hat{S}_p$.

Here, the circuit of Fig.~\ref{fig:states_and_circuit}e is implemented as follows. First, the sensing state is prepared by applying $\hat{H}_\mathrm{SDF}(t)$ with dynamically modulated phases $\phi_s(t)$ and $\phi_m(t)$. The phase modulation waveforms are numerically optimised to prepare grid states with varying target squeezing parameters, see Ref.~\cite{Matsos2024} and Materials and Methods. The waveforms are modelled as piecewise constant functions with 30 optimisable segments, and the resulting durations are in the range 0.3--1.5~ms. We also constrain the numerical optimisation such that the ancilla returns to the $\ket{\downarrow}$ state after applying the pulse and is disentangled from the motional mode. Second, the sensing states are subjected to a force which displaces the state by $\hat{D}(\frac{1}{\sqrt{2}}(\epsilon_x + i\epsilon_p))$. In our experiment, this force is controllably injected by applying a laser-driven interaction, $\hat{H}_\mathrm{SDF}$, for a duration $t = \sqrt{2}|\epsilon_x + i \epsilon_p|/\Omega$ with $\phi_s=0$ and $\phi_m = -\pi/2 - \mathrm{arg}(\epsilon_x + i \epsilon_p)$. The SDF is surrounded by pulses that rotate the ancilla in and out of a $\hat{\sigma}_x$ eigenstate, such that it remains disentangled from the motion after applying $\hat{H}_\mathrm{SDF}$. We then perform QPE and retrieve measurement outcomes $\mathrm{m}_{x}$ and $\mathrm{m}_{p}$ by performing ancilla measurements in the $\hat{\sigma}_z$ basis through state-dependent fluorescence. Detections of $\ket{\downarrow}$ and $\ket{\uparrow}$ correspond to measurement outcomes of 0 and 1, respectively. Photons scattered from measurement outcomes of 1 decohere the sensing state due to their recoil energy, hence the experiment only proceeds if the measurement outcome is 0. Measurement outcomes of 1 are instead obtained by randomly initialising the ancillary qubit in $\ket{\downarrow}$ or $\ket{\uparrow}$ with equal probability at the beginning of a QPE sub-routine. We then record an outcome of 0 or 1 if the qubit was initialised in $\ket{\downarrow}$ or $\ket{\uparrow}$, respectively.

In the first experiment, we characterise the metrological gain of the grid states by calculating the multi-parameter variance. To this end, we reconstruct the probability distributions of Eq.~\ref{eq:prob_distribution_a} and Eq.~\ref{eq:prob_distribution_b} using the QPE circuit of Fig.~\ref{fig:states_and_circuit}e with varying $\epsilon_{x}$, $\epsilon_{p}$. For each pair $\{\epsilon_{x}, \epsilon_{p}\}$, we set $\theta_x = \theta_p = 0$ and perform the QPE circuit $M$ times. The probabilities $\mathrm{P}_x(\mathrm{m}_x| \epsilon_x)$ and $\mathrm{P}_p(\mathrm{m}_p | \epsilon_p)$ are calculated from the mean of the outcomes $\mathrm{m}_x$ and $\mathrm{m}_p$. Repeating this over a range of $\{\epsilon_{x}, \epsilon_{p}\}$ gives 2-dimensional probability distributions. These are used to compute the $2\times 2$ Fisher information matrix $\mathbf{F}$, which quantifies the amount of information corresponding to $\{\epsilon_{x}, \epsilon_{p}\}$ that is contained in the measurement outcomes. From the Fisher information matrix, we compute the $2\times 2$ covariance matrix, $\mathbf{\Sigma} = \mathbf{F}^{-1}$. The multi-parameter variance is then bounded by minimizing the trace of the covariance matrix over the range $\{\epsilon_{x}, \epsilon_{p}\}$, $\mathrm{V}(\epsilon_x) + \mathrm{V}(\epsilon_p) \geq \mathrm{min}_{\epsilon_x, \epsilon_p} \mathrm{Tr}(\mathbf{\Sigma})$~\cite{Genoni2013optimaljoint}. This procedure is repeated for grid states with different average phonon numbers, $\langle \hat{n} \rangle$, with results summarised in Fig.~\ref{fig:grid_state_metro}. The experiment agrees well with theory, where the uncertainty decreases with $\langle \hat{n} \rangle$. This improvement is corroborated by the reconstructed characteristic functions of the grid states, whose peaks are more squeezed with larger $\langle \hat{n} \rangle$ (Figure~\ref{fig:grid_state_metro}a). The lowest variance is obtained at $\langle \hat{n} \rangle = 3.2$ ($\Delta = 0.37$), giving a gain of $\SI{5.1(5)}{dB}$ over the simultaneous SQL (here and throughout, the terminology SQL* is used to refer to the simultaneous standard quantum limit for two non-commuting quadrature measurements). We observe that the variance does not further decrease for $\langle \hat{n} \rangle > 3.2$, and attribute this to dephasing of the motional mode.

\begin{figure}
    \centering
    \includegraphics[]{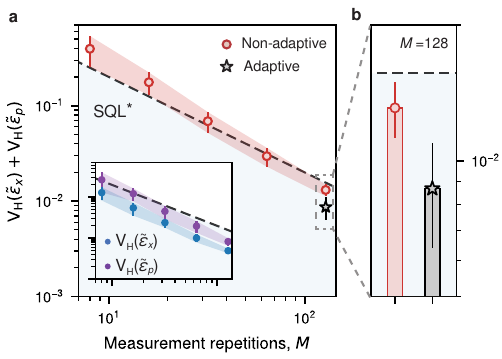}
    \caption{
    \textbf{Quantum Phase Estimation (QPE) with grid states.}
    \textbf{(a)} The multi-parameter variance of a grid state with $\Delta = 0.41$ is measured for increasing measurement repetitions $M \in [8, 128]$ with $N_S=1$ QPE sub-routine repetitions. Estimates ($\tilde{\epsilon}_x, \tilde{\epsilon}_p)$ of an unknown displacement $\hat{D}((\epsilon_x + i \epsilon_p)/\sqrt{2})$ are obtained from a QPE algorithm using a non-adaptive routine (red circles), where the control phases ($\theta_{x,p}$) are pre-determined prior to the experiment. Inset plots the individual variances, and shares the same x- and y-axes as the main plot. Dashed line plots SQL*, which is equal to $2/M$. The non-adaptive variance is below SQL* at $M=128$ measurements. The variance is further reduced by using an adaptive protocol (black star), where the control phases are optimised in real-time.
    \textbf{(b)} 
    Zoom-in examination at $M=128$, where adaptive QPE outperforms both SQL* and the non-adaptive measurement. 
    Error bars correspond to one standard deviation calculated from quantum projection noise.
    }
    \label{fig:grid_state_qpe}
\end{figure}

In the second experiment, we perform multi-parameter displacement sensing with a grid state to find estimates $\{\tilde{\epsilon}_{x}, \tilde{\epsilon}_{p}\}$ of a random signal $\{\epsilon_{x}, \epsilon_{p}\}$. The estimation procedure uses Bayesian inference as follows~\cite{Wiebe2016}. Starting from a prior distribution for $\{\epsilon_{x}, \epsilon_{p}\}$, the QPE circuit of Fig.~\ref{fig:states_and_circuit}e is performed once with $N_S = 1$ to give two measurement outcomes, $\mathrm{m}_{x}$ and $\mathrm{m}_{p}$, whose probability distributions follow Eq.~\ref{eq:prob_distribution_a} and Eq.~\ref{eq:prob_distribution_b}. A posterior distribution is then calculated from both the prior and the new measurement outcomes using Bayes' theorem. 
We determine $\{\tilde{\epsilon}_{x}, \tilde{\epsilon}_{p}\}$ by maximizing the posterior after several measurement iterations where the posterior distribution converges. 

The results of this Bayesian QPE are plotted in Fig.~\ref{fig:grid_state_qpe}. We measure the Holevo variance, defined as $\mathrm{V}_\mathrm{H}(\tilde{\epsilon}_{x,p}) = (|\langle e^{i l_s \tilde{\epsilon}_{x,p}} \rangle|)^{-2} - 1$~\cite{Bonato2015}, which we average over many randomly sampled signals $\{\epsilon_{x}, \epsilon_{p}\}$ and vary the number of measurement repetitions, $M$. We first perform non-adaptive QPE, where the phases of the ancilla rotation at the $m$th iteration are set to $\theta_{x, m} = \theta_{p, m} = \pi m/M$. We observe that $\mathrm{V}_\mathrm{H}(\tilde{\epsilon}_p)$ (purple circles) is larger than $\mathrm{V}_\mathrm{H}(\tilde{\epsilon}_x)$ (blue circles), which is due to $\epsilon_p$ being measured after $\epsilon_x$ and therefore suffers more decoherence. As there are no theoretical restrictions on the ordering of the measurement operations, one could alternate between first measuring $\epsilon_x$ or $\epsilon_p$ to balance their variances. The total variance (red circles) decreases with $M$ as expected and falls below SQL* at $M=128$. To further reduce the variance, we perform adaptive QPE where the phases $\theta_{x, m}$ and $\theta_{p, m}$ are optimised in real-time before each $m$th measurement iteration~\cite{Berry2000, Huang2017}. This protocol maximises the information gained, and we achieve a combined variance 2.6(1.1)~dB below SQL*. Variances at $M<128$ obtained from non-adaptive QPE are above SQL* due to small-sample size effects, and the variance is expected to reach the Cramer-Rao bound as $M$ increases~\cite{Huang2017}. 

\begin{figure}[h!]
    \centering
    \includegraphics[width=7cm]{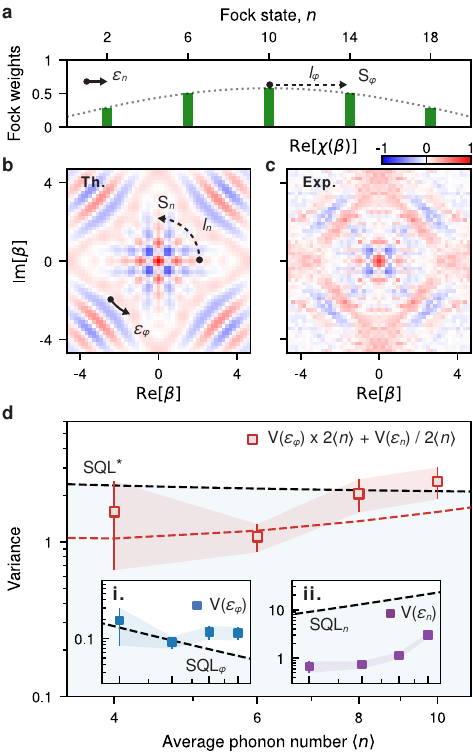}
    \caption{
    \textbf{Metrological gain of number--phase states for multi-parameter number and phase sensing.}
    \textbf{(a, b)} Theoretical Fock distribution and characteristic function of a number--phase (NP) state with spacing $N=4$, Fock cutoff $F=18$ and offset of 2 (lowest occupied Fock state is $\ket{n = 2}$). The energy of this state is constrained by damping the Fock coefficients with a sine envelope (dotted grey line), giving $\langle \hat n \rangle = 10$. The NP state has exact rotational phase-space symmetry of $l_n = 2\pi/4$ and approximate translational Fock symmetry of $l_\phi=4$, making it an exact eigenstate of $\hat{S}_n$ and an approximate eigenstate of $\hat{S}_\phi$. This spacing in phase and number can be used to simultaneously sense small unknown rotations by $\epsilon_\phi$ in phase space and shifts by $\epsilon_n$ in Fock space. 
    \textbf{(c)} Experimentally reconstructed characteristic function of the above NP state. 
    \textbf{(d)} The variances of number and phase are obtained from the classical Fisher information extracted from reconstructed probability distributions. Dashed red line is the expected multi-parameter variance of the target NP state. Insets (i., ii.) plot the individual variances for number and phase, with $\mathrm{SQL}_n = 2 \langle \hat{n}\rangle + 1$ and $\mathrm{SQL}_\phi = (2 \langle \hat{n}\rangle)^{-1} + 3 (8 \langle \hat{n}\rangle^2)^{-1}$, which correspond to a coherent state subjected to heterodyne measurements (see SM). SQL* is the sum of SQL for number and phase after rescaling by $(2\langle \hat{n} \rangle)^{-1}$ and $2 \langle \hat{n} \rangle$, respectively, such that both variances are equal to 1 and $\text{SQL*}=2$ at large $\langle n \rangle$. The combined variances (red squares) are the total of $\mathrm{V}(\epsilon_n)$ and $\mathrm{V}(\epsilon_\phi)$ after appropriately rescaling the covariance matrix.
    Error bars correspond to one standard deviation calculated from quantum projection noise.
    }
    \label{fig:np_state_metro}
\end{figure}

In a separate experiment, we investigate the potential of further leveraging backaction evasion by repeatedly measuring $\hat{S}_x$ and $\hat{S}_p$ for $N_S$ times within a single circuit iteration. We derive the joint probability distribution with arbitrary $N_S$ and find that it can be conveniently described by only a few points of the characteristic function lying on the lattice. Moreover, effects from backaction due to finite-energy grid states can be incorporated into the estimation analysis, allowing us to straightforwardly perform the same non-adaptive Bayesian estimation detailed above. With $M=32$, the combined variance is reduced by $\SI{1.4}{dB}$ from $N_S=1$ to $N_S=2$ (see circuit of Fig.~\ref{fig:states_and_circuit}e). No discernible improvement in metrological gain is observed for $N_S=3$, which we attribute primarily to the dephasing of the sensing state, evidenced by independent measurements of the lifetimes of $\hat{S}_x$ and $\hat{S}_p$ (see SM). The variance could be further reduced by minimizing backactions using finite-energy $C\hat{S}_x$ and $C\hat{S}_p$ operators tailored to the physical grid states~\cite{Royer2020, Matsos2024arxiv}. Repeating the experiment in Fig.~\ref{fig:grid_state_metro} with finite-energy operators gives an improved metrological gain of $\SI{6.0(5)}{dB}$ (see SM). 

Moving beyond position--momentum, we investigate simultaneous multi-parameter estimations of \textit{number} and \textit{phase}, which adhere to the uncertainty relation $\Delta n \Delta \phi \geq \hbar / 2$, where $(\Delta \phi)^2$ is the Holevo phase variance~\cite{holevo1979}. The corresponding sensing states are referred to as \textit{number--phase} (NP) states, $\ket{\npstate}$. In analogy to grid states, NP states are simultaneous eigenstates of two shift operators: the \textit{number} operator, $\hat{S}_n = e^{- i l_n \hat{n}}$, which is used to estimate shifts in number, and the \textit{phase} operator~\cite{Grimsmo2020}, $\hat{S}_\phi = \hat{E}^-_{l_\phi}$, which is used to estimate shifts in phase. The former is a rotation, and the latter is a phonon-shift operator defined as $\hat{E}^-_{l_\phi} = (\hat{E}^-)^{l_\phi}$, where $\hat{E}^- = \sum^{\infty}_{n=0} \ket{n} \bra{n+1}$ is the Susskind-Glogower phase operator, which lowers phonon number by one in an unweighted way (independent of the Fock state)~\cite{Susskind1964}. 
$\hat{S}_n$ and $\hat{S}_\phi$ commute by setting $l_n = 2\pi/N$ and $l_\phi = N$, where the positive integer $N$ can be freely chosen. The resulting ideal and unnormalised eigenstates, $\ket{\npstate} = \sum^{\infty}_{k=0} \ket{kN}$, are periodic in Fock space with a spacing given by $N$ and have a $2\pi/N$ rotational symmetry in phase space; see Fig.~\ref{fig:np_state_metro}(a-b). NP states can simultaneously sense a continuous rotation and a discrete shift in number: given an unknown rotation and phonon shift described by $\ket{\npstate_{\shiftparam}} = e^{i \epsilon_\phi \hat{n}} \hat{E}^+_{\epsilon_n} \ket{\npstate}$, the parameters $\epsilon_n$ and $\epsilon_\phi$ can be estimated from the eigenvalues of $\hat{S}_n$ and $\hat{S}_\phi$, respectively. Applying the number and phase operators results in $\hat{S}_n \ket{\npstate_{\shiftparam}} = e^{- i \epsilon_n l_n} \ket{\npstate_{\shiftparam}}$ and $\hat{S}_\phi \ket{\npstate_{\shiftparam}} = e^{- i \epsilon_\phi l_\phi} \ket{\npstate_{\shiftparam}}$, allowing one to estimate $\epsilon_n~\mathrm{mod}~N$ and $\epsilon_\phi~\mathrm{mod}~2\pi/N$.

Experiments with NP states follow the general structure shown in Fig.~\ref{fig:states_and_circuit}e by replacing the sensing states and associated measurement operators. We apply numerically optimised phase-modulated $\hat{H}_\mathrm{SDF}$ to create finite-energy NP states, $\ket{\tilde{\npstate}}$, which are made physical by applying a sinusoidal damping envelope with a finite cutoff to the Fock state probability distribution of idealised NP states. We verify the prepared NP states by reconstructing the characteristic functions and find good agreement with theory; see Fig.~\ref{fig:np_state_metro}b,c and SM for the data.

QPE routines for NP states are analogously performed by first applying a conditional number or phase operator and then measuring the ancilla. The conditional operations are implemented by driving ``blue-sideband'' (BSB) interactions (details in the SM) described by the Hamiltonian
\begin{align}
    & \hat{H}_{\mathrm{b}, k} = \frac{\Omega_{\mathrm{b}, k}}{2} \hat{\sigma}^+ (\hat{a}^\dagger)^k e^{- i \delta_{\mathrm{b}, k} t} e^{- i \phi_{\mathrm{b},k}} + \mathrm{h.c.},
\end{align}
where $k=1$ ($k=2$) gives the first (second) order interaction. A conditional number operator, $C \hat{S}_n = e^{- i l_n \hat{\sigma}_z \hat{n} / 2}$, is obtained by setting $\phi_{\mathrm{b}, k} = 0$ and $\delta_{\mathrm{b}, k} \gg \Omega_{\mathrm{b}, k}$, where $\phi_{\mathrm{b}, k}$, $\delta_{\mathrm{b}, k}$, and $\Omega_{\mathrm{b}, k}$ are the phase, detuning, and Rabi rate of the interaction. A weaker second-order interaction with $\Omega_{\mathrm{b}, 2} \ll \Omega_{\mathrm{b}, 1}$ is applied to counter-act parasitic interactions from the first-order field that would otherwise introduce errors.

The controlled phase operator, $C\hat{S}_\phi$, is implemented by alternately applying $\hat{H}_{\mathrm{b}, 1}$ and a qubit rotation~\cite{Um2016}. We set $\hat{H}_{\mathrm{b}, 2} = 0$ and $\delta_{\mathrm{b}, 1}=0$, and $\phi_{\mathrm{b}, 1}(t)$ is dynamically modulated with a numerically optimised waveform such that applying $\hat{H}_{\mathrm{b}, 1}$ enacts the transition $\ket{\downarrow, n} \rightarrow \ket{\uparrow, n+1}$ with coupling strength independent of $n$. Applying a $\hat{\sigma}_x$ $\pi$-pulse then returns the qubit to its original state, $\ket{\uparrow, n + 1} \rightarrow \ket{\downarrow, n + 1}$. A single iteration of BSB-$\hat{\sigma}_x$ sequence gives a Susskind-Glogower phase operator conditioned on the ancilla~\cite{Susskind1964}: the $\ket{\downarrow}$ state results in an upwards shift operator, $\ket{\downarrow}\bra{\downarrow} \hat{E}^+$, while the $\ket{\uparrow}$ state results in a downwards shift operator, $\ket{\uparrow}\bra{\uparrow} \hat{E}^-$. The overall $C\hat{S}_\phi$ operator is then implemented by applying the BSB-$\hat{\sigma}_x$ sequence $l_{\phi}/2$ times, giving $C \hat{S}_\phi = \ket{\downarrow}\bra{\downarrow}\hat{E}^+_{l_\phi/2} + \ket{\uparrow}\bra{\uparrow}\hat{E}^-_{l_\phi/2} $. Applying $\hat{E}^-$ to states in $\ket{n=0}$ causes unwanted rotations of the ancilla, which we avoid by using NP states with Fock state populations that are shifted by $\hat{E}^+_{l_{\phi}/2}$.

The choice of spacing, $N$, of the target NP state is subject to several tradeoffs. First, the theoretical variances of number and phase scale with $N$ as $\mathrm{V}(\epsilon_n) \propto N^2$ and $\mathrm{V}(\epsilon_\phi) \propto 1/N^2$. Second, increasing $N$ reduces the visibility parameter of the phase operator, $\eta_\phi = \langle \hat{E}^-_{l_\phi}\rangle$, which quantifies the metrological performance in the QPE algorithm and should ideally be 1. In contrast, the visibility parameter associated with the number operator is independent of $N$ and is $\eta_n = 1$, since NP states are exact eigenstates of $\hat{S}_n$. Third, the choice of $N$ influences the quality of the experimental implementations of number and phase operators. Large $N$ improves the quality of the controlled number operator, as the target phase parametrized by $l_n$ becomes smaller, thereby reducing higher-order terms and enabling shorter pulse duration. However, large $N$ degrades the quality of the controlled phase operator, requiring more applications of carrier and BSB pulses which lengthens the duration of the control. Overall, we empirically find that $N=4$ strikes a good balance, offering a favorable metrological gain in both number and phase while ensuring a sufficiently high quality of experimental controls.

The metrological gain of the NP states is characterised in a similar manner as grid states using the QPE circuit of Fig.~\ref{fig:states_and_circuit}e with the following changes. After preparation, the NP state is subjected to phonon shifts and rotations. The shifts are experimentally implemented by initializing the qubit in the $\ket{\downarrow}$ state and repeating $\hat{E}^+$ for $\epsilon_n$ times. A rotation $e^{i \epsilon_\phi \hat{n}}$ is then applied by offsetting the phases of all subsequent pulses in software, such that the blue-sideband phase of $\hat{H}_\mathrm{b}$ becomes $\phi_\mathrm{b} \rightarrow \phi_\mathrm{b} + \epsilon_\phi$. Measurements of the conditional operators require a Hadamard rotation applied to the ancilla before and after $C\hat{S}_n$ and $C\hat{S}_\phi$, as the operators act in the $\hat{\sigma}_z$ basis, and the ancillary rotations are also performed in the $\hat{\sigma}_z$ basis. 

The variances of number and phase, $\mathrm{V}(\epsilon_\phi)$ and $\mathrm{V}(\epsilon_n)$, are plotted in Fig.~\ref{fig:np_state_metro}d for increasing $\langle \hat{n} \rangle$. We first observe that $\mathrm{V}(\epsilon_\phi)$ (blue squares) decreases up to $\langle \hat{n} \rangle =6$, where it follows the SQL which scales with $1/\langle \hat{n}\rangle$. The variance $\mathrm{V}(\epsilon_n)$ (purple squares) remains significantly below its SQL, which scales as $\langle \hat{n} \rangle$. We also observe that the variance $\mathrm{V}(\epsilon_n)$ is much lower than its respective SQL compared to $\mathrm{V}(\epsilon_\phi)$; this is expected, since the experimentally prepared NP states are only approximate eigenstates of the phase operator, but are exact eigenstates of the number operator.  We further investigate the metrological gain for simultaneous estimation of number and phase by summing their variances, after rescaling $\mathrm{V}(\epsilon_\phi)$ and $\mathrm{V}(\epsilon_n)$ such that their respective SQLs are equal to 1 (Fig.~\ref{fig:np_state_metro}d). The smallest combined variance is obtained for $\langle \hat n \rangle = 6$, giving a gain of $\SI{3.1(9)}{dB}$ over the simultaneous SQL. 

\section{Discussion}

In summary, we demonstrated multi-parameter sensing of incompatible observables using a single mode with estimated parameter uncertainties simultaneously reduced below the SQL. This was achieved by a backaction-evading measurement scheme combined with a versatile, high-fidelity quantum control protocol that allows the preparation of highly non-classical states tailored for measuring small displacements. We prepared grid states and NP states to simultaneously measure position--momentum and number--phase, respectively. We achieved a metrological gain with the grid states of $\SI{5.1(5)}{dB}$ over the SQL, and a gain of $\SI{3.1(9)}{dB}$ for the NP states. In addition, we implemented a multi-parameter quantum phase estimation algorithm to estimate changes in the position and momentum caused by random displacements. The combined variance was reduced below the simultaneous SQL after a sufficient number of measurements, and the variance was further reduced by adaptively varying controllable phases during the experiment. 

In addition, this work constitutes the first experimental realization and control of number--phase states. A polar decomposition into number and phase is a natural representation for devices where intrinsic noise is dominated by dephasing, making these states prime candidates for novel metrology in the presence of noise. For example, NP states can be employed in error-corrected quantum metrology, where the precision of estimating a parameter is enhanced by protecting against noise in the other parameter~\cite{Kessler2014,ouyang2021tradeoffs}. Beyond quantum metrology, these states serve as code states of rotation symmetric codes for error-corrected quantum information processing~\cite{Grimsmo2020, leviant2022quantum}. Measurements of the NP states' stabiliser lifetimes reveal a strong bias (see SM), suggesting a favorable QEC performance~\cite{puri2020biased,Putterman2024}.  Opportunities exist to enhance the quality of NP states and their measurements. First, the required interactions for preparation and measurement could potentially be improved through light-atom Hamiltonian engineering and optimal control strategies. Second, better designs of NP states, such as Fock spacings and envelopes, could improve performance in the presence of specific noises in devices.

Looking forward, our multi-parameter displacement sensor can be used for phase-insensitive force sensing. The adaptive-QPE results of Fig.~\ref{fig:grid_state_qpe} give an equivalent phase-insensitive force sensitivity of $\SI{14.3}{yN/\sqrt{Hz}}$ ($\SI{1}{yN} = \SI{1e-24}{N}$), which is comparable to state-of-the-art sensors that rely on prior phase synchronizations between the force and the sensor~\cite{Affolter2020, Gilmore2021} (see SM). Our scheme is well suited for quantum-logic-enabled photon-recoil spectroscopy~\cite{Hempel2013,Wan2014} to drive narrow linewidth transitions in molecular ions~\cite{Wolf2016,Chou2017,Sinhal2020} and highly-charged ions~\cite{Kozlov2018,Micke2020}, which can benefit atomic clocks~\cite{King2022} and the search for new physics~\cite{Safronova2018}. Another exciting aspect is the further reduction of our force sensitivity by several orders of magnitude by increasing the sensing duration and employing larger ion crystals, which could benefit the search for dark matter~\cite{Gilmore2021,Budker2022}. Our demonstration is also compatible with photonic-linked trapped-ion devices~\cite{Nichol2022}, providing an exciting prospect to use an entangled network of sensors to perform long-baseline quantum-enhanced sensing~\cite{Brady2022}.

\section{Material and Methods}

\subsection{Experimental control toolbox}

The experiments in this work use a combination of red-sideband (RSB) and blue-sideband (BSB) interactions. These are obtained from stimulated Raman transitions with a 355~nm pulsed laser. A bichromatic field with a frequency difference of $\omega_0 - \omega_m + \delta_\mathrm{r}$ or $\omega_0 + \omega_m + \delta_\mathrm{b}$ gives an RSB or a BSB interaction, respectively, where $\omega_0$ is the qubit frequency and $\omega_m$ is the motional frequency. The frequencies $\delta_\mathrm{r}$ and $\delta_\mathrm{b}$ are the detunings from the RSB and BSB interactions, respectively. In an interaction picture with respect to the ion's spin and motion, and after several rotating wave approximations, the RSB and BSB Hamiltonians are~\cite{Wineland1998}
\begin{align}
    & \hat{H}_\mathrm{r} = \frac{\Omega_\mathrm{r}}{2} \hat{\sigma}^+ \hat{a} e^{- i \delta_\mathrm{r}} e^{ i \phi_\mathrm{r}} + \mathrm{h.c.}, \label{eq:general_rsb}\\
    & \hat{H}_\mathrm{b} = \frac{\Omega_\mathrm{b}}{2} \hat{\sigma}^+ \hat{a}^\dagger e^{- i \delta_\mathrm{b}} e^{ i \phi_\mathrm{b}} + \mathrm{h.c.}, \label{eq:general_bsb}
\end{align}
where $\hat{\sigma}^+ = \ket{\uparrow}\bra{\downarrow}$ is the Pauli raising operator, $\phi_r$ and $\phi_b$ are controllable red- and blue-sideband phases. A state-dependent force (SDF) is obtained by simultaneously applying $\hat{H}_\mathrm{r}$ and $\hat{H}_\mathrm{b}$ with $\delta_\mathrm{r} = \delta_\mathrm{b} = 0$ and $\Omega = \Omega_\mathrm{r} = \Omega_\mathrm{b}$, which results in
\begin{equation}
    \label{eq:general_H_sdf}
    \hat{H}_\mathrm{SDF} = \frac{\Omega}{2} \hat{\sigma}_{\phi_s} (\hat{a}^\dagger e^{- i \phi_m} + \hat{a} e^{i \phi_m}),
\end{equation}
where $\sigma_{\phi_s} = (\hat{\sigma}_x \cos(\phi_s) + \hat{\sigma}_y \sin(\phi_s))$, $\phi_s = (\phi_r + \phi_b)/2$ is a phase associated with the spin and $\phi_m = (\phi_r - \phi_b)/2$ is a phase associated with the motion. 
Applying $\hat{H}_\mathrm{SDF}$ for a duration $t$ gives a displacement conditioned on the state of the spin, $\hat{D}(\hat{\sigma}_{\phi_s} \alpha)$, whose operation can be written as
\begin{equation}
     \hat{D}(\hat{\sigma}_{\phi_s} \alpha) = \ket{+}\bra{+} \otimes \hat{D}(\alpha) + \ket{-} \bra{-} \otimes \hat{D}^\dagger(\alpha),
\end{equation}
where $\ket{\pm}$ are the eigenstates of $\hat{\sigma}_{\phi_s}$ and $\alpha = - i \Omega t e^{- i \phi_m}/2$. The magnitude of the displacement, $|\alpha|$, is set by varying the duration $t$ for a given Rabi frequency, $\Omega$. The phase of the displacement, $\mathrm{arg}(\alpha)$, is set by the motional phase, $\phi_m$, which can be controlled by adjusting the phase of the radio-frequency driving the acousto-optic modulator (AOM) in the path of the Raman beam.

\subsection{Preparing sensing states}
\label{sec:general_state_prep}

The sensing states in this work are prepared via optimal control by applying $\hat{H}_\mathrm{SDF}$ of Eq.~\ref{eq:general_H_sdf} with dynamically modulated phases $\phi_s(t)$ and $\phi_m(t)$~\cite{Matsos2024}. The modulation waveforms are obtained through a numerical optimisation procedure using Q-CTRL's graph based optimiser (Boulder Opal)~\cite{Ball2021, boulder_opal}. The numerical optimiser minimises the cost function,
\begin{equation}
    \label{eq:cost_function}
    \mathcal{C} = (1 - \mathcal{F}) + \epsilon \frac{T}{T_\mathrm{max}}, 
\end{equation}
where $\mathcal{F}$ is a fidelity, $T$ is the duration of the pulse and $T_\mathrm{max}$ is the maximum allowed duration. In this way, the duration of the pulse is minimised while the convergence criterion $\epsilon$ ensures that $1 - \mathcal{F} \lesssim \epsilon$. We define the fidelity as the overlap between the generated state and the target state $\ket{\psi_\mathrm{t}}$ (see Supplemental Material for definitions of the grid and number--phase states), $\mathcal{F} = |\bra{\downarrow, \psi_\mathrm{t}} \hat{U} \ket{\downarrow, 0}|^2$, with $\hat{U} = \mathrm{exp}(- i \int^T_0 dt \hat{H}_\mathrm{SDF}(t))$.

The phase modulations of $\phi_s(t)$ and $\phi_m(t)$ are represented as a piece-wise constant signal with $N_\mathrm{opt}$ optimizable segments. We add several constraints to the segments to avoid signal line distortions from the finite-bandwidth of the AOMs. First, the slew rate $|\phi_{i+1} - \phi_{i}| < \mathrm{SR}$ limits the maximum difference of the values of two consecutive segments. Second, the optimizable segments are filtered with a sinc filter, and $N_\mathrm{seg}$ segments are resampled from the filtered signal. We set $N_\mathrm{opt} = 50$ ($N_\mathrm{opt} = 30$) for the grid states (number--phase states), and resample $N_\mathrm{seg} = 150$ segments for all states. The slew rate is set to $\mathrm{SR} = 1.5$, and the cutoff frequency of the sinc filter is $f_c \times T = 2\pi \times 15$.

\subsection{Dephasing in the motional degrees of freedom}

The two common sources of decoherence in the motional modes of trapped ions are heating and dephasing (see Ref.~\cite{Wineland1998} for a detailed discussion). In our experiment, we measure a heating rate of 0.2~phonons/s, which is negligible given the typical duration of an experiment. The dominant source of noise is therefore motional dephasing, which we describe in greater detail below.

Motional dephasing, modeled by the noisy Hamiltonian term $\nu(t) \hat{a}^\dagger \hat{a}$ with stochastic $\nu(t)$, describes random fluctuations of the motional mode's frequency. 
In trapped ions, these fluctuations arise from noise in the confining potential which, for radial modes of motion, originates from noise in the amplitude of the trapping RF signal due to thermal fluctuations in the resonator and electronic noise in the controller hardware. In our setup, the amplitude of the trapping RF voltage is actively stabilised by a closed-loop feedback circuit. The amplitude is first measured from the RF signal that is capacitively coupled to opposing blades that are radially symmetric w.r.t to the ion. The amplitude is then stabilised by a custom-built analog PID, whose output regulates a variable voltage attenuator in the path of the RF trapping voltage connected to the trap. With this, we measure a motional coherence time of $T_2^* \approx \SI{50}{ms}$ from a Ramsey sequence between Fock states $\ket{0}$ and $\ket{1}$~\cite{Matsos2024}. Increasing the coherence time is ongoing work, with potential hardware improvements including better thermal stabilisation and lower electronic noise.

\section*{Acknowledgements}

We thank H.~Wiseman, B.~Terhal, R.~Blatt for fruitful discussions. This work was supported by the U.S. Office of Naval Research Global (N62909-24-1-2083), the U.S. Army Research Office Laboratory for Physical Sciences (W911NF-21-1-0003), the U.S. Air Force Office of Scientific Research (FA2386-23-1-4062), the Australian Research Council (FT220100359, FT230100571, DE230100144, CE170100009, CE170100012), Lockheed Martin, EPSRC Quantum Engineering Centre for Doctoral Training EP/S023607/1 and European Commission PEQEM ERC-2018-STG803665 (MPS), the Sydney Quantum Academy (MJM), the University of Sydney Postgraduate Award scholarship (VGM), the Australian Research Council, and H.\ and A.\ Harley.

\section*{Data availability}

All data needed to evaluate the conclusions in the paper are present in the paper; the data are also available in an online repository (\url{https://doi.org/10.5281/zenodo.14512033}).

\bibliography{bib.bib}

\end{document}


\title{Supplemental Material for: ``Quantum-Enhanced Multi-Parameter Sensing \linebreak in a Single Mode''}

\author{Christophe~H.~Valahu}
\email{christophe.valahu@sydney.edu.au}
\affiliation{School of Physics, University of Sydney, NSW 2006, Australia}
\affiliation{ARC Centre of Excellence for Engineered Quantum Systems, University of Sydney, NSW 2006, Australia}
\affiliation{Sydney Nano Institute, University of Sydney, NSW 2006, Australia}

\author{Matthew~P.~Stafford}
\affiliation{Quantum Engineering Technology Labs, H. H. Wills Physics Laboratory and Department of Electrical and Electronic Engineering, University of Bristol, Bristol, UK.}
\affiliation{Quantum Engineering Centre for Doctoral Training, University of Bristol, Bristol, UK.}

\author{Zixin~Huang}
\affiliation{School of Mathematical and
Physical Sciences, Macquarie University, NSW 2109, Australia}
\affiliation{Centre for Quantum Software and Information, Faculty of Engineering and
Information Technology, University of Technology Sydney, NSW 2007, Australia}

\author{Vassili~G.~Matsos}
\affiliation{School of Physics, University of Sydney, NSW 2006, Australia}
\affiliation{ARC Centre of Excellence for Engineered Quantum Systems, University of Sydney, NSW 2006, Australia}

\author{Maverick~J.~Millican}
\affiliation{School of Physics, University of Sydney, NSW 2006, Australia}
\affiliation{ARC Centre of Excellence for Engineered Quantum Systems, University of Sydney, NSW 2006, Australia}

\author{Teerawat~Chalermpusitarak}
\affiliation{School of Physics, University of Sydney, NSW 2006, Australia}

\author{Nicolas~C.~Menicucci}
\affiliation{Centre for Quantum Computation and Communication Technology,
School of Science, RMIT University, VIC 3000, Australia}

\author{Joshua~Combes}
\affiliation{University of Melbourne, VIC 3052, Australia}

\author{Ben~Q.~Baragiola}
\affiliation{Centre for Quantum Computation and Communication Technology,
School of Science, RMIT University, VIC 3000, Australia}

\author{Ting~Rei~Tan}
\email{tingrei.tan@sydney.edu.au}
\affiliation{School of Physics, University of Sydney, NSW 2006, Australia}
\affiliation{ARC Centre of Excellence for Engineered Quantum Systems, University of Sydney, NSW 2006, Australia}
\affiliation{Sydney Nano Institute, University of Sydney, NSW 2006, Australia}

\maketitle
\onecolumngrid

\tableofcontents
\newpage

\section{Notation}
\label{sec:notation}

The following conventions are used throughout the manuscript and the supplemental material. The position and momentum quadrature operators are
\begin{align}
    & \hat{x} := \frac{1}{\sqrt{2}}(\hat{a}^\dagger + \hat{a}), \\ 
    & \hat{p} := \frac{i}{\sqrt{2}}(\hat{a}^\dagger - \hat{a}),
\end{align}
where $\hat{a}^\dagger$ and $\hat{a}$ are the creation and annihilation operators, respectively. The displacement operator, $\hat{D}(\alpha)$, is related to the position and momentum operators via
\begin{align}
    \hat{D}(\alpha) := & ~\mathrm{exp}[\alpha \hat{a}^\dagger - \alpha^* \hat{a}] \nonumber \\
    = & ~\mathrm{exp}[i\sqrt{2}(\mathrm{Im}(\alpha) \hat{x} - \mathrm{Re}(\alpha) \hat{p})].
\end{align}
The quadrature squeezing operator, $\hat{S}(r)$, is given by
\begin{align}
    \hat{S}(r) &\coloneqq \exp{\left[r(\hat{a}^2 - {\hat{a}^{\dagger 2}}) /2\right]},
\end{align}
which squeezes the position quadrature by a factor of $\Delta \coloneqq e^{-r}$. 

The set of displacement operators $\hat{D}(\alpha)$ for a single mode form an operator basis. Therefore, any continuous-variable operator $\hat{O}$ can be expressed in terms of displacement operators, given by

\begin{equation}
    \hat{O} = \frac{1}{\pi}\int d^2\alpha \Tr\big[{\hat{O}\hat{D}^{\dagger}(\alpha)}\big] \hat{D}(\alpha).
\end{equation}
The characteristic function of $\hat{O}$, denoted $\bigchi_{\hat{O}}(\alpha)$, is defined by
\begin{equation}
    \bigchi_{\hat{O}}(\alpha) \coloneqq \Tr\big[\hat{O}\hat{D}^{\dagger}(\alpha)\big].
\end{equation}
The characteristic function of a state $\hat{\rho}$ has some additional properties which will be used throughout:
\begin{itemize}
    \item The value of the characteristic function at the origin is $\bigchi_{\hat{\rho}}(0) = \Tr\left[\hat{\rho}\right] = 1$.
    \item The characteristic function has symmetries in its real and imaginary parts. From $\hat{\rho} = \hat{\rho}^{\dagger}$, it follows that $\bigchi_{\hat{\rho}}^* (-\alpha) = \bigchi_{\hat{\rho}}(\alpha)$, hence, $\mathrm{Re}\big[\bigchi_{\hat{\rho}}(\alpha)\big] = \mathrm{Re}\big[\bigchi_{\hat{\rho}}(-\alpha)\big]$ and $\mathrm{Im}\big[\bigchi_{\hat{\rho}}(\alpha)\big] = -\mathrm{Im}\big[\bigchi_{\hat{\rho}}(-\alpha)\big]$.
    \item The characteristic function of a pure state, $\ket{\psi}$, is $\bigchi_{\hat{\rho}}(\alpha) = \braket{\psi | \hat{D}^{\dagger}(\alpha) | \psi}$.
\end{itemize}
A state $\hat{\rho}$ that is displaced by $\hat{D}(\beta)$ has the following useful property,
\begin{align}
    \bigchi_{\hat D \hat{\rho} \hat D^{\dagger}}(\alpha) &= \Tr\left[\hat{D}(\beta) \hat{\rho} \hat{D}^{\dagger}(\beta) \hat{D}^{\dagger}(\alpha) \right]\\
    &= e^{\alpha^*\beta - \alpha\beta^*} \bigchi_{\hat{\rho}}(\alpha).
\end{align}
In this way, the characteristic function of a displaced state is equal to the characteristic function of the original state up to a phase.

The Wigner function is finally obtained from the Fourier transform of the characteristic function,
\begin{equation}
    W_{\hat{\rho}}(\gamma) = \frac{1}{\pi^2} \int \bigchi_{\hat{\rho}}(\alpha) e^{\gamma \alpha^* - \gamma^* \alpha} d^2 \alpha.
\end{equation}

\section{Grid states}

\renewcommand{\arraystretch}{1.5}
\setlength\tabcolsep{8pt}
\begin{table}
    \begin{tabular}{lccccccccccc}
        \Xhline{1pt} 
        Target state & $\Delta_\mathrm{t}$ & 0.74 & 0.67 & 0.61 & 0.55 & 0.50 & 0.45 & 0.41 & 0.37 & 0.32 & 0.30\\
          & $\Delta_x$ & 0.77 & 0.69 & 0.62 & 0.56 & 0.50 & 0.45 & 0.41 & 0.37 & 0.32 & 0.30\\
         & $\Delta_p$ & 0.78 & 0.71 & 0.63 & 0.56 & 0.50 & 0.45 & 0.41 & 0.37 & 0.32 & 0.30\\
         & $\langle \hat{n} \rangle$ & 0.21 & 0.42 & 0.71 & 1.09 & 1.51 & 2.00 & 2.55 & 3.22 & 4.41 & 5.25 \\
        \hline
        \multirow{2}{*}{\shortstack{Optimised \\ pulse}} & $\Delta_x$ & 0.81 & 0.72 & 0.65 & 0.59 & 0.56 & 0.51 & 0.42 & 0.38 & 0.33 & 0.33\\
        & $\Delta_p$ & 0.82 & 0.72 & 0.63 & 0.58 & 0.53 & 0.53 & 0.42 & 0.38 & 0.33 & 0.33\\
        & $\langle \hat{n} \rangle $ & 0.14 & 0.36 & 0.67 & 1.01 & 1.28 & 1.62 & 2.47 & 3.14 & 4.28 & 4.68\\
        & Fidelity & 0.998 & 0.999 & 0.999 & 0.998 & 0.994 & 0.988 &  0.999 & 0.999 & 0.999 & 0.990 \\
        \hline 
        Experiment & $\Delta_x$ & - & - & 0.71 & - & - & - & - & 0.53 & - & 0.66 \\
         & $\Delta_p$ & - & - & 0.72 & - & - & - & - & 0.50 & - & 0.56 \\
         & $T$ / ms & 0.34 & 0.39 & 0.41 & 0.52 & 0.39 & 0.41 & 0.73 & 0.87 & 1.11 & 1.08 \\ 
        \Xhline{1pt} 
    \end{tabular}
    \caption{
    \textbf{Characteristics of finite-energy grid states prepared with optimal quantum control.}
    We target finite-energy grid states defined in Eq.~\ref{eq:finite_energy_grid_state} with a target squeezing parameter, $\Delta_t$. We first characterise the effective squeezing parameters of the target state along position and momentum, $\Delta_x$ and $\Delta_p$, calculated from Eq.~\ref{eq:delta_x_p}. We find that $\Delta_x \neq \Delta_p $ for low energy grid states with large $\Delta_t$ due to the asymmetry of the state. However, this discrepancy vanishes for better approximate grid states with smaller $\Delta_t$. We then report the characteristics of the grid states prepared from optimal control, which are obtained by numerically simulating the evolution of $\hat{H}_\mathrm{SDF}(t)$ of Eq.~7 with the optimised modulation waveforms. Along with the squeezing parameters $\Delta_x$ and $\Delta_p$, we report the average phonon number, $\langle \hat{n} \rangle$, and the fidelity. The fidelity is calculated from the overlap between the prepared state and the target state, as defined in Eq.~9. We finally report the experimentally measured squeezing parameters of three states, where $\Delta_x$ and $\Delta_p$ are computed from the reconstructed density matrices (see Fig.~\ref{fig:qunaught_chis_and_wigner}). The durations of the experimental pulses, $T$, are determined from the modulation waveforms of the numerically optimised pulses and the Rabi frequency of the SDF interaction is $\Omega/2\pi = \SI{2}{kHz}$.
    }
    \label{tab:grid_state_preparation}
\end{table}

\subsection{Theory}
\label{sec:grid_states_theory}

Ideal translation-symmetric states, $\ket{\gridstate}$, are joint eigenstates of the following shift operators,
\begin{subequations}
    \begin{alignat}{2}
        & \hat{S}_x = e^{- i l_x \hat{x} } &&= \hat{D}(- il_x / \sqrt{2}), \label{eq:Sx_shift_def}\\
        & \hat{S}_p = e^{-i l_p \hat{p} } &&= \hat{D}(l_p / \sqrt{2}),
    \end{alignat}
\end{subequations}
which shift $\hat{p}$ by $- l_x$ and $\hat{x}$ by $l_p$, respectively. Note that the shift operator $\hat{S}_x$ of Eq.~\ref{eq:Sx_shift_def} differs from the common convention in the literature by a minus sign~\cite{eczoo_gkp,Gottesman2001,Royer2020}; this choice is made such that the resulting probability distributions have the same functional form, without affecting the physics (see section~\ref{sec:grid_state_prob_distribution}). The \emph{modular lengths} $l_x$ and $l_p$ satisfy $l_x l_p =  2\pi d$ for positive integer $d$, ensuring that $\hat{S}_x$ and $\hat{S}_p$ commute. As such, $\ket{\gridstate}$ are eigenstates of modular position, $\hat{x}_{[2\pi/l_x]}$, and modular momentum, $\hat{p}_{[2\pi/l_p]}$, where a modular quadrature with respect to $m$ is defined as~\cite{Pantaleoni2020modular,Royer2020}
\begin{align}
    \hat{q}_{[m]}
    \coloneqq
    \hat{q}\text{ mod }m 
    = \sum_{k \in \mathbb{Z}} \int dx \, x \ket{x + k m}_{q} \bra{x + km}.
\end{align}
Ideal translation-symmetric states can be represented in the position and momentum bases as
\begin{align}
    \ket{\gridstate} = \sum_{k \in \mathbb{Z}} \ket{ k l_p }_x = \sum_{k \in \mathbb{Z}} \ket{ k l_x }_p
\end{align}
A prominent set of translation-symmetric states arising for $d>1$ are Gottesman-Kitaev-Preskill (GKP) states~\cite{Gottesman2001,brady2024review}, which have important applications in fault-tolerant quantum computation~\cite{fukui2018highthreshold,bourassa2021blueprint,noh2022lowoverhead,walshe2024totl}, quantum communication~\cite{fukui2021alloptical}, and more fundamental quantum information studies~\cite{takagi2024bridging,conrad2024thesis}. 

Translation-symmetric states with $d=1$ and symmetric modular lengths, $l_{x,p} = l_s = \sqrt{2\pi}$, are often referred to as sensor states~\cite{Duivenvoorden2017} or as qunaught states~\cite{Walshe2020}. Here and in the main text, we refer to them as \emph{grid states}. In addition to quantum sensing applications demonstrated here, grid states are also important for bosonic QEC, as they can be used to create logical GKP Bell states~\cite{Walshe2020} and have applications in multi-mode encodings~\cite{Royer2022}.

Ideal grid states are unphysical, as they have infinite energy. We experimentally target approximate grid states by constructing a superposition of displaced squeezed vacuum states, given by
\begin{equation}
    \label{eq:finite_energy_grid_state}
    \ket{\tilde{\gridstate}} = \frac{1}{\sqrt{N_{\Delta}}}\sum_{k \in \mathbb{Z}} e^{- \pi \Delta^2 |k|
    ^2} \hat{D} ( k \sqrt{\pi}) \hat{S}(r)\ket{0},
\end{equation}
where $N_{\Delta}$ is a normalisation constant. The single parameter $\Delta$ determines the quality of the state, with squeezing parameter given by $r = - \mathrm{log}{\Delta}$, such that a (normalised) ideal grid state is obtained in the limit $\Delta \rightarrow 0$. The position- and momentum-basis wavefunctions of the target state in Eq.~\ref{eq:finite_energy_grid_state} are given by
\begin{subequations}
\label{eq:wavefunctions_grid}
\begin{align}
    & \psi_{\tilde \gridstate} (s) 
    = {}_{\hat{x}}\langle s | \tilde \gridstate \rangle = 
    \sqrt{\frac{1}{N_\Delta \sqrt{\pi \Delta^2}}} \sum_{k \in \mathbb{Z}} e^{- \pi \Delta^2 k^2} e^{-\frac{(s - k\sqrt{2\pi})^2}{2\Delta^2}},\\
     & \tilde{\psi}_{\tilde \gridstate} (t) 
     = {}_{\hat{p}}\langle t | \tilde \gridstate \rangle
     = \sqrt{\frac{1}{N_\Delta \sqrt{\pi \Delta^2}}} \sum_{k \in \mathbb{Z}} e^{-\frac{\pi \Delta^2 k^2}{(1 + \Delta^4)}} e^{-\frac{1 + \Delta^4}{2\Delta^2}\left(t - \frac{k\sqrt{2\pi}}{(1 + \Delta^4)}\right)^2}.
\end{align}  
\end{subequations}
Each wavefunction is a superposition of equidistant identical Gaussians with an overall Gaussian envelope. Eq.~\ref{eq:wavefunctions_grid} reveals a slight asymmetry between the position and momentum wavefunctions, arising from the approximate definition of Eq.~\ref{eq:finite_energy_grid_state}. The Gaussian peaks in the position basis are ``on-grid'', meaning that they exactly lie on a grid of spacing $\sqrt{2\pi}$. In the momentum basis, the peaks are slightly ``off-grid,'' as they are scaled by $1/(1 + \Delta^4)$~\cite{Matsuura2020}, and further the variances of the individual peaks and of the envelope are also scaled by $1/(1 + \Delta^4)$. The asymmetry in position and momentum wavefunctions vanishes as the squeezing of the grid state increases ($\Delta \rightarrow 0$). However, for low energy states with large $\Delta$, the asymmetry can impact the metrological performance of the grid state for displacement sensing. Nevertheless, as discussed below, we find that this asymmetry has a negligible effect from the finite-energy states of this work.

The quality of an approximate grid state $\hat{\rho}$ can be characterised through the effective squeezing parameters~\cite{Duivenvoorden2017}
\begin{subequations}
    \label{eq:delta_x_p}
    \begin{align}
        & \Delta_x = \sqrt{\frac{1}{\pi}\mathrm{log}\left(\frac{1}{|\mathrm{Tr}(\hat{S}_x\hat{\rho})|^2}\right)}, \\
        & \Delta_p = \sqrt{\frac{1}{\pi}\mathrm{log}\left(\frac{1}{|\mathrm{Tr}(\hat{S}_p\hat{\rho})|^2}\right)},
    \end{align}
\end{subequations}
which are often expressed in units of dB using $\Delta_{x, p}~(\mathrm{dB}) = -10 \log_{10}(\Delta_{x, p}^2)$. The expectation values $\mathrm{Tr}(\hat{S}_{x}\hat{\rho})$ and $\mathrm{Tr}(\hat{S}_{p}\hat{\rho})$ in the denominators correspond to the characteristic function of the state evaluated at $\bigchi(i\sqrt{\pi})$ and $\bigchi(- \sqrt{\pi})$, respectively. For the target state of Eq.~\ref{eq:finite_energy_grid_state}, the characteristic function is 
\begin{multline}
    \label{eq:chi_func_grid}
    \bigchi_{\tilde{\gridstate}}(\beta) = \frac{1}{\bigchi_{\tilde{\gridstate}}(0)} e^{-\frac{1}{2\Delta^2}\big(\beta_R^2 + (1 + \Delta^4)\beta_I^2\big)}
    \Bigg( \vartheta
        \bigg[
    \setlength{\arraycolsep}{0.2em} 
    \begin{array}{c}
        0 \\ 0
    \end{array}
\bigg]
    \bigg(-\frac{i\beta_R}{\sqrt{\pi}\Delta^2}, \frac{2i(1+\Delta^4)}{\Delta^2}\bigg)
    \vartheta
    \bigg[
    \setlength{\arraycolsep}{0.2em} 
    \begin{array}{c}
        0 \\ 0
    \end{array}
    \bigg]
    \bigg(\frac{i\beta_I}{2\sqrt{\pi}\Delta^2}, \frac{i}{2\Delta^2}\bigg) \\
    +
    \vartheta
    \bigg[
    \setlength{\arraycolsep}{0.2em} 
    \begin{array}{c}
        \nicefrac{1}{2} \\ 0
    \end{array}
\bigg]
    \bigg(-\frac{i\beta_R}{\sqrt{\pi}\Delta^2}, \frac{2i(1+\Delta^4)}{\Delta^2}\bigg)
    \vartheta
    \bigg[
    \setlength{\arraycolsep}{0.2em} 
    \begin{array}{c}
        0 \\ \nicefrac{1}{2}
    \end{array}
\bigg]
    \bigg(\frac{i\beta_I}{2\sqrt{\pi}\Delta^2}, \frac{i}{2\Delta^2}\bigg)
    \Bigg),
\end{multline}
where $\beta_R = \mathrm{Re}[\beta]$ and $\beta_I = \mathrm{Im}[\beta]$, and the Jacobi theta functions with characteristics are defined as
\begin{align}
    \vartheta
    \bigg[
    \setlength{\arraycolsep}{0.2em} 
    \begin{array}{c}
        a \\ b
    \end{array}
\bigg]
    (z, \tau) & \coloneqq \sum_{n \in \mathbb{Z}} \exp \big[\left(\pi i \tau (n + a)^2 + 2 \pi i (z+b)(n+a)\right) \big].
\end{align}
The characteristic function is normalised such that $\bigchi_{\tilde{\gridstate}} (0) = 1$. 

We investigate the asymmetry of the approximate grid state by measuring the effective squeezing parameter from Eq.~\ref{eq:delta_x_p} using the characteristic function of Eq.~\ref{eq:chi_func_grid}. The effective squeezing parameters $\Delta_x$ and $\Delta_p$ are reported in Table~\ref{tab:grid_state_preparation}. We observe a slight difference in $\Delta_x$ and $\Delta_p$ for low quality states with a target squeezing parameter $\Delta_\mathrm{t} \geq 0.67$. However, for states where $\Delta_t < 0.67$, the asymmetry becomes negligible.

\subsection{State preparation}

\begin{figure}
    \centering
    \includegraphics[]{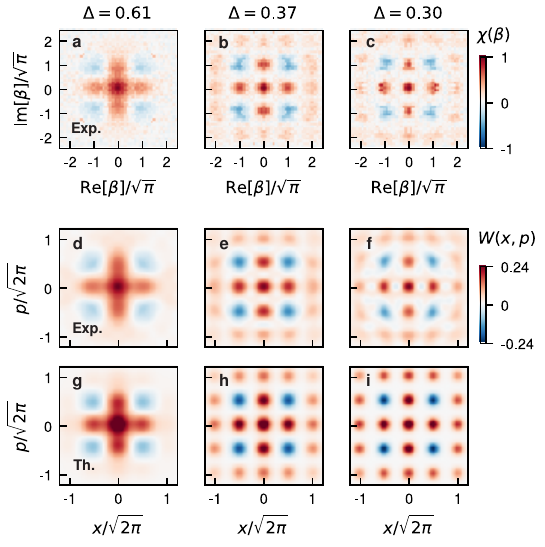}
    \caption{
    \textbf{Reconstructed characteristic and Wigner functions of grid states.}
    \textbf{(a-c)} Experimentally measured real part of the characteristic functions, $\mathrm{Re}[\bigchi(\beta)]$, of grid states with target squeezings \textbf{(a)} $\Delta_\mathrm{t} = 0.61$, \textbf{(b)} $\Delta_\mathrm{t} = 0.37$ and \textbf{(c)} $\Delta_\mathrm{t} = 0.30$. We directly reconstruct $\mathrm{Re}[\bigchi(\beta)]$ from measurements of the ancilla qubit after mapping information from the bosonic mode with state-dependent forces~\cite{Flhmann2020}. Only the top right quadrant is reconstructed, data in the remaining quadrants are obtained from symmetries of the characteristic function~\cite{Matsos2024}.
    \textbf{(d-f)} Wigner functions of the states calculated from the experimentally reconstructed Fock-basis density matrix. The density matrix is obtained from the characteristic function using a numerical optimisation procedure~\cite{Ahmed2021github, Ahmed2021, Strandberg2022} (see Ref.~\cite{Matsos2024} for detailed methods).
    Effective squeezing parameters computed from the reconstructed density matrices are shown in Table~\ref{tab:grid_state_preparation}.
    \textbf{(g-i)} Theoretical Wigner functions obtained from target states defined from Eq.~\ref{eq:finite_energy_grid_state}. Theoretical and experimental data show good agreement. The circular smearing in \textbf{(c,f)} is attributed to motional dephasing.
    }
    \label{fig:qunaught_chis_and_wigner}
\end{figure}

The grid states are experimentally prepared using a modulated spin-boson interaction. The target grid states are defined with Eq.~\ref{eq:finite_energy_grid_state}, and the target squeezing parameter is varied in the range $\Delta_\mathrm{t} \in [0.30, 0.74]$. The characteristics of the grid states obtained from the numerical optimiser for each $\Delta_\mathrm{t}$ are summarised in Table~\ref{tab:grid_state_preparation}.

We experimentally characterise the prepared grid states by reconstructing the real part of the characteristic functions, $\mathrm{Re}[\bigchi(\beta)]$, of three grid states with increasing quality parameterised by the target squeezings $\Delta_t = \{0.61, 0.37, 0.30\}$ (see Fig.~\ref{fig:qunaught_chis_and_wigner}(a-c))~\cite{Flhmann2020}. 
Imaginary parts of the characteristic functions are assumed to be zero due to the parity symmetry of grid states. The experimental Fock-basis density matrix $\hat{\rho}_\mathrm{exp}$ of each state is then obtained from $\mathrm{Re}[\bigchi(\beta)]$ using a numerical optimization procedure~\cite{Ahmed2021, Ahmed2021github, Strandberg2022, Matsos2024}. The Wigner functions calculated from the density matrices are also shown in Fig.~\ref{fig:qunaught_chis_and_wigner}(d-f), and shows good visual agreement with theory. The effective  squeezing parameters are calculated from $\hat{\rho}_\mathrm{exp}$ using Eq.~\ref{eq:delta_x_p} and are summarised in Table~\ref{tab:grid_state_preparation}. We confirm that the effective squeezing parameter decreases as the quality of the state improves from $\Delta_\mathrm{t} = 0.61$ to $\Delta_\mathrm{t} = 0.37$. However, the measured effective squeezing of the experimental state worsens from $\Delta_\mathrm{t} = 0.37$ to $\Delta_\mathrm{t} = 0.30$ even though the target squeezing decreased. This corroborates the results of Fig.~2 of the main text, where no additional metrological gain is observed at $\Delta_\mathrm{t} = 0.30$. We attribute this to motional dephasing which limits the achievable squeezing of the experimentally prepared grid state. The effects of motional dephasing are also observed in the reconstructed characteristic and Wigner functions of Fig.~\ref{fig:qunaught_chis_and_wigner}, where a circular smearing becomes apparent for states with smaller $\Delta_\mathrm{t}$.

\subsection{Conditional position and momentum operators}
\label{sec:conditional_x_p_stabilisers}

\begin{figure}
    \centering
    \includegraphics[]{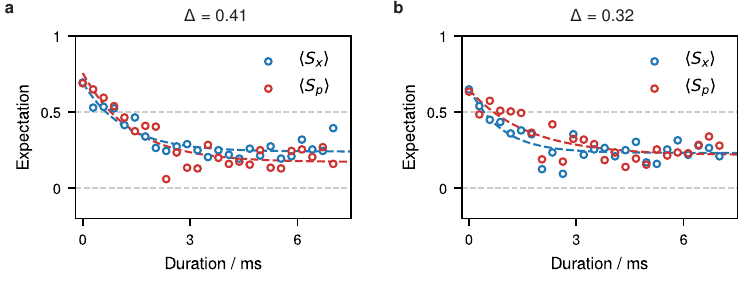}
    \caption{
    \textbf{Lifetime measurements of position and momentum stabilisers for grid states.}
    Lifetimes are measured for grid states with target squeezing parameters of \textbf{(a)} $\Delta_\mathrm{t} = 0.41$ and \textbf{(b)} $\Delta_\mathrm{t} = 0.32$. For both states, we measure expectation values $\langle \hat{S}_x\rangle$ (blue) and $\langle \hat{S}_p\rangle$ (red) of the position and momentum operators, respectively. Lifetimes are extracted from fits to the exponential decay, $f(t) = A e^{-T_\mathrm{wait}/T_1} + A_0$, where the amplitude $A$ and offset $A_0$ are set as free parameters to account for the approximate nature of the grid states. In \textbf{(a)}, we find $(T_1, A, A_0) = (1.2(3)~\mathrm{ms}, 0.45(4), 0.24(2))$ for measurements of $\langle \hat{S}_x \rangle$ and $(T_1, A, A_0) = (1.4(5)~\mathrm{ms}, 0.59(6), 0.17(3))$ for measurements of $\langle \hat{S}_p \rangle$. In \textbf{(b)}, we find $(T_1, A, A_0) = (0.9(3)~\mathrm{ms}, 0.43(4), 0.23(2))$ for measurements of $\langle \hat{S}_x \rangle$ and $(A, A_0) = (1.6(5)~\mathrm{ms}, 0.43(6), 0.22(3))$ for measurements of $\langle \hat{S}_p \rangle$.
    }
    \label{fig:grid_stabiliser_lifetime}
\end{figure}

The conditional position and momentum operators, $C\hat{S}_x$ and $C\hat{S}_p$, are experimentally implemented by applying the SDF Hamiltonian of Eq.~7 with $(\phi_s, \phi_m) = (0, 0)$ and $(\pi, \pi/2)$, respectively. Setting the duration to $t = \sqrt{\pi}/\Omega$ results in the operations $C\hat{S}_x = e^{- i \hat{\sigma}_x \hat{x} l_s/2}$ and $C\hat{S}_p = e^{- i \hat{\sigma}_x \hat{p} l_s/2}$. The Rabi frequency is calibrated to be $\Omega/2\pi = \SI{2}{kHz}$, corresponding to $t = \SI{141}{\micro s}$.

We further characterise the conditional operators by measuring their lifetime. We prepare the initial state $\ket{\downarrow} \otimes \ket{\tilde{\gridstate}}$, wait for a duration $T_\mathrm{wait}$ and finally measure the expectation value of the position or momentum operators. The measurement is performed by applying one of the conditional operators, $C\hat{S}_x$ or $C\hat{S}_p$, and measuring the state of the ancillary qubit in the $\hat{\sigma}_z$ basis. The measured expectation values, shown in Fig.~\ref{fig:grid_stabiliser_lifetime}, are then fit to an exponential decay from which we extract a lifetime, $T_x$ or $T_p$. For a grid state with a target squeezing parameter $\Delta_\mathrm{t} = 0.41$, the fitted lifetimes $(T_{x}, T_{x})$ are $(1.2(3), 1.4(5))~\mathrm{ms}$. For a grid state with a target squeezing of $\Delta_\mathrm{t}=0.32$, the lifetimes are $(0.9(3), 1.6(5))~\mathrm{ms}$. We first observe that, for each grid state, the lifetimes associated with $\hat{S}_x$ and $\hat{S}_p$ are not significantly different, which is expected from the symmetry of these states. We then observe that the lifetimes between the two states with different target squeezing parameters are also not significantly different.

\section{Number--phase states}
    \label{sec:num_phase_states}

\begin{table}
    \caption{
    \textbf{Specifications and characterisations of prepared number--phase states.}
    All states have a spacing of $N=4$ and an offset of $\lambda=2$. Number--phase states are prepared using optimal quantum control with a dynamically modulated light-atom interaction. Each target state is defined from a target average phonon number, $\langle \hat{n} \rangle_\mathrm{t}$. For sine states, the average phonon number is varied by setting the Fock cutoff, $F$. For airy states, the average phonon number is varied by setting the parameter $\mu$ of Eq.~\ref{eq:num_phase_airy_coeffs}. We also report the duration of the resulting modulation waveform and the state fidelity of the control pulse with the target state. Durations are calculated with the experimentally calibrated Rabi frequency of $\Omega/2\pi = \SI{2}{kHz}$.
    }
    \vspace{10pt}
    \centering
    \begin{tabular}{cccccccccc}
        \hline\hline 
        & \multicolumn{4}{c}{Sine state} & & \multicolumn{4}{c}{Airy state} \\
        \cline{2-5}\cline{7-10} 
        $\langle \hat{n} \rangle_\mathrm{t}$ & 4 & 6 & 8 & 10 &  & 4 & 6 & 8 & 10\\
        \hline        
        $F$ & 4 & 8 & 12 & 16 & & - & - & - & - \\
        $\mu$ & - & - & - & - & & 0.291 & 0.119 & 0.060 & 0.035 \\
        Duration / ms & 0.58 & 0.76 & 0.84 & 0.90 & & 0.76 & 1.21 & 1.43 & 1.34 \\
        Fidelity & 0.998 & 0.993 & 0.989 & 0.979 & & 0.995 & 0.993 & 0.982 & 0.978 \\
        \hline\hline
    \end{tabular}
    \vspace{20pt}
    \label{tab:num_phase_state_preparation}
\end{table}

\subsection{Theory}
\label{sec:num_phase_theory}

Ideal number-phase states, $\ket{\npstate}$, are mutual eigenstates of the number and phase operators~\cite{Grimsmo2020, albert2022bosoniccodes,xu2024multimoderotation}
\begin{subequations}
    \begin{alignat}{2}
        & \hat{S}_n = e^{- i l_n \hat{n} } = \hat{R}_{l_n}, \\
        & \hat{S}_\phi = (\hat{E}^-)^{l_\phi} = \hat{E}^-_{l_\phi},
    \end{alignat}
\end{subequations}
where $\hat{R}_\theta = e^{-i \theta \hat{n}}$ is a phase delay/rotation operator, and $\hat{E}^- = \widehat{\mathrm{exp}}(i \phi) = \sum_{n=0}^\infty \ket{n}\bra{n+1}$ is the Susskind-Glogower phase operator~\cite{Susskind1964}. The operator $\hat{E}^-$ and its conjugate $\hat{E}^+ = (\hat{E}^-)^\dagger$ serve to lower or raise the phonon number by 1, respectively. Taking $N$ powers lowers or raises the phonon number by $N$,
\begin{align} \label{eq:modularphaseops}
    & \hat{E}^-_N = \sum_{n=0}^\infty \ket{n}\bra{n+N},
    \\
    & \hat{E}^+_N = \sum_{n=0}^\infty \ket{n+N}\bra{n}. \label{eq:modularphaseops_Eplus}
\end{align}
The modular lengths $l_n = 2\pi/N$ and $l_\phi = N$ satisfy $l_n l_\phi = 2\pi$, ensuring that $\hat{S}_n$ and $\hat{S}_\phi$ commute. As such, $\ket{\npstate}$ are also eigenstates of modular number, $\hat{n}\text{ mod }N$, and have a definite modular phase $\phi_{[N]} = \phi\text{ mod }\frac{2\pi}{N}$ with $-\frac{\pi}{N} \leq \phi_{[N]} < \frac{\pi}{N}$, despite the nonexistence of a Hermitian phase operator~\cite{Susskind1964, Carruthers1968}. Ideal number-phase states are discrete superpositions of Susskind-Glogower phase states~\cite{Grimsmo2020}, and have the Fock-space representation
\begin{equation}
    \ket{\npstate} = \sum^\infty_{k = 0} \ket{k N + \lambda},
\end{equation}
with an offset $0<\lambda<N-1$ that gives the value of modular number.

\begin{figure}
    \centering
    \includegraphics[]{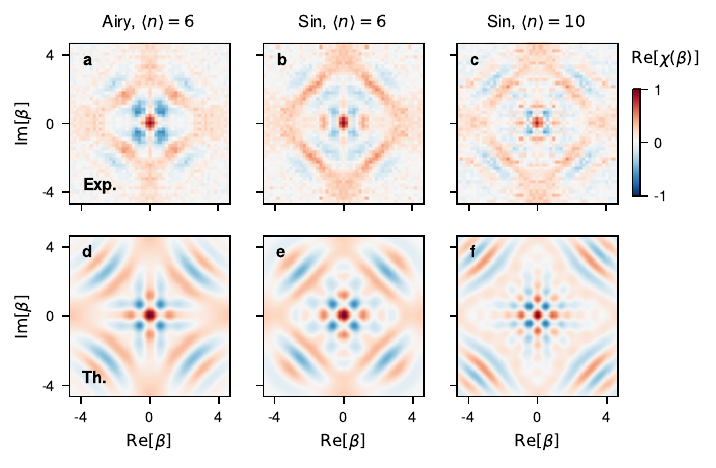}
    \caption{
    \textbf{Reconstructed characteristic functions of number--phase states.}
    The real part of the characteristic function $\mathrm{Re}[\bigchi(\beta)]$ is experimentally reconstructed for \textbf{(a)} an airy state with $\langle \hat{n} \rangle_\mathrm{t} = 6$, \textbf{(b)} a sine state with $\langle \hat{n} \rangle_\mathrm{t} = 6$ and \textbf{(c)} a sine state with $\langle \hat{n} \rangle_\mathrm{t} = 10$. Only the top right quadrant is reconstructed, data in the remaining quadrants are obtained from symmetries of the characteristic function~\cite{Matsos2024}.
    \textbf{(d-f)} The corresponding theoretical characteristic functions from numerical simulations are plotted below.
    }
    \label{fig:chis_num_phase}
\end{figure}

Just like ideal grid states, number-phase states $\ket{\npstate}$ are unphysical as they have infinite energy. They can be made physical by applying an envelope to the Fock coefficients in analogy to applying a (typically Gaussian) envelope to the Wigner function for grid states. We focus on two types of Fock-space envelope, inspired by the phase-optimised states of Summy and Pegg~\cite{Summy1990}. First, minimising the modular Holevo phase variance of the state~\cite{Grimsmo2020}, 
    \begin{align} \label{eq:modular_phase_variance}
        V_N(\phi) = \frac{1}{|\langle \hat{E}_N^- \rangle|^2} - 1
        ,
    \end{align}
while enforcing a maximum Fock cutoff, $F$, gives approximate number-phase states with a sine envelope, which we refer to as \emph{sine states},
\begin{align}
    \label{eq:ket_num_phase_sine}
    \ket{\tilde{\npstate}^{(\mathrm{sin})}} = \sum_{k=0}^{\lfloor F/N \rfloor} c_{kN}^{(\mathrm{sin})} \ket{kN + \lambda},
\end{align}
with Fock amplitudes 
\begin{equation}
    c_{kN}^{(\mathrm{sin})} = \frac{1}{ \sqrt{\lfloor F/N \rfloor / 2 + 1} } \sin \left[ \frac{ \pi N(k + 1)}{F + 2N - (F\text{ mod }N)} \right].
\end{equation}
The average energy of a sine state is $\langle \hat{n} \rangle^{(\mathrm{sin})} = \frac{N}{2} \lfloor F/N \rfloor + \lambda$. The sine state is related to the so-called $0N$ states~\cite{sabapathy2018ONstates}, with $\ket{0N} = \frac{1}{\sqrt{2}}(\ket{0} + \ket{N})$. The $0N$ state corresponds to a sine state by setting $F = N$ for a given $N$. While $0N$ states are the minimal-energy sine states for a given $N$, they have larger modular Holevo phase variance than other sine states in the family. Increasing $F$ to include higher multiples of $N$ increases support in Fock space (thus increasing energy) but decreases the modular Holevo variance.  In terms of the stabiliser expectation, $\langle{ \hat{E}_N^- \rangle} = \frac{1}{2}$ for an $0N$ state; for sine states, $\langle{ \hat{E}_N^- \rangle} \rightarrow 1$ as $F$ increases for a given $N$. 

Alternatively, one can minimise the Holevo variance while constraining the average phonon number without any restriction on Fock-space support, giving number--phase states with a Fock-space envelope determined by an Airy function.  We refer to this set as \emph{airy states},
\begin{align}
    \label{eq:ket_num_phase_airy}
    \ket{\tilde{\npstate}^{(\mathrm{airy})}} = \sum^{\infty}_{k=0} c_{kN}^{(\mathrm{airy})} \ket{kN + \lambda},
\end{align}
with amplitudes
\begin{equation}
    \label{eq:num_phase_airy_coeffs}
    c_{kN}^{(\mathrm{airy})} = \mathcal{N} \mathrm{Ai}\left[ \Big(\frac{\mu}{N^2} \Big)^{1/3} N (k + 1) - |z_1| \right],
\end{equation}
where $\mathcal{N}$ is a normalisation and $|z_1| \approx 2.338$ is the absolute value of the first zero of the Airy function $\mathrm{Ai}(x)$. In practice, we numerically optimise the value of $\mu$ to obtain an Airy state with a target mean phonon number $\langle \hat{n} \rangle_\mathrm{t}$.

\subsection{State preparation}

The number--phase states are experimentally prepared using a modulated spin-boson interaction. We prepare both sine and airy states, whose target states are defined in Eq.~\ref{eq:ket_num_phase_sine} and Eq.~\ref{eq:ket_num_phase_airy}, respectively. For each state, we target an average phonon number in the range $\langle \hat{n} \rangle_\mathrm{t} \in [4, 10]$. We further fix the offset of the states to $\lambda = 2$ to avoid parasitic interactions with the bosonic vacuum when applying the conditional phase operator (see section~\ref{sec:conditional_phase_stabiliser}). For sine states, a target $\langle \hat{n} \rangle_\mathrm{t}$ is achieved by varying the Fock cutoff. For Airy states, $\langle \hat{n} \rangle_\mathrm{t}$ is set by numerically optimising the parameter $\mu$. The characteristics of the number--phase states obtained from the numerical optimiser for each $\langle \hat{n} \rangle_\mathrm{t}$ are summarised in Table~\ref{tab:num_phase_state_preparation}. We also reconstruct the characteristic function of an airy state and two sine states (see Fig.~\ref{fig:chis_num_phase}), and find good general agreement with the theory according to the eyeball norm, $||\cdot ||_\text{\eye}$. 

\subsection{Conditional number operator}
\label{sec:conditional_rotation_op}

\begin{figure*}
    \centering
    \includegraphics[]{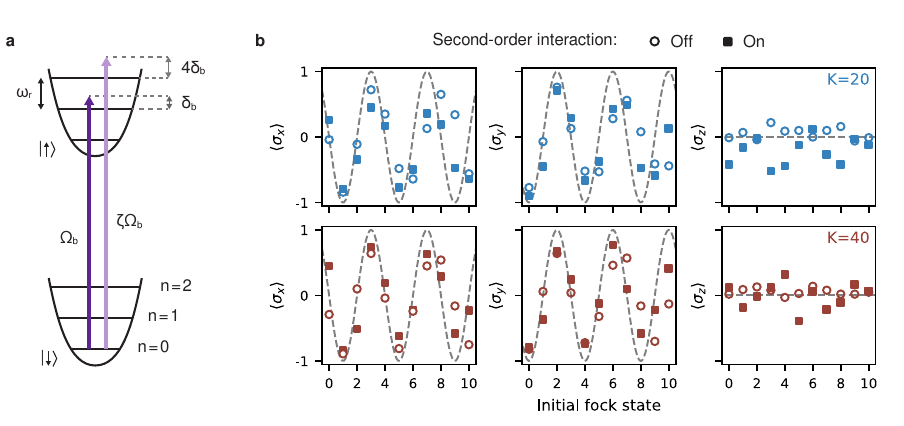}
    \caption{
    \textbf{Experimental implementation of the conditional number operator.}
    \textbf{(a)} The operator $\hat{S}_n = \mathrm{exp}(- i \phi_t \hat{\sigma}_z \hat{n}/2)$ is implemented by driving a blue-sideband transition (dark purple) with Rabi frequency $\Omega_\mathrm{b}$ and detuning $\delta_\mathrm{b}$. The desired interaction is obtained from the non-commutativity of the resulting time-dependent Hamiltonian. The target phase $\phi_t$ is set by choosing an appropriate duration, Rabi frequency and detuning. An optional second-order blue-sideband interaction (light purple) with a weaker Rabi frequency $\zeta \Omega_\mathrm{b}$ ($\zeta \ll 1$) can be introduced to increase the quality of the operation by negating the effects of parasitic higher order Magnus expansion terms.
    \textbf{(b)} The experimental implementation of the conditional number operator is verified by applying $\hat{S}_n$ with $\phi_t = - \pi/2$ to the state $\frac{1}{\sqrt{2}}(\ket{\downarrow} + e^{-i \pi/2} \ket{\uparrow})\ket{n}$. The ancillary qubit is then measured in the three Pauli bases for varying Fock states, $n \in [0, 10]$. The experiment is further repeated for $K=20$ (top, blue) and $K=40$ (bottom, red), and we investigate the action of omitting the second-order blue-sideband interaction (open circles) or including it (closed squares). Dashed grey lines plot theoretical expectation values obtained with an ideal conditional number operator. Measurements without the second-order interaction have $\Omega_b/2\pi = \SI{2.74}{kHz}$, giving $\delta_\mathrm{b}/2\pi = \SI{17.34}{kHz}$ and $t=\SI{1.15}{ms}$ for $K=20$, and $\delta_\mathrm{b}/2\pi = \SI{24.53}{kHz}$ and $t=\SI{1.63}{ms}$ for $K=40$. Measurements with the second-order sideband interaction have, for $K=20$, $\Omega_b/2\pi = \SI{1.11}{kHz}$, $\zeta = 0.18$, $\delta_\mathrm{b}/2\pi = \SI{7.02}{kHz}$ and $t = \SI{2.85}{ms}$, and, for $K=40$, $\Omega_\mathrm{b} / 2\pi = \SI{1.41}{kHz}$, $\zeta = 0.14$, $\delta_\mathrm{b} / 2\pi = \SI{12.62}{kHz}$ and $t=\SI{3.17}{ms}$.
    }
    \label{fig:num_op_rabi_flop}
\end{figure*}

The conditional number operator of number--phase states is given by

\begin{equation}
    \label{eq:Sn_target}
    C\hat{S}_{n} = e^{- i \phi_\mathrm{t} \hat{\sigma}_z \hat{n} / 2},
\end{equation}
where the target phase $\phi_\mathrm{t} = 2\pi/N$ is determined by the spacing $N$ of the number--phase state. Our approach to implementing $C\hat{S}_{n}$ is through simultaneous applications of a detuned first-order blue-sideband (BSB) interaction and a detuned second-order BSB interaction (see Fig.~\ref{fig:num_op_rabi_flop}a).

We begin by considering the blue-sideband Hamiltonian of Eq.~6 with $\phi_\mathrm{b}(t) = 0$, giving
\begin{equation} \label{eq:cs_r_blue_hamiltonian}
    \hat{H}_\mathrm{b} = \frac{\Omega_\mathrm{b}}{2} \hat{\sigma}^+ \hat{a}^\dagger e^{- i \delta_\mathrm{b} t} + \mathrm{h.c.}
\end{equation}
We derive its Magnus expansion,
\begin{equation}
    \hat{U}(t) = \mathrm{exp}\left( \sum_{k=1}^{\infty} \hat{Y}_k(t) \right),
\end{equation}
where the unitary $\hat{U}(t)$ approximates the operator obtained from applying $\hat{H}_\mathrm{b}$ for a duration $t$. Expanding up to the fourth order gives
\begin{subequations}
\begin{align}
    \hat{Y}_1(t) = & - i \int^{t}_{0} dt_1 \hat{H}(t_1), \\
    \hat{Y}_2(t) = & - \frac{1}{2} \int^{t}_{0}\int^{t_1}_{0} dt_1 dt_2 [\hat{H}(t_1), \hat{H}(t_2)], \\
    \hat{Y}_3(t) = & \frac{i}{6} \int^{t}_0 \int^{t_1}_0 \int^{t_2}_0 dt_1 dt_2 dt_3 [\hat{H}(t_1), [\hat{H}(t_2), \hat{H}(t_3)]] + [\hat{H}(t_3), [\hat{H}(t_2), \hat{H}(t_1)]],  \\
    \hat{Y}_4(t) = & \frac{1}{12} \int^{t}_0 \int^{t_1}_0 \int^{t_2}_0 \int^{t_3}_0 dt_1 dt_2 dt_3 dt_4 [[[\hat{H}(t_1), \hat{H}(t_2)], \hat{H}(t_3)], \hat{H}(t_4)] \nonumber \\ 
    & \hphantom{\frac{1}{12} \int^{t}_0 \int^{t_1}_0 \int^{t_2}_0 \int^{t_3}_0 } + [\hat{H}(t_1), [[\hat{H}(t_2), \hat{H}(t_3)], \hat{H}(t_4)]] \nonumber \\
    &   \hphantom{\frac{1}{12} \int^{t}_0 \int^{t_1}_0 \int^{t_2}_0 \int^{t_3}_0 }  + [\hat{H}(t_1), [\hat{H}(t_2), [\hat{H}(t_3), \hat{H}(t_4)]]] \nonumber \\ 
    & \hphantom{\frac{1}{12} \int^{t}_0 \int^{t_1}_0 \int^{t_2}_0 \int^{t_3}_0 } + [\hat{H}(t_2), [\hat{H}(t_3), [\hat{H}(t_4), \hat{H}(t_1)]]].
\end{align}
\end{subequations}
These expressions can be simplified by setting $t = K t_i$ and $\delta_\mathrm{b} = 2\pi / t_i$~\cite{Sutherland2021}, where $K \in \mathbb{Z^+}$ is a positive integer and $t_i$ is a shorter duration, giving
\begin{subequations}
\begin{align}
    \hat{Y}_1(t) & =  0, \label{eq:y1} \\
    \hat{Y}_2(t) & = - \frac{i t \Omega_\mathrm{b}^2}{4 \delta_\mathrm{b}} (\ket{\downarrow}\bra{\downarrow} \hat{a} \hat{a}^\dagger - \ket{\uparrow}\bra{\uparrow} \hat{a}^\dagger \hat{a}), \label{eq:y2} \\
    \hat{Y}_3(t) & = - \frac{i t \Omega_\mathrm{b}^3}{4 \delta_\mathrm{b}^2} (\ket{\downarrow}\bra{\uparrow} \hat{a} \hat{a}^\dagger \hat{a} + \ket{\uparrow}\bra{\downarrow} \hat{a}^\dagger \hat{a} \hat{a}^\dagger), \label{eq:y3} \\
    \hat{Y}_4(t) & = \frac{3 i t \Omega_\mathrm{b}^4}{16 \delta_\mathrm{b}^3} (\ket{\downarrow}\bra{\downarrow} (\hat{a} \hat{a}^\dagger)^2 - \ket{\uparrow}\bra{\uparrow} (\hat{a}^\dagger \hat{a})^2). \label{eq:y4}
\end{align}
\end{subequations}
The second-order term, $\hat{Y}_2(t)$, contains the required $\hat{\sigma}_z\hat{a}^\dagger \hat{a}$ interaction to implement the conditional number operator. The third- and fourth-order terms, $\hat{Y}_3(t)$ and $\hat{Y}_4(t)$, are parasitic terms that introduce errors in the interaction and should be mitigated. For large $\delta_\mathrm{b}$, their interaction strengths become negligible. As discussed below, this can be achieved by choosing a large $K$, at the cost of longer durations.

The second order term, $\hat{Y}_2(t)$, can be expressed as
\begin{align}
    \label{eq:y2_2}
    \hat{Y}_2(t) = - \frac{i t \Omega_\mathrm{b}^2}{4 \delta_\mathrm{b}} (\hat{\sigma}_z \hat{n} + \frac{1}{2}\hat{\sigma}_z + \frac{1}{2} \hat{\mathds{1}}),
\end{align}
and the resulting operator is
\begin{align}
    \label{eq:num_op_exp}
    \hat{U}_n = e^{\hat{Y}_2(t)} = \mathrm{exp}\left({- i\frac{\Phi}{2}\hat{\sigma}_z \hat{n}}\right) \mathrm{exp}\left({- i \frac{\Phi}{4} \hat{\sigma}_z}\right),
\end{align}
with $\Phi = \Omega_\mathrm{b}^2 t/(2\delta_\mathrm{b})$, and we have omitted a global phase associated with the identity operator of Eq.~\ref{eq:y2_2}. The first unitary of Eq.~\ref{eq:num_op_exp} implements the desired conditional number operator of Eq.~\ref{eq:Sn_target} after setting $\Phi = \phi_t$, while the second unitary gives an unwanted $\hat{\sigma}_z$ rotation of the ancillary qubit. This additional rotation can be undone by applying the inverse rotation, $\mathrm{exp}\left({ i \frac{\Phi}{4} \hat{\sigma}_z}\right) $, after application of $\hat{U}_n$. 

The phase of the conditional number operator, $\Phi$, can be set by appropriately choosing the parameters of the BSB interaction. To implement a target phase $\phi_t$ of Eq.~\ref{eq:Sn_target}, we set the BSB detuning to be $\delta_\mathrm{b} = \Omega_\mathrm{b} \sqrt{K \pi / \phi_t}$. The choice of the integer $K$ is subject to a trade-off: small $K$ is desired to reduce the overall duration of the operation, whereas a large $K$ is wanted to minimise the strength of the parasitic higher order terms of the Magnus expansion. To better illustrate the latter, we replace $\delta_\mathrm{b}$ with $\Omega_\mathrm{b} \sqrt{K \pi / \phi_t}$ in $\hat{Y}_3(t)$ and $\hat{Y}_4(t)$ (Eq.~\ref{eq:y3} and~\ref{eq:y4}), giving
\begin{align}
    \hat{Y}_3(t) = & - \frac{i t \Omega_\mathrm{b}\phi_t}{4 \pi K} (\ket{\downarrow}\bra{\uparrow} \hat{a} \hat{a}^\dagger \hat{a} + \ket{\uparrow}\bra{\downarrow} \hat{a}^\dagger \hat{a} \hat{a}^\dagger), \label{eq:magnus_Y3_b}\\
    \hat{Y}_4(t) = & \frac{3 i t \Omega_\mathrm{b}}{16} \left(\frac{\phi_t}{\pi K} \right)^{3/2} (\ket{\downarrow}\bra{\downarrow} (\hat{a} \hat{a}^\dagger)^2 - \ket{\uparrow}\bra{\uparrow} (\hat{a}^\dagger \hat{a})^2). \label{eq:magnus_Y4_b}
\end{align}
With this, we obtain the following scalings: the interaction strengths of $\hat{Y}_3(t)$ and $\hat{Y}_4(t)$ scale with $\propto K^{-1}$ and $\propto K^{-3/2}$, respectively, while the overall duration scales with $t \propto K$. The integer $K$ should, therefore, be carefully chosen to minimise errors from higher-order terms while ensuring that the duration of the interaction is sufficiently short to minimise effects arising from decoherence. We make two additional observations from Eq.~\ref{eq:magnus_Y3_b} and Eq.~\ref{eq:magnus_Y4_b}. First, the interaction strengths of $\hat{Y}_3(t)$ and $\hat{Y}_4(t)$ increase with $\phi_t$, hence we expect the quality of the interaction to worsen for number--phase states with smaller spacings. Second, the term $\hat{Y}_4(t)$ is quadratic in the number operator, $\hat{n}$, and thus we expect the interaction to worsen for number--phase states with larger energies that populate higher Fock states.

We experimentally demonstrate the conditional number operator by applying $\hat{U}_n$ of Eq.~\ref{eq:num_op_exp} to the state $\frac{1}{\sqrt{2}}(\ket{\downarrow} + e^{-i \pi/2} \ket{\uparrow})\ket{n}$ and measuring the ancillary qubit in the three Pauli bases, $\langle \hat{\sigma}_{x,y,z}\rangle$. The motional mode is prepared in an initial Fock state $\ket{n}$ by applying a sequence of $n$ BSB and carrier pulses after cooling the mode to its ground state. The resulting expectation values with increasing $n$ are plotted in Fig.~\ref{fig:num_op_rabi_flop}b (open circles). In general, the experimental data shows good agreement with theory, reproducing the oscillatory behavior of the expectation values. We find that discrepancies become more apparent for larger $n$, which is due to the fourth-order term in the Magnus expansion that scales non-linearly with $\hat{n}$. The experimental data with $K=40$ (blue) shows better agreement with theory than for $K=20$ (red). This can be seen from the mean square error (MSE): the average MSE over measurements of $\langle \hat{\sigma}_x \rangle$, $\langle \hat{\sigma}_y \rangle,$ and $\langle \hat{\sigma}_z \rangle$ is reduced from 0.84 to 0.52 when increasing $K=20$ to $K=40$, demonstrating that the quality of the conditional number operator can be improved by increasing $K$.

From experimental investigations, we find that the errors induced by higher-order Magnus expansion terms for the number--phase states considered in this work become unacceptably large for moderate values of $K$. To remedy this, we introduce an additional weaker interaction near the second-order BSB transition with a detuning of $4 \delta$ (see Fig.~\ref{fig:num_op_rabi_flop}a). We find that that the leading terms in the Magnus expansion from this additional interaction can be tuned to mitigate the parasitic terms from the original interaction. 

Adding a second-order blue-sideband to the Hamiltonian of Eq.~\ref{eq:cs_r_blue_hamiltonian} gives
\begin{equation}
    \hat{H}'_\mathrm{b}(t) = \frac{\Omega_\mathrm{b}}{2} \hat{\sigma}^+ \hat{a}^\dagger e^{- i \delta_\mathrm{b} t} + \frac{\zeta \Omega_\mathrm{b}}{2} \hat{\sigma}^+ (\hat{a}^\dagger)^2 e^{-i 4 \delta_\mathrm{b} t} + \mathrm{h.c.},
\end{equation}
with $\zeta \ll 1$. The second-order detuning of $4\delta_\mathrm{b}$ is chosen to avoid driving unwanted interactions. After performing the Magnus expansion of $\hat{H}'_\mathrm{b}(t)$, we find that $Y_2'(t)$ contains two additional terms,
\begin{align}
    \hat{Y}_2'(t) = \hat{Y}_2(t) & - \frac{i t \zeta^2 \Omega_\mathrm{b}^2}{16\delta_\mathrm{b}} \left(\ket{\downarrow}\bra{\downarrow}(a a^\dagger) - \ket{\uparrow}\bra{\uparrow} (a^\dagger a) \right) \nonumber \\
    & -\frac{i t \zeta^2 \Omega_\mathrm{b}^2}{16\delta_\mathrm{b}} \left(\ket{\downarrow}\bra{\downarrow}(a a^\dagger)^2 - \ket{\uparrow}\bra{\uparrow} (a^\dagger a)^2 \right) + \mathcal{O} \left( \frac{\zeta^2\Omega^3}{\delta_\mathrm{b}^2}\right),
    \label{eq:y2'}
\end{align}
where $\hat{Y}_2(t)$ is given by Eq.~\ref{eq:y2_2}. The first term of Eq.~\ref{eq:y2'} is akin to the original $\hat{Y}_2(t)$ term of Eq.~\ref{eq:y2}, and modifies the phase $\Phi$ of Eq.~\ref{eq:num_op_exp} as $\Phi \rightarrow \Phi + t \zeta^2\Omega_\mathrm{b}^2/(8 \delta_\mathrm{b})$. The second term of $Y'_2(t)$ is similar to the parasitic term of $\hat{Y}_4(t)$ and can therefore be used to cancel out unwanted effects. 

Improvements from the application of the second-order sideband are verified by repeating the above experiment in which the ancillary qubit is measured in different Pauli bases after applying the conditional number operator to a range of initial Fock states, $n$ (full squares of Fig.~\ref{fig:num_op_rabi_flop}). The measurements show better agreement with theory for a wider range of Fock states $n$ than the previous measurements that did not include the second-order BSB interaction. This is apparent from the mean square error (MSE): the MSE averaged over measurements of $\langle \hat{\sigma}_x \rangle$, $\langle \hat{\sigma}_y \rangle$ and $\langle \hat{\sigma}_z \rangle $ for $K=40$ is reduced from $0.52$ to $0.25$ when including the second-order BSB interaction. 

The parameters of the conditional number operator performed in Fig.~4 of the main text are chosen to maximise the visibility parameter of the resulting probability distribution, and hence the metrological gain, of a number--phase state. We consider states with a spacing of $N=4$, giving a target phase of $\phi_t = \pi/2$. We set $K=40$ and calibrate the Rabi frequency of the first-order interaction to be $\Omega_\mathrm{b}/2 \pi = \SI{1.34}{kHz}$. The value of $\zeta$ is empirically chosen from numerical simulations of the quantum phase estimation circuit, and we set $\zeta = 0.14$, giving $\zeta \Omega_\mathrm{b}/2 \pi = \SI{0.18}{kHz}$. The detuning is set to $\delta_\mathrm{b} / 2\pi = \SI{11.95}{kHz}$, resulting in a total duration of $\SI{3.35}{ms}$. The Rabi frequency of the second-order blue-sideband interaction is calibrated by setting $\delta_\mathrm{b} = 0$, which drives the transition $\ket{\downarrow}\ket{n} \rightarrow \ket{\uparrow}\ket{n+2}$, and measuring the probability of the ancilla being in $\ket{\uparrow}$. Fitting the oscillation frequency, $\sqrt{2}\zeta \Omega_\mathrm{b}$, gives the second-order Rabi frequency.

\subsection{Conditional phase operator}
\label{sec:conditional_phase_stabiliser}

\begin{figure*}
    \centering
    \includegraphics[width=15cm]{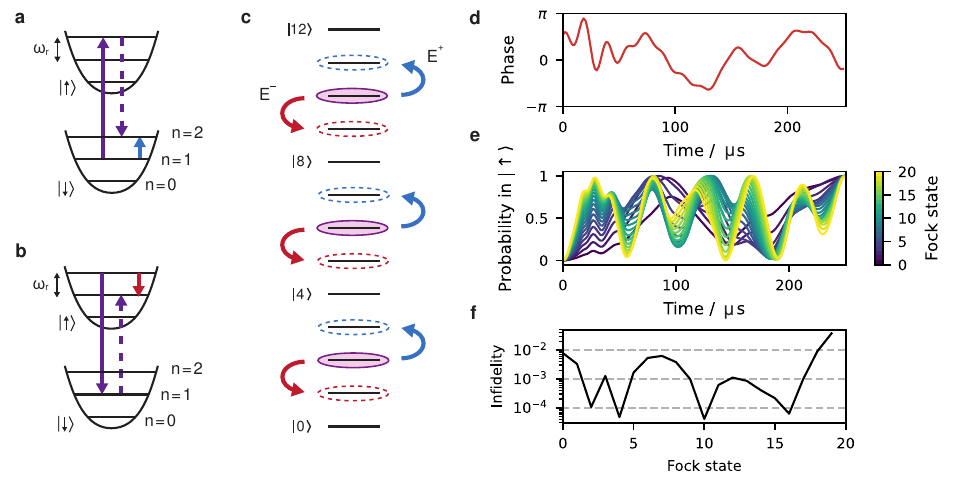}
    \caption{
    \textbf{Experimental implementation of the conditional phase operator.}
    \textbf{(a,b)} The conditional phase operator is obtained by sequentially applying a blue-sideband pulse (solid purple arrow) followed by a carrier pulse (dashed purple arrow). The phase of the blue-sideband interaction is dynamically modulated such that the resulting operation gives $\ket{n} \rightarrow \ket{n+1}$ (blue arrow in \textbf{(a)}) if the qubit is in $\ket{\downarrow}$, and $\ket{n+1} \rightarrow \ket{n}$ (red arrow in \textbf{(b)}) if the qubit is in $\ket{\uparrow}$, independent of the Fock number $n$.
    \textbf{(c)} The resulting operation implements phonon up and down ladder operators, $\hat{E}^+$ or $\hat{E}^-$, conditioned on the ancilla state.
    \textbf{(d)} Modulation waveform of the blue-sideband phase that is obtained from a numerical optimiser, with $\Omega / 2\pi  = \SI{4.34}{kHz}$.
    \textbf{(e)} Evolution of the probability for initial states $\ket{\downarrow}\ket{n}$ with $n \in [0, 20]$ under the dynamically modulated blue-sideband interaction. After a time $t=\SI{288}{\micro s}$, all states evolve to $\ket{\uparrow}\ket{n+1}$, independent of $n$.
    \textbf{(f)} Infidelity of the final state for initial Fock states $n$. The infidelity is calculated as $\bra{\langle \uparrow, n+1 } \hat{U}_\mathrm{b} \ket{\downarrow, n} |^2$, where $\hat{U}_\mathrm{b}$ is the operator that is obtained from the blue-sideband interaction. Infidelities for all Fock states $n < 19$ are less than $10^{-2}$.
    }
    \label{fig:conditional_phase_op}
\end{figure*}

The phase operator outlined in section~\ref{sec:num_phase_theory} is defined from the phonon ladder operator, $\hat{E}^-$, which shifts the phonon number by 1. We experimentally implement a phase operator conditioned on the ancilla by sequentially applying blue-sideband and carrier pulses (see Fig.~\ref{fig:conditional_phase_op}(a-c)). The blue-sideband pulse first increases or decreases the phonon number by one conditioned on the ancillary qubit being in the $\ket{\downarrow}$ or $\ket{\uparrow}$ state, respectively. The blue-sideband pulse also changes the state of the ancilla, which is undone by applying the carrier pulse. The challenge of implementing a phonon ladder operator in this way lies in the unequal interaction strengths of the blue-sideband interaction with the Fock state: creation and annihilation operators acting on $\ket{n}$ have an interaction strength proportional to $\sqrt{n+1}$ or $\sqrt{n}$, respectively~\cite{Wineland1998}. Here, we overcome this challenge by employing optimal quantum control to implement a blue-sideband pulse with coupling strength that is independent of the Fock state.

We first consider the blue-sideband Hamiltonian of Eq.~6 and set the detuning to $\delta_\mathrm{b} = 0$, giving
\begin{equation}
    \hat{H}_\mathrm{b}(t) = \frac{\Omega_\mathrm{b}}{2} \hat{\sigma}^+\hat{a}^\dagger e^{i \phi_\mathrm{b}(t)} + \mathrm{h.c.},
\end{equation}
with $\Omega_\mathrm{b} / 2\pi = \SI{4.34}{kHz}$. Applying $\hat{H}_\mathrm{b}(t)$ drives a transition between $\ket{\downarrow, n}$ and $\ket{\uparrow, n+1}$ with an interaction strength that is proportional to $\sqrt{n+1}$. We seek to obtain an operation that increases or decreases the phonon number by one without residual spin-motion coupling.  To this end, we employ optimal control by modulating the phase $\phi_\mathrm{b}(t)$. The modulation waveform of $\phi_\mathrm{b}(t)$ is numerically optimised such that the resulting operator $\hat{U}_\mathrm{b} = e^{- i \int^{T}_{t=0}dt \hat{H}_b(t)}$ enacts the transition $\hat{U}_\mathrm{b} \ket{\downarrow, n} = \ket{\uparrow, n+1}$, independent of the initial Fock state $n$. We employ a similar optimisation procedure as outlined in Methods, and set $N_\mathrm{opt} = 30$ optimisable segments, $N_\mathrm{seg} = 150$ resampled segments after filtering and a maximum duration of $T_\mathrm{max} = \SI{1}{ms}$. The resulting waveform, shown in Fig.~\ref{fig:conditional_phase_op}d, has a duration of $\SI{288}{\micro s}$. We characterise the quality of the resulting control by computing the average fidelity,
\begin{equation}
    \label{eq:phonon_ladder_fidelity}
    \mathcal{F} = \frac{1}{N_\mathrm{max}} \sum_{n=0}^{N_\mathrm{max} - 1} |\bra{\uparrow, n+1} \hat{U}_\mathrm{b} \ket{\downarrow, n}|^2.
\end{equation}
For $N_\mathrm{max}=18$, this gives $\mathcal{F} = 0.998$ (see Fig.~\ref{fig:conditional_phase_op}(e-f)).

The conditional ladder operator is obtained by alternating the application of $\hat{H}_\mathrm{b}(t)$ and a carrier pulse. This is made apparent by first writing the operator resulting from the dynamically modulated blue-sideband interaction as
\begin{align}
    \hat{U}_\mathrm{b} & = \mathrm{exp}\left(- i \int^T_{t=0} dt \hat{H}_\mathrm{b}(t) \right) \nonumber \\
    & \approx \ket{\uparrow}\bra{\uparrow} \otimes {\ket{0}\bra{0}} + \sum_{n=0} \ket{\uparrow}\bra{\downarrow} \otimes \ket{n+1} \bra{n} +  \sum_{n=0} \ket{\downarrow}\bra{\uparrow} \otimes \ket{n}\bra{n+1} .
\end{align}
Applying a carrier pulse $\hat{U}_\mathrm{c} = \ket{\uparrow}\bra{\downarrow} +\ket{\downarrow}\bra{\uparrow}$ after $\hat{U}_\mathrm{b}$ then gives
\begin{align}
    \label{eq:U_c_U_b}
    \hat{U}_\mathrm{c}\hat{U}_\mathrm{b} & = \ket{\downarrow}\bra{\uparrow} \otimes {\ket{0}\bra{0}} + \sum_{n=0} \ket{\downarrow}\bra{\downarrow} \otimes \ket{n+1} \bra{n} +  \sum_{n=0} \ket{\uparrow}\bra{\uparrow} \otimes \ket{n}\bra{n+1}  \nonumber \\
    & = \ket{\downarrow}\bra{\uparrow} \otimes \ket{0}\bra{0} + \ket{\downarrow}\bra{\downarrow} \otimes \hat{E}^+ + \ket{\uparrow}\bra{\uparrow} \otimes \hat{E}^-,
\end{align}
where the phonon raising and lowering operators, $\hat{E}^+$ and $\hat{E}^-$ of Eq.~\ref{eq:modularphaseops} and \ref{eq:modularphaseops_Eplus} with $N=1$, increase or decrease the phonon number by one, respectively. Repeatedly applying an alternating sequence of blue-sideband and carrier pulses $N$ times results in the conditional phase operator,
\begin{equation}
    \label{eq:sn_op_a}
    C\hat{S}_\phi = (\hat{U}_\mathrm{c}\hat{U}_\mathrm{b})^N = \ket{\downarrow}\bra{\downarrow} \otimes \hat{E}^+_N + \ket{\uparrow}\bra{\uparrow} \otimes \hat{E}^-_N.
\end{equation}
In Eq.~\ref{eq:sn_op_a}, we have omitted interactions with the $n=0$ Fock state which would otherwise appear from the term $\ket{\downarrow}\bra{\uparrow} \otimes \ket{0}\bra{0}$ of Eq.~\ref{eq:U_c_U_b} and cause unwanted ancilla rotations and subsequent phonon shifts. In the experiment, we circumvent this issue by preparing number--phase states with a spacing of $N=4$ that are offset by $\lambda = 2$ from $n=0$ (see section~\ref{sec:num_phase_states}). Since applying the conditional ladder operator shifts the phonons down by at most $l_\phi/2 = 2$, we avoid unwanted interactions from applying $C\hat{S}_\phi$ to states with population in $n=0$.

The phonon ladder operation shown here is faster than previous demonstrations. For comparison, Ref.~\cite{Um2016} demonstrated phonon addition and subtraction in trapped ions using a STIRAP pulse, whose duration was $7\times$ larger than the duration of a $\pi$-pulse from $\ket{\downarrow, n=0}\rightarrow \ket{\uparrow, n=1}$. Here, the duration of the phase-modulated pulse of Fig.~\ref{fig:conditional_phase_op}d is $2.5\times$ greater than the duration of a $\pi$-pulse. Moreover, the duration can be further reduced at the cost of a lower average fidelity. For example, we numerically optimise a separate waveform with a smaller duration constraint and obtain a pulse that is $1.8\times$ greater than the duration of a $\pi$-pulse, with an average fidelity of 0.988 (calculated from Eq.~\ref{eq:phonon_ladder_fidelity}).

\begin{figure}
    \centering
    \includegraphics[]{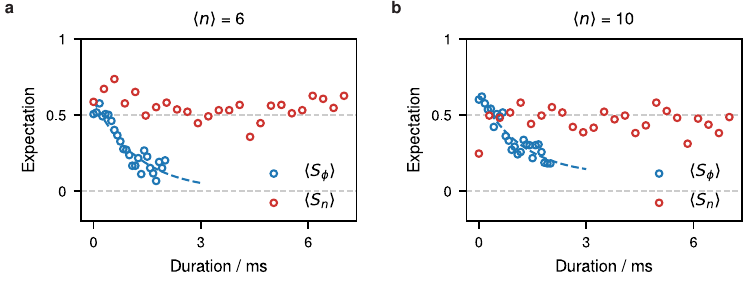}
    \caption{
    \textbf{Lifetime measurements of number and phase operators for number--phase states with a sine envelope.}
    Lifetimes are measured for number--phase states with a spacing of $N=4$ and an offset of $\lambda = 2$, and target average phonon numbers of \textbf{(a)} $\langle \hat{n} \rangle = 6$ and \textbf{(b)} $\langle \hat{n} \rangle = 10$. For both states, we measure expectation values $\langle \hat{S}_\phi\rangle$ (blue) and $\langle \hat{S}_n\rangle$ (red) of the phase and number operators, respectively. Lifetimes are extracted from fits to exponential decay, $f(t) = A e^{-T_\mathrm{wait}/T_1} + A_0$, where the amplitude $A$ and offset $A_0$ are set as free parameters to account for the approximate nature of the number--phase states. Fits to the decay of the phase operator give \textbf{(a)} $(T_\phi, A, A_0) = (1.3(5)~\mathrm{ms}, 0.59(10), 0.0(1))$ and \textbf{(b)} $(T_\phi, A, A_0) = (1.2(5)~\mathrm{ms}, 0.54(9), 0.10(10))$. 
    }
    \label{fig:num_phase_stabiliser_lifetime}
\end{figure}

\subsection{Number and phase operator lifetime}

We further characterise the conditional number and phase operators of the number--phase states by measuring their lifetime. We perform a similar experiment as detailed in section~\ref{sec:conditional_x_p_stabilisers}: the initial state $\ket{\downarrow} \otimes \ket{\tilde{\npstate}^{(\mathrm{sin})}}$ is prepared, and the expectation value of the number or phase operator is measured after an increasing duration $T_\mathrm{wait}$. The measurement is performed by applying one of the conditional operators, $C\hat{S}_n$ or $C\hat{S}_\phi$, in between Hadamard rotations and measuring the state of the ancillary qubit in the $\hat{\sigma}_z$ basis. The measured expectation values, shown in Fig.~\ref{fig:num_phase_stabiliser_lifetime}, are then fit to an exponential decay from which we extract lifetimes $T_n$ and $T_\phi$. For $\langle \hat{n} \rangle = 6$, the fitted lifetime associated with $C\hat{S}_\phi$ is $T_{\phi} = \SI{1.3(5)}{ms}$. We observe no decay in the expectation value of the number operator for durations up to $T_\mathrm{wait} = \SI{7}{ms}$. For $\langle \hat{n} \rangle = 10$, we measure a lifetime $T_{\phi} = \SI{1.2(5)}{ms}$, and again observe no decay of the number operator within the measured timescale. We first note that the lifetimes of the phase operator between the two number--phase states of different energies are not statistically different. We then observe a strong bias in the noise, where the decay associated with the phase operator is much larger than the decay associated with the number operator. This is due to a strong bias in the intrinsic hardware noise: dephasing, corresponding to $\hat{a}^\dagger \hat{a}$ noise which is aligned with the phase operator, has a measured decay rate of $\Gamma_\mathrm{d} \sim \SI{20}{s^{-1}}$, whereas phonon heating, corresponding to $\hat{a}$ and $\hat{a}^\dagger$ noise which is aligned with the number operator, has a much weaker decay of $\Gamma_\mathrm{h} = \SI{0.2}{s^{-1}}$.

%
%

\section{Multi-parameter probability distributions for quantum phase estimation}

Quantum Phase Estimation (QPE) aims to estimate a phase $\phi$ that results from applying a unitary $\hat{U}$ to its eigenstate, $\hat{U} \ket{\psi} =e^{i\phi} \ket{\psi}$. In Bayesian estimation, the phase is estimated by obtaining measurement outcomes whose posterior probability distribution gives an estimate for $\phi$. 
In the following, we derive probability distributions of measurement outcomes for both the grid and number--phase sensing states. In both cases, the probability distributions for a single measurement outcome $\mathrm{m} \in \{0, 1\}$ are well described by functions of the form
\begin{equation}
    \label{eq:general_prob_dist}
    \mathrm{P}(\mathrm{m}|\epsilon, \theta) = \frac{1}{2}\left(1 + (-1)^\mathrm{m} \eta \cos(\theta + l \epsilon)\right),
\end{equation}
where $\eta \in [0, 1]$ is a visibility parameter, $\theta\in [0, 2\pi]$ is a controllable rotation angle, $l$ is the modulus of the modular variable and $\epsilon$ is the unknown signal that we wish to determine.

Moving beyond conventional QPE, the sensing protocol used in this work performs several measurements after a single preparation of a sensing state, where two commuting operators are applied in an alternating manner. In the following, we extend the probability distribution of Eq.~\ref{eq:general_prob_dist} to a joint-probability distribution given a list of measurement outcomes. We find that the joint-probability distribution contains correlations between measurements and, hence, can not be exactly described by a product of the independent distributions since prior measurements exhibit backaction that modifies the sensing state. Nevertheless, we find that one can approximate the joint-probability distributions by products of modified independent distributions, significantly simplifying the model. The validity of this approximation increases for sensing states with higher energy. 

In section~\ref{sec:general_prob_distribution}, we derive independent and dependent probability distributions for an arbitrary sensing state with measurements of commuting modular variables. These generalised results are applied to grid states in section~\ref{sec:grid_state_prob_distribution} and to number--phase states in section~\ref{sec:num_phase_prob_distribution}.

\subsection{Probability distributions for arbitrary sensing states}
\label{sec:general_prob_distribution}

We first derive general probability distributions for measurement outcomes of a QPE circuit such as the one depicted in Fig.~1e. We consider a general sensing state, $\ket{\psi}$, that is the eigenstate of two commuting operators, $\hat{S}_a$ and $\hat{S}_b$.
These operators correspond to modular position and momentum for grid states and modular number and phase for number--phase states. Measurements of $\hat{S}_a$ or $\hat{S}_b$ give an outcome represented by a single bit, $\mathrm{m}_{a} \in \{0, 1\}$ or $\mathrm{m}_{b} \in \{0, 1\}$. In a scenario where the measurements are independent, i.e. preparation of a sensing state is followed by a single measurement of $\hat{S}_a$ or $\hat{S}_b$, the probability distributions of the measurement outcomes are $\mathrm{P}_{a}(\mathrm{m}_{a})$ or $\mathrm{P}_{b}(\mathrm{m}_{b})$ and have the form of Eq.~\ref{eq:general_prob_dist}. We also consider a scenario where the measurements are dependent: after preparing the sensor state, the operators $\hat{S}_a$ and $\hat{S}_p$ are measured sequentially, giving a string of measurement outcomes $\mathbf{m} = \mathrm{m}_a \mathrm{m}_b$, and the joint probability distribution is given by $\tilde{\mathrm{P}}_{a,b}(\mathbf{m})$. In the following, we aim to derive both the independent and dependent probability distributions for an arbitrary sensing state.

The probability distributions are derived from the operations of the QPE circuit. The initial state is $\ket{\Psi_\mathrm{i}} = \ket{\downarrow} \otimes \ket{\psi}$, where the ancilla is initialised in its ground state and the bosonic mode is prepared in a sensing state, $\ket{\psi}$. The sensing state is then subjected to a signal that we wish to determine, whose action is modeled by an operator, $\hat{\mathcal{E}}$. The state after this interaction is
\begin{equation}
    \ket{\Psi_{\boldsymbol{\epsilon}}} = \hat{\mathcal{E}} \ket{\Psi_\mathrm{i}} =  \ket{\downarrow} \otimes \ket{\psi_{\boldsymbol{\epsilon}}},
\end{equation}
with $\ket{\psi_{\boldsymbol{\epsilon}}} = \hat{\mathcal{E}} \ket{\psi}$. The operator $\hat{\mathcal{E}}$ corresponds to a displacement for sensing with grid states and corresponds to a rotation and a phonon shift for sensing with number--phase states. In the following, we remain general and parametrise $\hat{\mathcal{E}}$ by two parameters, $(\epsilon_a, \epsilon_b)$, which we wish to estimate. We further assume that $\hat{\mathcal{E}}$ has the following commutation relations with $\hat{S}_a$ and $\hat{S}_b$, which are used extensively in the derivations below,
\begin{subequations}
    \label{eq:general_commutation_relations}
    \begin{align}
        & (\hat{S}_a)^{1/2} \hat{\mathcal{E}} = e^{- i \epsilon_a l_a/2} \hat{\mathcal{E}} (\hat{S}_a)^{1/2}, \\
        & (\hat{S}_b)^{1/2} \hat{\mathcal{E}} = e^{- i \epsilon_b l_b/2} \hat{\mathcal{E}} (\hat{S}_b)^{1/2},
    \end{align}
\end{subequations}
where $l_{a}$ and $l_b$ are the modular lengths of $\hat{S}_a$ and $\hat{S}_b$.

After initialization of the sensing state and interaction with a signal to be measured, a conditional operation $C\hat{S}_A$ or $C\hat{S}_B$ is applied, which couples the bosonic mode with the ancilla in the $\hat{\sigma}_z$ basis. The conditional operators can be generalised as
\begin{subequations}
    \label{eq:general_conditional_stabilisers}
    \begin{align}
        & C\hat{S}_a = \ket{\downarrow}\bra{\downarrow} \otimes (\hat{S}_a)^{1/2} + \ket{\uparrow}\bra{\uparrow} \otimes (\hat{S}_a^\dagger)^{1/2}, \\
        & C\hat{S}_b = \ket{\downarrow}\bra{\downarrow} \otimes (\hat{S}_b)^{1/2} + \ket{\uparrow}\bra{\uparrow} \otimes (\hat{S}_b^\dagger)^{1/2}.
    \end{align}
\end{subequations}
Following the QPE protocol, the controlled operators are surrounded by Hadamard rotations to map the interaction in and out of the $\hat{\sigma}_x$ basis. The QPE circuit also applies a $\hat{\sigma}_z$ rotation to the ancilla parameterised by rotation angles $\theta_{a,b}$, with the operator $\hat{Z}(\theta_{a,b})  = e^{- i \theta_{a,b} \hat{\sigma}_z}$. With this, the action of the conditional operators is
\begin{subequations}
    \label{eq:general_controlled_displacements_2}
    \begin{align}
        \tilde{C}\hat{S}_a \ket{\Psi_{\boldsymbol{\epsilon}}} = \hat{H}\hat{Z}(\theta_{a})  C\hat{S}_a \hat{H} \ket{\Psi_{\boldsymbol{\epsilon}}} = & 
        \frac{1}{2} \ket{\downarrow} \otimes  \left( e^{-i\theta_a/2} (\hat{S}_a)^{1/2}  + e^{i\theta_a/2} (\hat{S}_a^\dagger)^{1/2} \right)\ket{\psi_{\boldsymbol{\epsilon}}} \nonumber \\
        + & \frac{1}{2} \ket{\uparrow}\otimes  \left( e^{-i\theta_a/2}(\hat{S}_a)^{1/2}  - e^{i\theta_a/2} (\hat{S}_a^\dagger)^{1/2} \right) \ket{\psi_{\boldsymbol{\epsilon}}}, \\
        \tilde{C}\hat{S}_b \ket{\Psi_{\boldsymbol{\epsilon}}} = \hat{H}\hat{Z}(\theta_{b})  C\hat{S}_b \hat{H} \ket{\Psi_{\boldsymbol{\epsilon}}}= & 
        \frac{1}{2} \ket{\downarrow} \otimes  \left( e^{-i\theta_b/2} (\hat{S}_b)^{1/2}  + e^{i\theta_b/2} (\hat{S}_b^\dagger)^{1/2} \right)\ket{\psi_{\boldsymbol{\epsilon}}} \nonumber \\
        + & \frac{1}{2} \ket{\uparrow}\otimes  \left( e^{-i\theta_b/2}(\hat{S}_b)^{1/2}  - e^{i\theta_b/2} (\hat{S}_b^\dagger)^{1/2} \right) \ket{\psi_{\boldsymbol{\epsilon}}},
    \end{align}
\end{subequations}
In sections~\ref{sec:grid_state_prob_distribution} and \ref{sec:num_phase_prob_distribution}, we show that the action of the conditional operators of both grid and number--phase states can be written in this way.

In the following, we consider two cases: first, where conditional operators are applied separately, and we derive the independent probability distributions associated with measurement outcomes of $\tilde{C}\hat{S}_a \ket{\Psi}$ and $\tilde{C}\hat{S}_b \ket{\Psi}$; second, where the conditional operators are applied sequentially, and we derive the joint probability distribution.

\subsubsection{Independent probability distributions from separate measurements}

The independent probability distributions are derived by taking the expectation value of the ancilla after applying $\tilde{C}\hat{S}_a$ or $\tilde{C}\hat{S}_b$ to the state $\ket{\Psi_{\boldsymbol{\epsilon}}}$. We consider the positive operator-valued measurements (POVM) $\hat{Q}_0 = \ket{\downarrow}\bra{\downarrow}\otimes \hat{\mathds{1}}$ and $\hat{Q}_1 = \ket{\uparrow}\bra{\uparrow} \otimes \hat{\mathds{1}}$ with outcomes 0 and 1, respectively. The probability of measuring an outcome $\mathrm{m}_{a,b} \in \{0, 1\}$ is given by
\begin{subequations}
    \label{eq:general_prob_dist_independent}
    \begin{align}
        & \mathrm{P}_a(\mathrm{m}_a) = |\bra{\Psi_{\shiftparam}} ( \tilde{C}\hat{S}_a)^\dagger \hat{Q}_{\mathrm{m}_a} \tilde{C}\hat{S}_a \ket{\Psi_{\boldsymbol{\epsilon}}}|^2 =  \frac{1}{2}(1 +  (-1)^{\mathrm{m}_a} \langle \hat{O}_a \rangle_{\shiftparam}), \\
        & \mathrm{P}_b(\mathrm{m}_b) =|\bra{\Psi_{\shiftparam}} ( \tilde{C}\hat{S}_b)^\dagger \hat{Q}_{\mathrm{m}_b} \tilde{C}\hat{S}_b \ket{\Psi_{\boldsymbol{\epsilon}}}|^2  = \frac{1}{2}(1 +  (-1)^{\mathrm{m}_b} \langle \hat{O}_b \rangle_{\shiftparam}),
    \end{align}
\end{subequations}
where $\langle \cdot \rangle_{\shiftparam} = \bra{\psi_{\boldsymbol{\epsilon}}} \cdot \ket{\psi_{\boldsymbol{\epsilon}}}$ and we've defined
\begin{subequations}
    \label{eq:operators_oa_ob}
    \begin{align}
        & \hat{O}_a = \frac{1}{2}\left(e^{- i \theta_a} \hat{S}_a + e^{i \theta_a} \hat{S}_a^\dagger \right), \\
        & \hat{O}_b = \frac{1}{2} \left( e^{- i \theta_b} \hat{S}_b + e^{i \theta_b} \hat{S}_b^\dagger \right).
    \end{align}
\end{subequations}
Using the commutation relations of Eq.~\ref{eq:general_commutation_relations}, the expectation values of $\langle\hat{O}_a \rangle_{\shiftparam}$ and $\langle\hat{O}_b \rangle_{\shiftparam}$ are simplified to
\begin{subequations}
    \label{eq:general_expectation_O}
    \begin{align}
        &\langle \hat{O}_a\rangle_{\shiftparam} = \cos(\theta_a + l_a\epsilon_a) \frac{1}{2} \langle \hat{S}_a + \hat{S}_a^\dagger \rangle -  \sin(\theta_a + l_a\epsilon_a) \frac{i}{2} \langle \hat{S}_a - \hat{S}_a^\dagger \rangle,  \label{eq:general_expectation_O_a} \\
        &\langle \hat{O}_b\rangle_{\shiftparam} = \cos(\theta_b + l_b\epsilon_b) \frac{1}{2}\langle \hat{S}_b + \hat{S}_b^\dagger \rangle  -  \sin(\theta_b + l_b\epsilon_b)  \frac{i}{2}\langle \hat{S}_b - \hat{S}_b^\dagger \rangle. \label{eq:general_expectation_O_b}
    \end{align}
\end{subequations}

From Eq.~\ref{eq:general_prob_dist_independent}, obtaining the independent probability distributions $\mathrm{P}_a(\mathrm{m}_a)$ and $\mathrm{P}_b(\mathrm{m}_b)$ requires the calculation of $\langle \hat{O}_a\rangle_{\shiftparam}$ and $\langle \hat{O}_b\rangle_{\shiftparam}$ which, from Eq.~\ref{eq:general_expectation_O}, involves calculating the expectation values of the sum and difference of $\hat{S}_{a,b}$ and their conjugate transpose. In sections~\ref{sec:grid_state_prob_distribution} and \ref{sec:num_phase_prob_distribution}, it is shown that $\langle \hat{S}_a - \hat{S}_a^\dagger \rangle$ and $\langle \hat{S}_b - \hat{S}_b^\dagger\rangle$ reduce to zero for the grid states and (approximately) for the number--phase states. Hence, the right-hand sides of Eq.~\ref{eq:general_expectation_O_a} and Eq.~\ref{eq:general_expectation_O_b} vanish, and one retrieves probability distributions of the same form as Eq.~\ref{eq:general_prob_dist},
\begin{subequations}
    \begin{align}
        & \mathrm{P}_a(\mathrm{m}_a|\epsilon_a, \theta_a) = \frac{1}{2}\left(1 + (-1)^{\mathrm{m}_a} \eta_a \cos(\theta_a + l_a \epsilon_a)\right), \\
        & \mathrm{P}_b(\mathrm{m}_b|\epsilon_b, \theta_b) = \frac{1}{2}\left(1 + (-1)^{\mathrm{m}_b} \eta_b \cos(\theta_b + l_b \epsilon_b)\right).
    \end{align}
\end{subequations}
The visibility parameters $ \eta_a = \langle \hat{S}_a + \hat{S}_a^\dagger \rangle/2$ and $\eta_b = \langle \hat{S}_b + \hat{S}_b^\dagger \rangle / 2$ are determined from the expectation values of $\hat{S}_a$ and $\hat{S}_b$ and should be maximised to improve the metrological gain.

\subsubsection{Joint probability distribution from sequential measurements}

We now derive the joint probability distribution from sequential measurement of $\hat{S}_a$ followed by $\hat{S}_b$. We first aim to gain insights into the potential backaction that appears from the first measurement. To this end, we consider the state that is obtained from the application of $\tilde{C}\hat{S}_a$ followed by a measurement of the ancilla. Using the POVMs $\hat{Q}_{0}$ and $\hat{Q}_{1}$ defined above, the state after a measurement outcome $\mathrm{m}_a$ is 
\begin{align} 
    \label{eq:post_meas_state}
    \ket{\Psi}_{\mathrm{m}_a} = \frac{\hat{Q}_{\mathrm{m}_a} \tilde{C}\hat{S}_a  \ket{\Psi_{\boldsymbol{\epsilon}}} }{\mathrm{P}_a(\mathrm{m}_a)} \propto &  \left((1-\mathrm{m}_a)\ket{\downarrow} + \mathrm{m}_a \ket{\uparrow} \right) \otimes e^{- i \theta/2}  (\hat{S}_a)^{1/2} (\hat{\mathds{1}} +  (-1)^{\mathrm{m}_a} e^{i \theta } \hat{S}_a ) \hat{\mathcal{E}} \ket{\psi}.
\end{align}
For ideal sensing states, the operation $(\hat{\mathds{1}} +  (-1)^{\mathrm{m}_a} e^{i \theta } \hat{S}_a )$ in the right-hand side of Eq.~\ref{eq:post_meas_state} leaves the state unchanged since $\ket{\psi}$ is an eigenstate of $\hat{S}_a$. However, $(\hat{S}_a)^{1/2}$ will cause backaction on subsequent measurements. This can be seen from the commutation relation, $\hat{S}_b (\hat{S}_a)^{1/2} = - (\hat{S}_a)^{1/2} \hat{S}_b$: a first measurement of $\hat{S}_a$ will flip the outcome of a subsequent measurement of $\hat{S}_b$.

The joint probability distribution is found by calculating $\tilde{\mathrm{P}}(\mathrm{m}_a\mathrm{m}_b) = |\bra{\Psi}_{\mathrm{m}_a} (\tilde{C}\hat{S}_b)^\dagger \hat{Q}_{\mathrm{m}_b} \tilde{C}\hat{S}_b \ket{\Psi}_{\mathrm{m}_a}|^2$, which simplifies to
\begin{equation}
    \label{eq:joint_prob_dist_general}
    \tilde{\mathrm{P}}(\mathrm{m}_a\mathrm{m}_b) = \frac{1}{16}\bra{\psi_{\shiftparam}}  \hat{\alpha}^\dagger_{\mathrm{m}_a} \hat{\beta}^\dagger_{1 - \mathrm{m}_b} \hat{\beta}_{1 - \mathrm{m}_b} \hat{\alpha}_{\mathrm{m}_a} \ket{\psi_{\shiftparam}},
\end{equation}
where we've introduced the following notation,
\begin{subequations}
    \begin{align}
        & \hat{\alpha}_k = \hat{\mathds{1}} + (-1)^{k}e^{i \theta_a} \hat{S}_a^\dagger, \\
        & \hat{\beta}_k = \hat{\mathds{1}} + (-1)^k e^{i \theta_b} \hat{S}_b^\dagger.
    \end{align}
\end{subequations}
The joint-probability distribution is further simplified by noting that $\hat{\alpha}_k$ and $\hat{\beta}_k$ commute and that they are related to the operators $\hat{O}_a$ and $\hat{O}_b$ of Eq.~\ref{eq:operators_oa_ob} through
\begin{subequations}
    \begin{align}
        \label{eq:def_alpha_beta}
        & \hat{\alpha}^\dagger_k \hat{\alpha}_k =  2 (\hat{\mathds{1}} + (-1)^{k} \hat{O}_a), \\
        & \hat{\beta}^\dagger_k \hat{\beta}_k = 2 (\hat{\mathds{1}} + (-1)^k \hat{O}_b), 
    \end{align}
\end{subequations}
with which Eq.~\ref{eq:joint_prob_dist_general} becomes
\begin{align}
    \label{eq:general_joint_prob_dist}
    & \tilde{\mathrm{P}}(\mathrm{m}_a\mathrm{m}_b) = \frac{1}{4}(1 + (-1)^{\mathrm{m}_a} \langle \hat{O}_a\rangle_{\shiftparam} + (-1)^{\mathrm{m}_b+1} \langle \hat{O}_b \rangle_{\shiftparam} + (-1)^{\mathrm{m}_a + \mathrm{m}_b + 1} \langle \hat{O}_b \hat{O}_a\rangle_{\shiftparam}).
\end{align}
In this way, the probability distribution can be entirely characterised by calculating the expectation values $\langle \hat{O}_a\rangle_{\shiftparam}$, $\langle \hat{O}_b\rangle_{\shiftparam}$ and $\langle \hat{O}_b \hat{O}_a\rangle_{\shiftparam}$. In the following sections, we calculate these expectation values for the grid and number--phase states. We find that, for both sensing states, the joint expectation value $\langle \hat{O}_b \hat{O}_a \rangle_{\shiftparam}$ can be expressed as the product of individual expectation values. From this, the joint probability distribution can be approximated by a product of independent probability distributions, where the measurement outcome $\mathrm{m}_b$ is appropriately changed to take into account backaction from the first measurement.

\subsection{Grid-state probability distributions}
\label{sec:grid_state_prob_distribution}

We derive the independent and joint probability distributions associated with grid states by using the results of section~\ref{sec:general_prob_distribution} and replacing the labelling with $a \rightarrow x$ and $b \rightarrow p$. A measurement of position or momentum operators, $\hat{S}_x = e^{- i l_s \hat{x}}$ or $\hat{S}_p = e^{- i l_s \hat{p}}$, gives a measurement outcome $\mathrm{m}_x$ or $\mathrm{m}_p$, respectively. Interactions with the sensing signal is described by the operator $\hat{\mathcal{E}} = e^{i \epsilon_p \hat{x}}e^{-  i \epsilon_x \hat{p}}$, which is equivalent to a displacement $\hat{D}(\frac{\epsilon_x + i \epsilon_p}{\sqrt{2}})$ up to a phase. The initial state of the QPE circuit is
\begin{equation}
    \ket{\Psi_{\boldsymbol{\epsilon}}} = \ket{\downarrow} \otimes \ket{\tilde{\gridstate}_{\boldsymbol{\epsilon}}},
\end{equation}
where $\ket{\tilde{\gridstate}_{\boldsymbol{\epsilon}}} = \hat{\mathcal{E}} \ket{\tilde{\gridstate}}$. The conditional position and momentum operators are given by $C\hat{S}_x = e^{- i l_s \hat{\sigma}_x \hat{x}/2}$ and $C\hat{S}_p = e^{- i l_s \hat{\sigma}_x \hat{p}/2}$, and act on the ancilla qubit in the $\hat{\sigma}_x$ basis. Therefore, the Hadamard rotations that were implemented to obtain the results of Eq.~\ref{eq:general_controlled_displacements_2} are not required. Moreover, the ancilla rotation $\hat{R}_z(\theta_{a,b})$ is instead performed in the $\hat{\sigma}_x$ basis, $\hat{R}_x(\theta_{a,b}) = e^{- i \theta_{a,b} \hat{\sigma}_x}$. The resulting conditional position and momentum operators modify the state in the following way,
\begin{subequations}
    \begin{align}
        \tilde{C}\hat{S}_x \ket{\Psi_{\boldsymbol{\epsilon}}} =& \ket{\downarrow} \otimes \frac{1}{\sqrt{2}}(e^{-i\theta_x/2} (\hat{S}_x)^{1/2}  + e^{i\theta_x/2} (\hat{S}_x^\dagger)^{1/2})\ket{\tilde{\gridstate}_{\boldsymbol{\epsilon}}}  \nonumber \\
            + & \ket{\uparrow} \otimes \frac{1}{\sqrt{2}}(e^{-i\theta_x/2} (\hat{S}_x)^{1/2}  - e^{i\theta_x/2} (\hat{S}_x^\dagger)^{1/2} )\ket{\tilde{\gridstate}_{\boldsymbol{\epsilon}}}, \\
        \tilde{C}\hat{S}_p  \ket{\Psi_{\boldsymbol{\epsilon}}} =& \ket{\downarrow} \otimes \frac{1}{\sqrt{2}}(e^{-i\theta_p/2} (\hat{S}_p)^{1/2} + e^{i\theta_p/2} (\hat{S}_p^\dagger)^{1/2})\ket{\tilde{\gridstate}_{\boldsymbol{\epsilon}}} \nonumber \\
            + & \ket{\uparrow} \otimes \frac{1}{\sqrt{2}}(e^{-i\theta_p/2} (\hat{S}_p)^{1/2} - e^{i\theta_p/2}(\hat{S}_p^\dagger)^{1/2})\ket{\tilde{\gridstate}_{\boldsymbol{\epsilon}}}.
    \end{align}
\end{subequations}
which corresponds to Eq.~\ref{eq:general_controlled_displacements_2}.

We verify that the grid state operators and sensing operator follow the commutation relations of Eq.~\ref{eq:general_commutation_relations}, where, using the relation $e^{i \alpha \hat{x}} e^{i \beta \hat{p}} = e^{-  i \alpha \beta} e^{i \beta \hat{p}} e^{i \alpha \hat{x}}$,
\begin{subequations}
    \begin{align}
        & (\hat{S}_x)^{1/2} \hat{\mathcal{E}} = e^{- i l_s \hat{x}/2} e^{i \epsilon_p \hat{x}}e^{-  i \epsilon_x \hat{p}} =  e^{- i l_s \epsilon_x/2} e^{i \epsilon_p \hat{x}}e^{-  i \epsilon_x \hat{p}} e^{- i l_s \hat{x}/2} = e^{- i l_s \epsilon_x/ 2} \hat{\mathcal{E}} (\hat{S}_x)^{1/2}, \\
        & (\hat{S}_p)^{1/2} \hat{\mathcal{E}} = e^{- i l_s \hat{p}/2} e^{i \epsilon_p \hat{x}}e^{-  i \epsilon_x \hat{p}} =  e^{- i l_s \epsilon_p / 2} e^{i \epsilon_p \hat{x}}e^{-  i \epsilon_x \hat{p}} e^{- i l_s \hat{p}/2} = e^{- i l_s \epsilon_p / 2} \hat{\mathcal{E}} (\hat{S}_p)^{1/2}.
    \end{align}
\end{subequations}

\subsubsection{Independent probability distributions}

We derive the independent probability distributions of outcomes $\mathrm{m}_x$ and $\mathrm{m}_p$, which are given by separate measurements of $\hat{S}_x$ and $\hat{S}_p$ with the displaced grid state, $\ket{\tilde{\gridstate}_{\boldsymbol{\epsilon}}}$. To this end, we calculate the expectation values $\langle \hat{O}_x \rangle $ and $\langle \hat{O}_p \rangle $ from Eq.~\ref{eq:general_expectation_O}, giving 
\begin{subequations}
    \label{eq:expectation_o_grid_state_1}
    \begin{align}
        &\langle \hat{O}_x\rangle_{\shiftparam} = \cos(\theta_x + l_s\epsilon_x) \frac{1}{2} \langle \hat{S}_x + \hat{S}_x^\dagger \rangle -  \sin(\theta_x + l_s\epsilon_x) \frac{i}{2} \langle \hat{S}_x - \hat{S}_x^\dagger \rangle,  \\
        &\langle \hat{O}_p\rangle_{\shiftparam} = \cos(\theta_p + l_s\epsilon_p) \frac{1}{2}\langle \hat{S}_p + \hat{S}_p^\dagger \rangle  -  \sin(\theta_p + l_s\epsilon_p)  \frac{i}{2}\langle \hat{S}_p - \hat{S}_p^\dagger \rangle.
    \end{align}
\end{subequations}
The expectation values of $\hat{S}_x$ and $\hat{S}_p$ correspond to points of the characteristic function of the grid state,
\begin{subequations}
    \begin{align}
        & \frac{1}{2} \langle \hat{S}_x + \hat{S}_x^\dagger \rangle = \frac{1}{2} \langle \hat{D}(-i \sqrt{\pi}) + \hat{D}^\dagger(-i \sqrt{\pi})) = \mathrm{Re}[\bigchi(i \sqrt{\pi})], \\
        & \frac{i}{2} \langle \hat{S}_x - \hat{S}_x^\dagger \rangle = \frac{i}{2} \langle \hat{D}(-i \sqrt{\pi}) - \hat{D}^\dagger(-i \sqrt{\pi})) = \mathrm{Im}[\bigchi(i \sqrt{\pi})], \\
        & \frac{1}{2} \langle \hat{S}_p + \hat{S}_p^\dagger \rangle = \frac{1}{2} \langle \hat{D}( \sqrt{\pi}) + \hat{D}^\dagger( \sqrt{\pi})) = \mathrm{Re}[\bigchi(-\sqrt{\pi})] = \mathrm{Re}[\bigchi(\sqrt{\pi})], \\
        & \frac{i}{2} \langle \hat{S}_x - \hat{S}_x^\dagger \rangle = \frac{i}{2} \langle \hat{D}( \sqrt{\pi}) - \hat{D}^\dagger( \sqrt{\pi}) = \mathrm{Im}[\bigchi(-\sqrt{\pi})], 
    \end{align}
\end{subequations}
where $\mathrm{Re}[\bigchi(\beta)]$ and $\mathrm{Im}[\bigchi(\beta)]$ are the real and imaginary parts of the characteristic function, respectively, and we've made use of their symmetries (see section~\ref{sec:notation}). The imaginary part, $\mathrm{Im}[\bigchi(\beta)]$,  reduces to zero for the grid states. With this, the expectation values of Eq.~\ref{eq:expectation_o_grid_state_1} become
\begin{subequations}
    \begin{align}
        &\langle \hat{O}_x\rangle_{\shiftparam} = \eta_x \cos(\theta_x + l_s\epsilon_x),  \\
        &\langle \hat{O}_p\rangle_{\shiftparam} = \eta_p \cos(\theta_p + l_s\epsilon_p),
    \end{align}
\end{subequations}
where we've defined the visibility parameters $\eta_x = \mathrm{Re}[\bigchi(i \sqrt{\pi})]$ and $\eta_p = \mathrm{Re}[\bigchi( \sqrt{\pi})]$. From this, we find the independent probability distributions,
\begin{subequations}
    \label{eq:independent_prob_dist_grid}
    \begin{align}
        \mathrm{P}_x(\mathrm{m}_x|\epsilon_x, \theta_x) = \frac{1}{2}(1 + (-1)^{\mathrm{m}_x} \eta_{x}\cos(\theta_x + l_s \epsilon_x)), \\
        \mathrm{P}_p(\mathrm{m}_p|\epsilon_p, \theta_p) = \frac{1}{2}(1 + (-1)^{\mathrm{m}_p} \eta_{p}\cos(\theta_p + l_s \epsilon_p)),
    \end{align}
\end{subequations}
and their joint probability distribution assuming independent measurements is 
\begin{align}
    \label{eq:joint_prob_distribution_grid_independent}
    \mathrm{P}(\mathrm{m}_x\mathrm{m}_p|\epsilon_x, \epsilon_p, \theta_x, \theta_p)   = & \mathrm{P}_x(\mathrm{m}_x|\epsilon_x, \theta_x)\mathrm{P}_p(\mathrm{m}_p|\epsilon_p, \theta_p) \nonumber \\ 
      = & \frac{1}{4}\Big(1 + (-1)^{\mathrm{m}_x} \eta_x \cos(\theta_x + l_s \epsilon_x) +  (-1)^{\mathrm{m}_p} \eta_p \cos(\theta_p + l_s \epsilon_p)  \nonumber \\
    & + (-1)^{\mathrm{m}_x + \mathrm{m}_p} \eta_x\eta_p \cos(\theta_x + l_s \epsilon_x)\cos(\theta_p + l_s \epsilon_p)\Big).
\end{align}

\subsubsection{Joint probability distribution for two sequential measurements}

We now calculate the joint-probability distribution from a measurement of $\hat{S}_x$ followed by a measurement of $\hat{S}_p$ using the generalised result of Eq.~\ref{eq:general_joint_prob_dist}. The joint expectation value is
\begin{align}
    \langle \hat{O}_p \hat{O}_x\rangle_{\shiftparam} = \frac{1}{4} \Big( & e^{-i (\theta_x + \theta_p)}e^{- i l_s ( \epsilon_x + \epsilon_p)} \langle \hat{S}_p\hat{S}_x \rangle +   e^{i (\theta_x + \theta_p)}e^{ i l_s (\epsilon_x + \epsilon_p)} \langle \hat{S}_p^\dagger\hat{S}_x^\dagger\rangle \nonumber \\
    &+  e^{- i (\theta_x - \theta_p)} e^{ - i l_s (\epsilon_x - \epsilon_p)} \langle \hat{S}_p^\dagger\hat{S}_x \rangle  +  e^{ i (\theta_x - \theta_p)} e^{ i l_s (\epsilon_x - \epsilon_p)} \langle \hat{S}_p \hat{S}_x^\dagger \rangle \Big).
\end{align}
This expression is simplified by again noting that the expectation values of the displacement operators correspond to values of the characteristic function, and that the imaginary part of the characteristic function is zero for grid states, giving
\begin{align}\label{eq:expect_Op_Ox}
    \langle \hat{O}_p \hat{O}_x\rangle_{\shiftparam} = \frac{\eta_{x,p}}{2} \cos(\theta_x + \theta_p + l_s\epsilon_x + l_s \epsilon_x) + \frac{\eta_{x,p}'}{2} \cos(\theta_x - \theta_p + l_s \epsilon_x - l_s \epsilon_p )),
\end{align}
where the joint visibility parameters are $\eta_{x,p} = \frac{1}{2}\langle \hat{S}_p \hat{S}_x + \hat{S}_p^\dagger \hat{S}_x^\dagger \rangle = - \mathrm{Re}[\bigchi\left(\sqrt{\pi}(1-i)\right)]$ and $\eta_{x,p}' =  \frac{1}{2}\langle \hat{S}_p^\dagger \hat{S}_x + \hat{S}_p \hat{S}_x^\dagger \rangle = - \mathrm{Re}[\bigchi\left(\sqrt{\pi}(1+i)\right)]$. We further simplify Eq.~\ref{eq:expect_Op_Ox} by approximating $\eta_{x,p} \approx \eta_{x,p}'$, due to the symmetry of the characteristic function of the grid state. With this, the joint expectation value becomes
\begin{equation}
    \langle \hat{O}_p \hat{O}_x\rangle_{\shiftparam} =  \eta_{x,p} \cos(\theta_x + l_s \epsilon_x) \cos(\theta_p + l_s \epsilon_p).
\end{equation}
Plugging this back into the generalised result of  Eq.~\ref{eq:general_joint_prob_dist}, the joint probability distribution from sequential measurements is
\begin{align}
    \label{eq:joint_prob_distribution_grid_sequential_a}
    \tilde{\mathrm{P}}(\mathrm{m}_x\mathrm{m}_p| \epsilon_x, \epsilon_p, \theta_x, \theta_p) =  \frac{1}{4} \Big( & 1 + (-1)^{\mathrm{m}_x} \eta_x \cos(\theta_x + l_s \epsilon_x) + (-1)^{\mathrm{m}_p + 1} \eta_p \cos(\theta_p + l_s \epsilon_p)  \nonumber \\
    + & (-1)^{\mathrm{m}_x + \mathrm{m}_p + 1} \eta_{x, p} \cos(\theta_x + l_s \epsilon_x) \cos(\theta_p + l_s \epsilon_p) \Big).
\end{align}
The distribution $\tilde{\mathrm{P}}(\mathrm{m}_x\mathrm{m}_p| \epsilon_x, \epsilon_p, \theta_x, \theta_p)$ has the same form as the distribution $\mathrm{P}(\mathrm{m}_x\mathrm{m}_p| \epsilon_x, \epsilon_p, \theta_x, \theta_p)$ of Eq.~\ref{eq:joint_prob_distribution_grid_independent} obtained from independent measurements, with a few differences outlined below. We first observe that the joint visibility parameter is modified as $\eta_{x,p} \rightarrow \eta_x \eta_p $. The visibility parameters can be related by $\eta_{x,p} \approx  \eta_{x}\eta_p$ due to the Gaussian nature of the damping envelope, and the validity of this approximation increases for grid states with higher energies where $\eta_x \approx \eta_p \approx \eta_{x,p} \approx 1$. For the QPE demonstration of Fig.~3 which used a grid state with $\Delta = 0.41$, we find that the difference between $\eta_{x, p}$ and $\eta_{x}\eta_p$ is less than $10^{-4}$, validating this approximation within the finite-energy constraint of the experiment. After replacing $\eta_{x,p} \rightarrow \eta_x \eta_p$ in Eq.~\ref{eq:joint_prob_distribution_grid_sequential_a}, $\tilde{\mathrm{P}}(\mathrm{m}_x\mathrm{m}_p| \epsilon_x, \epsilon_p, \theta_x, \theta_p)$ can be made equal to $\mathrm{P}(\mathrm{m}_x\mathrm{m}_p| \epsilon_x, \epsilon_p, \theta_x, \theta_p)$ by replacing $\mathrm{m}_p \rightarrow \mathrm{m}_p + 1 $. This difference arises from backaction of the first measurement of $\hat{S}_x$ which flips the measurement outcome of the subsequent $\hat{S}_p$ measurement.

With the previous changes, we can therefore approximate the joint probability distribution by the product of independent probability distributions,  
\begin{equation}
    \tilde{\mathrm{P}}(\mathrm{m}_x \mathrm{m}_p|\epsilon_x, \epsilon_p, \theta_x, \theta_p) \approx \mathrm{P}_x(\mathrm{m}_x|\epsilon_x, \theta_x) \mathrm{P}_p(\mathrm{m}_p + 1|\epsilon_p, \theta_p).
\end{equation}
This approximation is extensively used in the QPE demonstration with Bayesian inference of Fig.~3, as it allows one to separately perform phase estimation to independently estimate the parameters $\epsilon_x$ and $\epsilon_p$. Moreover, writing the the joint distribution as a product of independent distributions makes the expression straightforwardly describable by a Fourier series which speeds up calculations and optimizations for adapative QPE (see section~\ref{sec:qpe}).

\subsubsection{Joint probability distribution for many sequential measurements}

\begin{figure}
    \centering
    \includegraphics[]{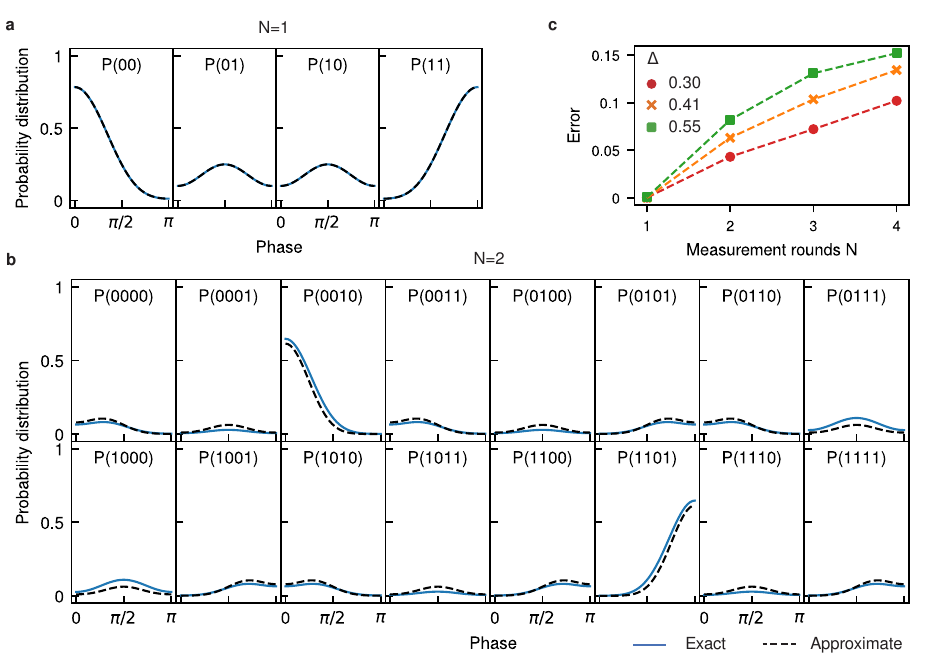}
    \caption{
    \textbf{Probability distributions of grid states with increasing sequential measurements.}
    \textbf{(a, b)} Probability distributions $\mathrm{P}(\mathbf{m}|\epsilon_x, \epsilon_p)$, with $N_S=1$ and $N_S=2$ rounds of sequential measurements, giving bitstrings $\mathbf{m} = \mathrm{m}_0 ... \mathrm{m}_{2N_S-1}$ of length 2 and 4, respectively. Exact probabilities (solid blue line) are calculated from Eq.~\ref{eq:general_prob_distribution_grid_sequential}. Expectation values $\langle \hat{O}_x^a \hat{O}_p^b \rangle$ are numerically computed from the finite-energy grid state of Eq.~\ref{eq:finite_energy_grid_state} with $\Delta=0.41$. Approximate distributions (dashed black line) are calculated from the product of independent probability distributions (see Eq.~\ref{eq:joint_prob_dist_general_approx}). The controllable phases are varied in the range $\theta_x=\theta_p \in [0, \pi]$ and the displacement signals are $\epsilon_x = \epsilon_p = 0$.
    \textbf{(c)} Scaling of the error between exact and approximate probability distributions for increasing $N_S$. The error is calculated as the largest difference between the exact and approximate probability distributions, $\tilde{P}$ and $P$, over all possible measurement outcomes, $\mathrm{max}_\mathbf{m}|\tilde{P} - P|$.
    }
    \label{fig:prob_distribution_gkp_comparison}
\end{figure}

The joint probability distribution for dependent measurements derived above is now generalised to an arbitrary number of sequential measurements. Specifically, we consider a scenario where the sensing state is prepared once, interacts with the sensing signal, and $N_S$ rounds of sequential $\hat{S}_x$ and $\hat{S}_p$ measurements are performed. This gives $2N_S$ measurement outcomes described by the bitstring $\mathbf{m} = \mathrm{m}_{0}\mathrm{m}_1\mathrm{m}_2 ... \mathrm{m}_{2N_S-1}$, where even and odd bits correspond to measurements associated with $\hat{S}_x$ and $\hat{S}_p$, respectively. Without loss of generality, we assume that the phases $\theta_x$ and $\theta_p$ are equal for all measurements of $\hat{S}_x$ or $\hat{S}_p$. Following the notation and derivations of section~\ref{sec:general_prob_distribution}, the generalised probability distribution is
\begin{align}
    \tilde{\mathrm{P}}(\mathbf{m}| \epsilon_x, \epsilon_p, \theta_x, \theta_p) = \frac{1}{2^{4N_S}} \langle \prod_{n=0}^{N_S-1} \hat{\beta}_{\mathrm{m}_{2n+1} + s(n+1)}^\dagger \hat{\beta}_{\mathrm{m}_{2n+1} + s(n+1)} \hat{\alpha}^\dagger _{\mathrm{m}_{2n} + s(n)}\hat{\alpha}_{\mathrm{m}_{2n} + s(n)} \rangle_{\shiftparam},
\end{align}
and the measurement outcomes are modified by the function $s(n) = n ~\mathrm{mod}~2$. This generalised probability distribution is further simplified using the relations of Eq~\ref{eq:def_alpha_beta}, and can be expressed as a sum of expectation values of powers of $\hat{O}_x$ and $\hat{O}_p$,
\begin{align}
    \label{eq:general_prob_distribution_grid_sequential}
    \tilde{\mathrm{P}}(\mathbf{m}| \epsilon_x, \epsilon_p, \theta_x, \theta_p) = \frac{1}{2^{4N_S-2}} \langle \prod^{N_S-1}_{n=0} ( \hat{\mathds{1}} + (-1)^{\mathrm{m}_{2n+1} + s(n+1)} \hat{O}_{p}  ) \times (\hat{\mathds{1}} + (-1)^{\mathrm{m}_{2n} + s(n)} \hat{O}_{x})  \rangle_{\shiftparam},
\end{align}
where the joint expectation values are given by
\begin{align}
    \langle \hat{O}_{x}^a \hat{O}_{p}^b\rangle_{\shiftparam} = \frac{1}{2^{(a + b)}}  \sum^a_{j=0} \sum^b_{k=0} \binom{a}{j} \binom{b}{k}  & \mathrm{exp}\left[- i\left((a - 2j)\theta_x + (b - 2k)\theta_p \right)\right] \langle \hat{S}_x^{(a-2j)} \hat{S}_p^{(b-2k)} \rangle_{\shiftparam},
\end{align}
with
\begin{align}
    \label{eq:joint_expectation_sx_sp}
     \langle \hat{S}_x^{(a-2j)} \hat{S}_p^{(b-2k)} \rangle_{\shiftparam} = & (-1)^{(a-2j)(b-2k)} \mathrm{exp}[-i (a - 2j) \epsilon_x l_s] \mathrm{exp}[- i (b - 2k) \epsilon_p l_s] \nonumber \\
     \times  & \mathrm{Re}[\bigchi(\sqrt{\pi}\left( i (a - 2j) - (b-2k)\right)].
\end{align}
From Eq.~\ref{eq:joint_expectation_sx_sp}, one can see that the joint-probability distribution can be entirely characterised by points of the real part of the characteristic function lying on a square grid of spacing $\sqrt{\pi}$. This provides an efficient characterization tool, where the metrological gain of a grid state for displacement sensing can be entirely estimated from only a few measurements.

The generalised probability distribution of Eq.~\ref{eq:general_prob_distribution_grid_sequential} can be simplified by noting that, for an approximate grid state that is subject to a Gaussian damping envelope, the value of points of the characteristic function lying on a square lattice of spacing $\sqrt{\pi}$ can be approximately related to $\eta_x = \mathrm{Re}[\bigchi(i \sqrt{\pi})]$ and $\eta_p = \mathrm{Re}[\bigchi(\sqrt{\pi})]$. To see this, we first approximate $\mathrm{Re}[\bigchi\left(i \sqrt{\pi}(a - 2j) - \sqrt{\pi} (b-2k)\right)] \approx (-1)^{(a-2j)(b-2k)}\mathrm{Re}[\bigchi(i\sqrt{\pi}(a - 2j))] \mathrm{Re}[\bigchi(\sqrt{\pi} (b-2k))]$. Then, noting that the points of the characteristic function are approximately given by a Gaussian envelope, we approximate $\mathrm{Re}[\bigchi(i \sqrt{\pi}(a-2j))] = \mathrm{Re}[\bigchi(i \sqrt{\pi})]^{(a-2j)^2}$ and $\mathrm{Re}[\bigchi(\sqrt{\pi}(b-2k))] = \mathrm{Re}[\bigchi(\sqrt{\pi})]^{(b-2k)^2}$. With this, any point of the characteristic function lying on the square lattice can be approximated as\linebreak $\mathrm{Re}[\bigchi\left(i \sqrt{\pi} (a - 2j) - \sqrt{\pi} (b-2k)\right)] \approx (-1)^{(a-2j)(b-2k)} \mathrm{Re}[\bigchi(i \sqrt{\pi})]^{(a-2j)^2} \mathrm{Re}[\bigchi(\sqrt{\pi} )]^{(b-2k)^2} = (-1)^{(a-2j)(b-2k)} \eta_x^{(a-2j)^2} \eta_p^{(b-2k)^2}$. The joint expectation value of Eq.~\ref{eq:joint_expectation_sx_sp} then becomes
\begin{align}
     \langle \hat{S}_x^{(a-2j)} \hat{S}_p^{(b-2k)} \rangle_{\shiftparam} = &  \mathrm{exp}[-i (a - 2j) \epsilon_x l_s] \mathrm{exp}[- i (b - 2k) \epsilon_p l_s] \eta_x^{(a-2j)^2} \eta_p^{(b-2k)^2}.
\end{align}

As the squeezing parameter of the grid states approaches $\Delta \rightarrow 0$ and the visibility parameters approach $\eta_x^m \approx \eta_p^n \approx 1$, the joint-probability distribution of Eq.~\ref{eq:general_prob_distribution_grid_sequential} can be approximated by the product of independent probability distributions, after appropriately changing the measurement outcomes to account for the backaction of the measurements,
\begin{equation}
    \label{eq:joint_prob_dist_general_approx}
    \tilde{\mathrm{P}}(\mathbf{m}|\epsilon_x, \epsilon_p, \theta_x, \theta_p) \approx \prod_{n=0}^{N_S-1} \mathrm{P}_x(\mathrm{m}_{2n} + s(n)|\epsilon_x, \theta_x) \mathrm{P}_p(\mathrm{m}_{2n+1} + s(n+1)| \epsilon_p, \theta_p).
\end{equation}
The quality of this approximation is investigated in Fig.~\ref{fig:prob_distribution_gkp_comparison}. The exact and approximate probability distributions for $N_S=1$ show great agreement (Fig.~\ref{fig:prob_distribution_gkp_comparison}a), whereas small discrepancies arise for $N_S=2$ (Fig.~\ref{fig:prob_distribution_gkp_comparison}b). We further verify the scaling of the error from this approximation by computing the maximum difference between the exact and approximate probability distributions (Fig.~\ref{fig:prob_distribution_gkp_comparison}c). As expected, the error grows for larger $N_S$, where the values of the characteristic function at larger distances from the origin are badly approximated by-products of $\eta_x$ and $\eta_p$. We further verify that the error is reduced for grid states with larger squeezings, where the values of the characteristic function better approximate an ideal grid state. Altogether, we deem the quality of the approximation up to $N_S=3$ sufficient, and model the joint distribution as a product of independent distributions for the QPE experiments in this work.

\subsection{Number--phase state probability distributions}
\label{sec:num_phase_prob_distribution}

We derive the independent and joint probability distributions associated with number--phase states by using the general results of section~\ref{sec:general_prob_distribution} and replacing the labelling with $a \rightarrow \phi$ and $b \rightarrow n$. Measuring the phase operator $\hat{S}_\phi$ and number operator $\hat{S}_n$ gives measurement outcomes $\mathrm{m}_\phi$ and $\mathrm{m}_n$, respectively. We first relate these operations to the generalised conditional operators of Eq.~\ref{eq:general_conditional_stabilisers}. The conditional phase and number operators are (see section~\ref{sec:conditional_phase_stabiliser} and section~\ref{sec:conditional_rotation_op}) 
\begin{subequations}
    \begin{align}
        & C\hat{S}_\phi = \ket{\downarrow}\bra{\downarrow} \otimes \hat{E}^+_{l_\phi/2} + \ket{\uparrow}\bra{\uparrow} \otimes \hat{E}^-_{l_\phi/2}, \\
        & C\hat{S}_n = \ket{\downarrow} \bra{\downarrow} \otimes \hat{R}_{l_n/2} + \ket{\uparrow} \bra{\uparrow} \otimes \hat{R}^\dagger_{l_n/2},
    \end{align}
\end{subequations}
and we recall that $\hat{E}^+_N = \sum_k \ket{k+N} \bra{k}$ is a phonon shift and $\hat{R}_\theta = e^{- i \theta \hat{n}}$ is a rotation. The number and phase conditional operators correspond to the generalised operators of Eq.~\ref{eq:general_conditional_stabilisers} with $(\hat{S}_\phi)^{1/2} = \hat{E}^+_{l_\phi/2}$ and $(\hat{S}_n)^{1/2} = \hat{R}_{l_n/2}$. The number--phase state is subjected to a sensing signal which is described by a phonon shift by an integer $\epsilon_n$, followed by a rotation by $\epsilon_\phi \in \mathbb{R}$, $\hat{\mathcal{E}} = \hat{R}_{-\epsilon_\phi} \hat{E}^+_{\epsilon_n} $, and the initial state of the QPE circuit is
\begin{equation}
    \ket{\Psi_{\boldsymbol{\epsilon}}} = \ket{\downarrow} \otimes \ket{\tilde{\npstate}_{\boldsymbol{\epsilon}}},
\end{equation}
where $\ket{\tilde{\npstate}_{\boldsymbol{\epsilon}}} =\hat{\mathcal{E}} \ket{\tilde{\npstate}}$. We verify that the number--phase state operators and sensing operator follow the commutation relations of Eq.~\ref{eq:general_commutation_relations}, where, using the relation $\hat{R}_{\theta} \hat{E}^+_n  = e^{- i \theta n} \hat{E}^+_n \hat{R}_{\theta}$ , 
\begin{subequations}
    \begin{align}
        & (\hat{S}_\phi)^{1/2} \hat{\mathcal{E}} = \hat{E}^+_{l_\phi/2} \hat{R}_{-\epsilon_\phi} \hat{E}^+_{\epsilon_n} = e^{- i l_\phi \epsilon_\phi/2}  \hat{R}_{-\epsilon_\phi} \hat{E}^+_{\epsilon_n} \hat{E}^+_{l_\phi/2} =  e^{- i l_\phi \epsilon_\phi/2} \hat{\mathcal{E}} (\hat{S}_\phi)^{1/2}, \\
        & (\hat{S}_n)^{1/2} \hat{\mathcal{E}} = \hat{R}_{l_n/2} \hat{R}_{-\epsilon_\phi} \hat{E}^+_{\epsilon_n} = e^{- i l_n \epsilon_n/2}   \hat{R}_{-\epsilon_\phi} \hat{E}^+_{\epsilon_n} \hat{R}_{l_n/2} =  e^{- i l_n \epsilon_n/2} \hat{\mathcal{E}} (\hat{S}_n)^{1/2}.
    \end{align}
\end{subequations}
The action of the conditional number and phase operators on the sensing state after applications of Hadamards and ancilla rotations are given by the generalised results of Eq.~\ref{eq:general_controlled_displacements_2}, with $\theta_a \rightarrow \theta_\phi$ and $\theta_b \rightarrow \theta_n$. 

\subsubsection{Independent probability distributions}

We derive the independent probability distributions for measurement outcomes $\mathrm{m}_\phi$ or $\mathrm{m}_n$, associated with measurements of $\hat{S}_\phi$ or $\hat{S}_n$, respectively. Following the results of section~\ref{sec:general_prob_distribution}, we calculate the expectation values $\langle \hat{O}_\phi \rangle $ and $\langle \hat{O}_n \rangle $ of Eq.~\ref{eq:general_expectation_O}, giving 
\begin{subequations}
    \label{eq:expectation_o_np_state_1}
    \begin{align}
        &\langle \hat{O}_\phi \rangle_{\shiftparam} = \cos(\theta_\phi + l_\phi \epsilon_\phi) \frac{1}{2} \langle\hat{E}^+_{l_\phi} + \hat{E}^-_{l_\phi} \rangle -  \sin(\theta_\phi + l_\phi \epsilon_\phi) \frac{i}{2} \langle \hat{E}^+_{l_\phi} - \hat{E}^-_{l_\phi} \rangle,  
        \label{eq:O_ell_np_states}
        \\
        &\langle \hat{O}_n \rangle_{\shiftparam} = \cos(\theta_n + l_n \epsilon_n) \frac{1}{2}\langle \hat{R}_{l_n} + \hat{R}_{l_n}^\dagger \rangle - \sin(\theta_n + l_n \epsilon_n)  \frac{i}{2}\langle\hat{R}_{l_n} - \hat{R}_{l_n}^\dagger \rangle. \label{eq:O_r_np_states}
    \end{align}
\end{subequations}
The sine and airy approximations to number-phase states have real Fock-space coefficients $c_{kN}$, so $\text{Im} [\langle \hat{E}^-_{l_\mathrm{\phi}} \rangle ] = 0$, and the second term in Eq.~\ref{eq:O_ell_np_states} vanishes.
These states are also eigenstates of $\hat{R}_{l_n}$ with eigenvalue given by a root of unity, $e^{i  l_n \lambda}$, determined by the offset $\lambda$, recalling that $ l_n = \frac{2 \pi}{N}$. This gives $\frac{1}{2}\langle \hat{R}_{l_n} + \hat{R}_{l_n}^\dagger \rangle = \cos  l_n \lambda$ and $-\frac{i}{2}\langle \hat{R}_{l_n} - \hat{R}_{l_n}^\dagger \rangle = \sin l_n \lambda$. With these considerations, the expectation values in Eq.~\ref{eq:expectation_o_np_state_1} become
\begin{subequations}
    \label{eq:O_phi_O_n_simplified}
    \begin{align}
        \langle \hat{O}_\phi \rangle_{\shiftparam} &= \eta_\phi \cos(\theta_\phi + l_\phi \epsilon_\phi)  \\
        \langle \hat{O}_n \rangle_{\shiftparam}  &= \cos \big(\theta_n + l_n (\epsilon_n - \lambda) \big) 
    \end{align}
\end{subequations}
with visibility parameter $\eta_\phi = \Re[\langle \hat{E}^-_{l_\phi} \rangle] = \langle \hat{E}^-_{l_\phi} \rangle = \sum_{k=0}^\infty c^*_{k l_\phi} c_{(k+1) l_\phi}$~\cite{Grimsmo2020} for number--phase states with real Fock-basis coefficients. The independent probability distributions are then
\begin{subequations}
    \begin{align}
        & \mathrm{P}_\phi(\mathrm{m}_\phi|\epsilon_\phi, \theta_\phi) 
        = \frac{1}{2}\Big(1 + (-1)^{\mathrm{m}_\phi} \eta_{\phi} \cos(\theta_\phi + l_\phi \epsilon_\phi)\Big), \\
        & \mathrm{P}_n(\mathrm{m}_n|\epsilon_n, \theta_n) 
        = \frac{1}{2} \Big( 1 + (-1)^{\mathrm{m}_n} \cos(\theta_n + l_n \big(\epsilon_n - \lambda) \big) \Big).
    \end{align}
\end{subequations}

\subsubsection{Sequential probability distributions}

We now calculate the joint probability distribution from sequential measurements of the phase and number operators. The joint expectation value of Eq.~\ref{eq:general_joint_prob_dist} is
\begin{align}
    \langle \hat{O}_n \hat{O}_\phi \rangle = \frac{1}{4} \Big( & e^{-i (\theta_\phi + \theta_n)}e^{-i (l_\phi  \epsilon_\phi + l_n \epsilon_n)} \langle \hat{S}_n\hat{S}_\phi \rangle +  e^{i (\theta_\phi + \theta_n)}e^{i (l_\phi  \epsilon_\phi + l_n \epsilon_n)}  \langle \hat{S}_n^\dagger\hat{S}_\phi^\dagger\rangle  \nonumber \\
    + & e^{- i (\theta_\phi - \theta_n)}e^{-i ( l_\phi  \epsilon_\phi - l_n \epsilon_n)}  \langle \hat{S}_n^\dagger\hat{S}_\phi \rangle  + e^{ i (\theta_\phi - \theta_n)}e^{ i (l_\phi  \epsilon_\phi - l_n \epsilon_n)}  \langle \hat{S}_n \hat{S}_\phi^\dagger \rangle \Big).
\end{align}
This expression is simplified with the same relations that were used to obtain Eq.~\ref{eq:O_phi_O_n_simplified}, giving 
\begin{equation}
    \langle \hat{O}_n \hat{O}_\phi \rangle =  \eta_{\phi} \cos(\theta_\phi + l_\phi \epsilon_\phi) \cos(\theta_n + l_n (\epsilon_n - \lambda)).
\end{equation}
Plugging this back into the generalised result of Eq.~\ref{eq:general_joint_prob_dist}, we find the joint probability distribution of NP states,
\begin{align}\label{eq:joint_prob_distribution_grid_sequential}
    \tilde{\mathrm{P}}(\mathrm{m}_\phi \mathrm{m}_n| \epsilon_\phi, \epsilon_n, \theta_\phi, \theta_n) =  \frac{1}{4} \Big( & 1 + (-1)^{\mathrm{m}_\phi} \eta_\phi \cos(\theta_\phi + l_\phi \epsilon_\phi) + (-1)^{\mathrm{m}_n + 1} \cos(\theta_n + l_n (\epsilon_n - \lambda)) \nonumber \\
    + & (-1)^{\mathrm{m}_\phi + \mathrm{m}_n + 1} \eta_{\phi} \cos(\theta_\phi + l_\phi \epsilon_\phi) \cos(\theta_n + l_n (\epsilon_n - \lambda) ) \Big).
\end{align}
Similarly to the grid states, the joint probability distribution of the number--phase state can be expressed from the product of independent distributions, after adjusting the second measurement outcome to take into account backaction from the first measurement. With this, the joint distribution is approximated as
\begin{equation}
    \tilde{\mathrm{P}}(\mathrm{m}_\phi \mathrm{m}_n|\epsilon_\phi, \epsilon_n, \theta_\phi, \theta_n) = \mathrm{P}_\phi(\mathrm{m}_\phi | \epsilon_\phi, \theta_\phi) \mathrm{P}_n(\mathrm{m}_n + 1| \epsilon_n, \theta_n)
\end{equation}

%
%

\section{Characterizing the metrological gain}

The metrological gain of the grid and number--phase states is characterised from the classical Fisher information of the experimentally reconstructed probability distributions, and we adapt the procedure of Ref.~\cite{Wolf2019} to the multiparameter case. We estimate the multi-parameter uncertainty from the classical Cram\'er-Rao bound, $\mathbf{\Sigma} \geq \mathbf{F}^{-1}$, which bounds the covariance matrix $\mathbf{\Sigma}$ by the inverse of the Fisher information matrix (FIM), $\mathbf{F}$~\cite{Cramr1946, Rao1945}. The experimental analysis, which was carried out for the results shown in Fig.~2b and Fig.~4d of the main text, is summarised in Fig.~\ref{fig:sm_fim_analysis_grid}. In short, we first calculate the $2\times 2$ FIM from the experimentally measured probability distributions $\mathrm{P}_a(\mathrm{m}_a)$ and $\mathrm{P}_b(\mathrm{m}_b)$ associated with measurements of $\hat{S}_a$ and $\hat{S}_b$. The FIM is then inverted to obtain the two-parameter covariance matrix,  $\mathbf{\Sigma} \geq \mathbf{F}^{-1}$, whose trace gives the sum of variances, $\mathrm{Tr}(\mathbf{\Sigma}) = \mathrm{V}(\epsilon_a) + \mathrm{V}(\epsilon_b)$. We report the trace minimised over the range of $\epsilon_{a,b}$ for which the distributions were measured, $\mathrm{V}(\epsilon_a) + \mathrm{V}(\epsilon_b) \geq \mathrm{min}_{\epsilon_a, \epsilon_b}\mathrm{Tr}(\mathbf{F}^{-1})$, as a figure of merit that quantifies the metrological performance of the sensing states. This analysis is repeated for grid states and number--phase states of increasing energy.

\subsection{Fisher information}

We first provide expressions for the FIM as a function of the probability distributions $\mathrm{P}_a(\mathrm{m}_a)$ and $\mathrm{P}_b(\mathrm{m}_b)$, which allows calculating the classical Fisher information from the experimental data. The FIM is given by~\cite{paris2009quantum, sidhu2020geometric}%
\begin{equation}
    \mathbf{F} = \begin{pmatrix}
            F_{a,a} & F_{a,b} \\
            F_{b,a} & F_{b,b}
        \end{pmatrix},
\end{equation}
with elements
\begin{equation}
    \label{eq:F_ij_general}
    F_{i,j} =  \sum_{\mathrm{m}_a\in \{0,1\}} \frac{1}{\mathrm{P}_a(\mathrm{m}_a)} \frac{\partial \mathrm{P}_a(\mathrm{m}_a)}{\partial \epsilon_i} \frac{\partial \mathrm{P}_a(\mathrm{m}_a)}{\partial \epsilon_j} + \sum_{\mathrm{m}_b \in \{0, 1\}}\frac{1}{\mathrm{P}_b(\mathrm{m}_b)} \frac{\partial \mathrm{P}_b(\mathrm{m}_b)}{\partial \epsilon_i} \frac{\partial \mathrm{P}_b(\mathrm{m}_b)}{\partial \epsilon_j},
\end{equation}
where $i,j \in \{a, b\}$. The diagonal elements of the FIM, $F_{a,a}$ and $F_{b,b}$, quantify the amount of information corresponding to $\epsilon_a$ or $\epsilon_b$ that is contained in the measurement outcomes. The off-diagonal elements, $F_{a,b}$ and $F_{b,a}$, quantify the degree of correlation between the parameters. Each element of the FIM can be simplified using the relation
\begin{align}
    & \sum_{\mathrm{m}_a\in \{0,1\}} \frac{1}{\mathrm{P}_a(\mathrm{m}_a)} \frac{\partial \mathrm{P}_a(\mathrm{m}_a)}{\partial \epsilon_i} \frac{\partial \mathrm{P}_a(\mathrm{m}_a)}{\partial \epsilon_j} =  \frac{1}{\mathrm{P}_a(0) (1 - \mathrm{P}_a(0))} \frac{\partial \mathrm{P}_a(0)}{\partial \epsilon_i} \frac{\partial \mathrm{P}_a(0)}{\partial \epsilon_j},
\end{align}
which gives
\begin{subequations}
    \label{eq:fim_elements_simplified}
    \begin{align}
        & F_{a,a} = \frac{1}{\mathrm{P}_a(0)(1-\mathrm{P}_a(0))} \left(\frac{\partial \mathrm{P}_a(0)}{\partial\epsilon_a}\right)^2 + \frac{1}{\mathrm{P}_b(0)(1-\mathrm{P}_b(0))} \left(\frac{\partial \mathrm{P}_b(0)}{\partial\epsilon_a}\right)^2, \label{eq:fim_element_Faa} \\
        & F_{b,b} = \frac{1}{\mathrm{P}_a(0)(1-\mathrm{P}_a(0))} \left(\frac{\partial \mathrm{P}_a(0)}{\partial\epsilon_b}\right)^2 + \frac{1}{\mathrm{P}_b(0)(1-\mathrm{P}_b(0))} \left(\frac{\partial \mathrm{P}_b(0)}{\partial\epsilon_b}\right)^2, \label{eq:fim_element_Fbb} \\
        & F_{a,b} = F_{b,a} = \frac{1}{\mathrm{P}_a(0)(1-\mathrm{P}_a(0))} \frac{\partial \mathrm{P}_a(0)}{\partial\epsilon_a}\frac{\partial \mathrm{P}_a(0)}{\partial\epsilon_b} + \frac{1}{\mathrm{P}_b(0)(1-\mathrm{P}_b(0))} \frac{\partial \mathrm{P}_b(0)}{\partial\epsilon_a}\frac{\partial \mathrm{P}_b(0)}{\partial\epsilon_b}.
    \end{align}
\end{subequations}

The trace of the covariance matrix can then be calculated as
\begin{equation}
    \label{eq:sum_variances_fim}
    \mathrm{V}(\epsilon_a) + \mathrm{V}(\epsilon_b) = \mathrm{Tr}(\mathbf{\Sigma}) \geq \mathrm{Tr}(\mathbf{F}^{-1}) = \frac{1}{\mathrm{det(\mathbf{F})}} (F_{a,a} + F_{b,b}),
\end{equation}
where $\mathrm{det(\mathbf{F})} = F_{a,a} F_{b,b} - F_{a,b}F_{b,a}$ is the determinant of the FIM. Minimizing the multi-parameter variance of Eq.~\ref{eq:sum_variances_fim} involves minimizing the off-diagonal elements, $F_{a,b} = F_{b,a} = 0$, and maximizing $F_{a,a}$ and $F_{b,b}$, which are proportional to the gradient of the probability distributions.

\subsection{Theoretical metrological gain}
\label{sec:theoretical_metro_gain}

The theoretical metrological gain that can be achieved by the sensing states is found by calculating the trace of the covariance matrix (see Eq.~\ref{eq:sum_variances_fim}) using the analytical probability distributions of the grid and number--phase states derived in sections~\ref{sec:grid_state_prob_distribution} and \ref{sec:num_phase_prob_distribution}. We recall that their probability distributions are given by functions of the form,
\begin{subequations}
    \begin{align}
        & \mathrm{P}_a(\mathrm{m}_a| \epsilon_a) = \frac{1}{2} (1 + (-1)^{\mathrm{m}_a} \eta_a \cos(l_a \epsilon_a + \theta_a)), \\
        & \mathrm{P}_b(\mathrm{m}_b| \epsilon_b) = \frac{1}{2} (1 + (-1)^{\mathrm{m}_b} \eta_b \cos(l_b \epsilon_b + \theta_b)),
    \end{align}
\end{subequations}
where $\eta_{a,b}$ are visibility parameters, $l_{a,b}$ are the modular lengths and $\theta_{a,b}$ are controllable rotation angles. Since the probability distribution $\mathrm{P}_a(\mathrm{m}_a)$ and $\mathrm{P}_b(\mathrm{m}_b)$ have no dependence on $\epsilon_b$ and $\epsilon_a$, respectively, $\partial \mathrm{P}_a  / (\partial \epsilon_b) = \partial \mathrm{P}_b  / (\partial \epsilon_a) = 0$ and the off-diagonal elements of the FIM reduce to zero, $F_{a,b} = F_{b,a} = 0$. The diagonal elements of the FIM are then calculated from Eq.~\ref{eq:fim_element_Faa} and Eq.~\ref{eq:fim_element_Fbb}, and are made large by maximizing the gradients $\partial \mathrm{P}_a/\epsilon_a$ and $\partial \mathrm{P}_b/\epsilon_b$. The maximum Fisher information is found by setting $\theta_{a,b} = 0$ and $l_a\epsilon_a = l_b\epsilon_b = \pi/2$, giving
\begin{align}
    \mathrm{max}(F_{a,a}) = l_a^2 \eta_a^2, \\
    \mathrm{max}(F_{b,b}) = l_b^2 \eta_b^2.
\end{align}
The multi-parameter variance of Eq.~\ref{eq:sum_variances_fim} is then bounded by
\begin{equation}
    \label{eq:th_multi_param_var_bound}
    \mathrm{V}(\epsilon_a) + \mathrm{V}(\epsilon_b) \geq \mathrm{min}\mathrm{Tr}(\mathbf{F}^{-1}) =  \frac{1}{\mathrm{max}(F_{a,a})} + \frac{1}{\mathrm{max} (F_{b,b})} = \frac{1}{l_a^2 \eta_a^2} + \frac{1}{l_b^2 \eta_b^2},
\end{equation}
and, for an ideal (infinite-energy) sensing state where the visibility parameter is $\eta_a = \eta_b = 1$, the minimum variance is
\begin{equation}
    \mathrm{V}(\epsilon_a) + \mathrm{V}(\epsilon_b) \geq \frac{1}{l_a^2 } + \frac{1}{l_b^2}.
\end{equation}

\subsubsection*{Grid-state theory}

We now consider the multi-parameter variance for approximate grid states. The modular lengths are $l_a = l_b = l_s = \sqrt{2 \pi}$, and the visibility parameter is determined from the characteristic function, $\eta_a = \eta_x = \mathrm{Re}[\bigchi( i \sqrt{\pi})]$ and $\eta_b = \eta_p = \mathrm{Re}[\bigchi(\sqrt{\pi})]$ (see section~\ref{sec:grid_state_prob_distribution}). We further approximate the visibility parameters as equal, $\eta_x \approx \eta_p $, and the multi-parameter variance of Eq.~\ref{eq:th_multi_param_var_bound} gives
\begin{equation}
    \mathrm{V}(\epsilon_x) + \mathrm{V}(\epsilon_p) \geq \frac{1}{\pi \mathrm{Re}[\bigchi( \sqrt{\pi})]^2}.
\end{equation}
The minimum variance for an ideal (infinite-energy) state where $\mathrm{Re}[\bigchi(\sqrt{\pi})] = 1$ is
\begin{equation}
    \label{eq:min_var_grid_fim}
    \mathrm{V}(\epsilon_x) + \mathrm{V}(\epsilon_p) \geq \frac{1}{\pi}.
\end{equation}
As discussed below, this limit is imposed by the measurement protocol considered here, and can be reduced by considering alternate schemes. For example, the bound can be reduced by measuring higher powers of position and momentum operators, by performing many sequential measurements, or by performing measurements of finite-energy operators tailored for the approximate nature of the grid states.

\subsubsection*{Number-phase-state theory}

We similarly consider the multi-parameter variance for approximate NP states. The modular lengths are $l_a = l_\phi = N$ and $l_b = l_n = 2\pi / N$, and the visibility parameters are $\eta_a = \eta_\phi = \langle \hat{E}^-_{l_\phi} \rangle$ and $\eta_b = \eta_n = 1$ (see section~\ref{sec:num_phase_prob_distribution}). We recall that measurements of the phase operator give information on the phase of an unknown rotation, $\epsilon_\phi$, and measurements of the number operator give information on the integer number of an unknown phonon shift, $\epsilon_n$. The multi-parameter variance of Eq.~\ref{eq:th_multi_param_var_bound} is then 
\begin{equation}
    \mathrm{V}(\epsilon_\phi) + \mathrm{V}(\epsilon_n) \geq \frac{1}{\left( N \eta_\phi \right)^2} + \left( \frac{N}{2\pi} \right)^2.
\end{equation}
For an ideal NP state where $\eta_\phi = 1$, the minimum variance is given by
\begin{equation}
    \label{eq:min_var_num_phase_fim}
    \mathrm{V}(\epsilon_\phi) + \mathrm{V}(\epsilon_n) \geq \frac{1}{N^2} + \left( \frac{N}{2\pi} \right)^2.
\end{equation}
We observe a trade-off in the variances of number and phase with the spacing $N$: increasing $N$ reduces the variance of phase but increases the variance in number.

\subsubsection*{Metrological gain from powers of operators}

The bounds of the multi-parameter variances for grid and NP states given by Eq.~\ref{eq:min_var_grid_fim} and Eq.~\ref{eq:min_var_num_phase_fim} are obtained from the classical Fisher information and are therefore applicable only to the particular measurement scheme that is considered here. Alternate measurement schemes that use high-quality sensing states could further reduce the variance. For example, one can employ the phase estimation protocol originally prescribed by Kitaev~\cite{kitaev1995quantum}, in which powers of the operators, $\hat{S}_{a,b}^k$, are applied sequentially with increasing $k$. The probability distribution of a measurement outcome associated with $\hat{S}_{a}^k$ or $\hat{S}_{b}^k$ is
\begin{subequations}
    \label{eq:prob_dist_powers_stab}
    \begin{align}
        & \mathrm{P}_{a,k}(\mathrm{m}_a|\epsilon_b, \theta_a) = \frac{1}{2} (1 + (-1)^{\mathrm{m}_a}\eta_{a,k} \cos(\theta_a + k l_a \epsilon_a)), \\
        & \mathrm{P}_{b,k}(\mathrm{m}_b|\epsilon_a, \theta_b) = \frac{1}{2} (1 + (-1)^{\mathrm{m}_b} \eta_{b,k} \cos(\theta_b + k l_b \epsilon_b)),
    \end{align}
\end{subequations}
where the visibility parameters are $\eta_{a,k} \sim \langle \hat{S}_{a}^k\rangle $ and $\eta_{b,k} \sim \langle \hat{S}_{b}^k\rangle $. Following the same derivations as outlined above, the maximum Fisher information is $\mathrm{max}(F_{a,a}) = (kl_a \eta_{a,k})^2$ and $\mathrm{max}(F_{b,b}) = (kl_b \eta_{b,k})^2$. For ideal sensing states where $\eta_{a,k} = \eta_{b,k} = 1$, the uncertainty can, therefore, be decreased with larger $k$, which increases the Fisher information and hence reduces the variance quadratically. However, for finite-energy sensing states where $\eta_{a,k} = \eta_{b,k} < 1$, the visibility parameter approaches zero with large $k$, giving vanishing Fisher information. For example, the visibility parameter associated with grid states is determined by a point of its characteristic function, $\eta_{x,k} = \mathrm{Re}[\bigchi(k i \sqrt{\pi})]$. Due to the Gaussian envelope of the approximate grid states, the characteristic function approaches zero for points further from the origin, hence $\eta_{x,k} \rightarrow 0$ with large $k$. Moreover, in Eq.~\ref{eq:prob_dist_powers_stab}, we do not consider the effects of backaction on subsequent measurements, which may be worsened with larger $k$.

We explored the measurement of higher powers of operators with grid states to increase the Fisher information. However, from numerical simulations, it was found that the visibility parameters for $k>1$ were sufficiently small that no additional metrological gain was found for the approximate grid states considered in this work. Nevertheless, an additional metrological gain is expected for higher-quality states.

\subsubsection*{Metrological gain from sequential operator measurements}

We now consider the metrological gain from measuring $\hat{S}_a$ and $\hat{S}_b$ many times in an alternating manner (see circuit of Fig.~1e of the main text). Performing $N_S$ alternating QPE measurements of $\hat{S}_a$ and $\hat{S}_b$ gives $2N_S$ measurement outcomes. Ignoring backaction from measurements and approximating the probabilitiy distributions as independent (see section~\ref{sec:grid_state_prob_distribution}), the total Fisher information is obtained by summing over the Fisher information of each $j$th measurement outcome,
\begin{align}
    & \mathrm{max}(F_{a,a}) = \sum_{j=0}^{N_S-1} l_a^2 \eta_{a,j}^2, \\
    & \mathrm{max}(F_{b,b}) = \sum_{j=0}^{N_S-1} l_b^2 \eta_{b,j}^2, 
\end{align}
and the off-diagonal elements are $F_{a,b} = F_{b,a} = 0$. The parameters $\eta_{a,j}$ and $\eta_{b,j}$ correspond to the visibility parameters of the $j$th measurement. The multi-parameter variance of Eq.~\ref{eq:th_multi_param_var_bound} becomes
\begin{equation}
    \label{eq:var_multi_N_th}
    \mathrm{V}(\epsilon_a) + \mathrm{V}(\epsilon_b) \geq  \frac{1}{ l_a^2 \sum_{j=0}^{N_S-1} \eta_{a,j}^2} + \frac{1}{l_b^2 \sum_{j=0}^{N_S-1} \eta_{b,j}^2}.
\end{equation}

\subsection{Standard quantum limit}
\label{sec:sql}

In the following, we derive the standard quantum limit (SQL) for grid and number-phase states in a multi-parameter setting. In both cases, we calculate the simultaneous standard quantum limit (SQL*) from the uncertainties of heterodyne measurements using a coherent state. The vacuum variances along position and momentum are $1/2$. The corresponding POVM is $\hat{E}_\beta = \frac{1}{\pi}\ket{\beta} \bra{\beta}$, and the distribution given a coherent state input is
\begin{equation}
    \label{eq:sql_povm_prob}
    P(\beta|\alpha) = \mathrm{Tr}[\hat{E}_\beta \ket{\alpha}\bra{\alpha}] = \frac{1}{\pi}|\langle \alpha | \beta \rangle|^2 = \frac{1}{\pi} e^{- |\alpha - \beta|^2}.
\end{equation}

\subsubsection*{SQL* of position--momentum}

The simultaneous SQL of position and momentum is found from Eq.~\ref{eq:sql_povm_prob} after replacing $\alpha = \frac{1}{\sqrt{2}}(\mathrm{Re}\alpha + i \mathrm{Im}\alpha)$ and $\beta=  \frac{1}{\sqrt{2}}(\mathrm{Re}\beta + i \mathrm{Im}\beta)$. The probability distribution can then be expressed as a function of position and momentum with $x = \mathrm{Re}\beta - \mathrm{Re}\alpha$ and $p = \mathrm{Im}\beta - \mathrm{Im}\alpha$, giving
\begin{equation}
    P(\beta | \alpha) = \frac{1}{2\pi} \mathrm{exp}\left(- \frac{x^2}{2} - \frac{p^2}{2}\right),
\end{equation}
From this, we find the variances
\begin{subequations}
    \begin{align}
        & \mathrm{V}({x}) = 1, \\
        & \mathrm{V}({p}) = 1, 
    \end{align}
\end{subequations}
and the simultaneous SQL is given by
\begin{equation}
    \text{SQL*} = \mathrm{V}({x}) + \mathrm{V}({p}) = 2.
\end{equation}

In Fig.~2 of the main text, we also plot the combined variance obtained from heterodyne measurements with a squeezed state. There, the variances of position and momentum are
\begin{subequations}
    \begin{align}
        & \mathrm{V}(x) = e^{-r} \cosh(r), \\
        & \mathrm{V}(p) = e^{r} \cosh(r), 
    \end{align}
\end{subequations}
where $r$ is the squeezing parameter and is related to the average phonon number of the squeezed state via $\langle \hat{n} \rangle = \cosh(r)^2 - 1$. With this, the total variance of a squeezed state becomes
\begin{equation}
    \mathrm{V}(x) + \mathrm{V}(p) = 2(1 + \langle \hat{n} \rangle).
\end{equation}

\subsubsection*{SQL* of number--phase}

The simultaneous SQL of number and phase is found from Eq.~\ref{eq:sql_povm_prob} after replacing $\beta = |\beta| e^{i \phi}$, giving
\begin{equation}
    P(|\beta|^2, \phi|\alpha) = \frac{1}{\pi} \mathrm{exp}(- |\alpha|^2 - |\beta|^2 + 2\cos(\phi - \theta)),
\end{equation}
where the corresponding change of measure from $d^2\beta$ is $|\beta| d|\beta| d\phi$ and $\alpha = |\alpha|e^{i\theta}$.  
The variances of number and phase from heterodyne measurements are~\cite{shapiro1984heterodyne}
\begin{subequations}
    \begin{align}
        & \mathrm{V}(\phi) = \frac{1}{2\langle \hat{n} \rangle} + \frac{3}{8 \langle \hat{n} \rangle^2} + \mathcal{O}\left(\frac{1}{\langle \hat{n} \rangle^3}\right), \\
        & \mathrm{V}(n) = 2 \langle \hat{n} \rangle + 1.
    \end{align}
\end{subequations}
In the limit of large $\langle \hat{n} \rangle$, the phase variance is well approximated by the first two terms. 
The variances of number and phase have different scalings and magnitudes, where $\mathrm{V}(\phi) \ll \mathrm{V}(n)$ for large $\langle \hat{n} \rangle$. Therefore, to obtain a meaningful simultaneous SQL where both number and phase variances have equal weightings and the metrological gain is comparable to displacement sensing, we normalise the individual variances such that $\mathrm{V}(\phi) = \mathrm{V}(n) = 1$ at large $\langle \hat{n}\rangle$. The rescaled simultaneous SQL is then
\begin{equation}
    \text{SQL*} = 2 \langle \hat{n}\rangle \mathrm{V}({\phi}) + \frac{1}{2 \langle \hat{n}\rangle}\mathrm{V}({n}) = 2 + \frac{5}{4\langle \hat{n}\rangle} + \mathcal{O}\left(\frac{1}{\langle \hat{n} \rangle^2}\right).
\end{equation}

\subsection{Experimental analysis}
\label{sec:metro_gain_exp_analysis}

\begin{figure}
    \centering
    \includegraphics[width=\textwidth]{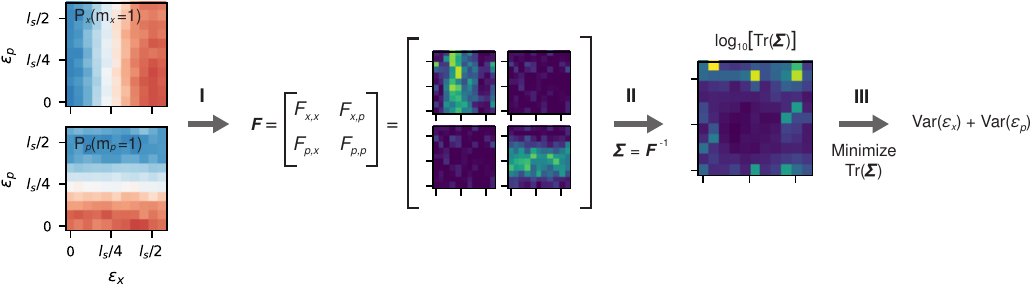}
    \caption{
    \textbf{Characterizing the metrological gain of a sensing state.} 
    As an example we consider the results of a grid state with an average phonon number of $\langle \hat{n} \rangle = 3.14$, but the characterization method is applicable to all states considered in this work. 
    We obtain the probability distributions of both modular observables, here $\hat{S}_x$ and $\hat{S}_p$, as a function of the sensing signal, $\epsilon_x$ and $\epsilon_p$, using the QPE circuit of Fig.~1e of the main text.
    \textbf{I.} The Fisher Information Matrix (FIM) is calculated for each pair ($\epsilon_x, \epsilon_p$). This involves computing the gradient of the probability distribution at each point. 
    \textbf{II.} The covariance matrix is calculated from the inverse of the FIM. The plot shows the logarithm of the trace of the covariance matrix for each pair ($\epsilon_x, \epsilon_p$). 
    \textbf{III.} The trace of the covariance matrix corresponds to the sum of the variances. We finally minimise $\mathrm{Tr}(\Sigma)$ over all pairs ($\epsilon_x, \epsilon_p$) and report the minimum variance. 
    Axes labels and tick values are identical for all plots.}
    \label{fig:sm_fim_analysis_grid}
\end{figure}

The experimentally reconstructed probability distributions for the grid states and number--phase states are shown in the Figures~\ref{fig:grid_probs_non_finite_raw_data} and \ref{fig:np_probs_raw_data}, respectively, and the detailed pulse sequence is shown in Figure~\ref{fig:sm_detailed_pulse_sequence}. Their analysis is summarised in Fig.~\ref{fig:sm_fim_analysis_grid}. The probability distributions are reconstructed by averaging measurement outcomes of the QPE circuit over $M=10^3$ repetitions and varying $\epsilon_{a,b} = \epsilon_{x,p}$ for grid states and $\epsilon_{a,b} = \epsilon_{\phi, n}$ for number--phase state. The phases of the ancilla rotations are set to zero. The elements of the FIM are then calculated from the probability distributions for each pair $(\epsilon_a, \epsilon_b)$ using Eq.~\ref{eq:fim_elements_simplified}. The gradients along $\epsilon_a$ and $\epsilon_b$ are computed from the central difference for interior points, and from forward and backward differences for outer points~\cite{Wolf2019}. The covariance matrix, $\boldsymbol{\Sigma}$, is then computed for each pair $(\epsilon_a, \epsilon_b)$ by inverting the FIM. The uncertainties in the variances are obtained by propagating uncertainties in the probability distribution throughout the analysis, where the standard deviation from quantum projection noise is $\sqrt{\mathrm{P}(1-\mathrm{P})/M}$.

The variances of number and phase plotted in Fig.~4d are rescaled such that their metrological gain relative to their individual SQL is comparable. The scaled covariance matrix is obtained with $\boldsymbol{\Sigma}' = \mathrm{diag}\left(2 \langle \hat{n} \rangle, \frac{1}{2 \langle \hat{n}\rangle}\right)\boldsymbol{\Sigma}$, such that the variances of phase and number are individually scaled by $2 \langle \hat{n} \rangle$ and $\frac{1}{2 \langle \hat{n}\rangle}$, respectively. The insets of Fig.~4d also plot the individual unscaled variances of number and phase. These are obtained from the diagonal elements of $\boldsymbol{\Sigma}$ for the pair $(\epsilon_\phi, \epsilon_n)$ that minimise $\mathrm{Tr}(\boldsymbol{\Sigma}')$.

\subsection{Metrological gain of grid states with finite-energy measurement operators}
\label{sec:metrological_gain_bsb_grid}

\begin{figure}
    \centering
    \includegraphics[]{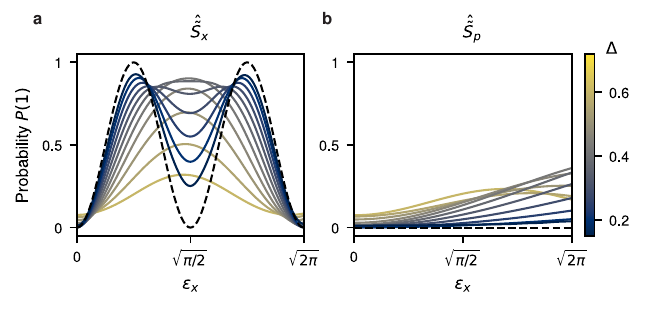}
    \caption{
    \textbf{Probability distributions for the Big-small-Big approximate stabiliser of grid states.}
    Probabilities are numerically simulated using the circuit of Fig.~1 with $N_S=1$ stabiliser repetitions, and setting the control phases to $\theta_x = \theta_p = 0$. We set $\epsilon_p=0$ and vary $\epsilon_x \in [0, \sqrt{2\pi}]$. Probabilities are measured for outcomes of \textbf{(a)} the conditional position stabiliser of Eq.~\ref{eq:CSx_finite} and \textbf{(b)} the conditional momentum stabiliser of Eq.~\ref{eq:CSp_finite}, for grid states with squeezing parameters varying in the range $\Delta \in [0.15, 0.6]$. The dashed black line plots the ideal probability distribution in the limit $\Delta \rightarrow 0$.
    }
    \label{fig:big_small_big_prob_dist}
\end{figure}

The above analysis of the metrological gain of grid states considered measurements of position and momentum operators designed for ideal grid states with infinite energy. However, finite-energy grid states are only approximate eigenstates of these operators, hence measurements will distort the Gaussian envelope and increase the energy of the state. Importantly, this will reduce the metrological gain from performing many sequential measurements. One can instead consider measurements of finite-energy position and momentum operators, which conserve the energy of the approximate grid state and minimise distortions of the Gaussian envelope. One such operator is given by the Big-small-Big protocol (BsB)~\cite{Royer2020}, in which a ``small'' conditional displacement is surrounded by two ``big'' conditional displacements. The BsB position and momentum operators are 
\begin{subequations}
    \label{eq:BsB_operators}
    \begin{align}
        & C\hat{S}_x^{\mathrm{(BsB)}} = \mathrm{exp}\Big({- \frac{i}{2}l_s c_\Delta \hat{\sigma}_x \hat{x} }\Big) \mathrm{exp}\Big({- i l_s \Delta^2 \hat{\sigma}_y \hat{p}}\Big) \mathrm{exp}\Big(-\frac{i}{2} l_s c_\Delta \hat{\sigma}_x\hat{x}\Big), \label{eq:CSx_finite}\\
        & C\hat{S}_p^{\mathrm{(BsB)}} = \mathrm{exp}\Big(\frac{i}{2} c_\Delta \hat{\sigma}_x \hat{p} \Big) \mathrm{exp}\Big(i l_s \Delta^2 \hat{\sigma}_y \hat{x} \Big) \mathrm{exp} \Big(\frac{i}{2} l_s c_\Delta \hat{\sigma}_x\hat{p} \Big), \label{eq:CSp_finite}
    \end{align}
\end{subequations}
where $c_\Delta = \cosh \Delta^2$. Analytical probability distributions from measurements of $ C\hat{S}_x^{\mathrm{(BsB)}}$ and $ C\hat{S}_p^{\mathrm{(BsB)}}$ are not straightforwardly obtained due to the non-commutativity of the ``big'' and ``small'' conditional displacements. We, therefore, numerically simulate the QPE circuit using BsB operations, and the resulting probability distributions are plotted in Fig.~\ref{fig:big_small_big_prob_dist}. We first observe that, as the energy of the grid state increases and $\Delta$ approaches zero, the probability distributions resemble those of Eq.~\ref{eq:independent_prob_dist_grid} with a period that is twice as small, i.e. with the replacement $l_s \rightarrow 2 l_s $. This is apparent from the BsB operators of Eq.~\ref{eq:BsB_operators} which, for $\Delta = 0$, gives $C\hat{S}_x^{\mathrm{(BsB)}} = (C\hat{S}_x)^2$ and $C\hat{S}_p^{\mathrm{(BsB)}} = (C\hat{S}_p^\dagger)^2$. We also find that the probability distributions at larger $\Delta$ deviate from a sinusoidal oscillation. Moreover, both measurements of $C\hat{S}_x^{\mathrm{(BsB)}}$ and $C\hat{S}_p^{\mathrm{(BsB)}}$ are sensitive to displacements along position by $\epsilon_x$, giving rise to correlations in the two measurement outcomes. 

\begin{figure}
    \centering
    \includegraphics[]{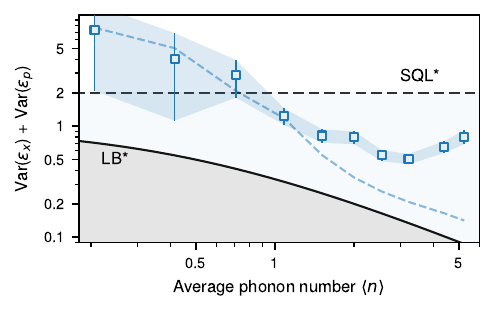}
    \caption{ 
    \textbf{Metrological gain of grid states with finite-energy position and momentum measurements.}
    The metrological gain is characterised with the same methods and analysis as the results of Fig.~2b of the main text. Measurements of the finite-energy Big-small-Big (BsB) operators are performed instead. The dashed blue line plots the theoretical performance of the finite-energy operators and is obtained from numerical simulations of the experimental sequence.
    Error bars correspond to one standard deviation calculated from quantum projection noise.
    }
    \label{fig:metrological_gain_grid_bsb}
\end{figure}

We experimentally investigate the metrological gain from measurements of finite-energy position and momentum operators by repeating the methods outlined in section~\ref{sec:metro_gain_exp_analysis}. We perform the circuit of Fig.~1e and apply the finite-energy operators $C\hat{S}_x^{\mathrm{(BsB)}}$ and $C\hat{S}_p^{\mathrm{(BsB)}}$ in each QPE sub-routine. The experimentally reconstructed probability distributions are plotted in Figure~\ref{fig:grid_probs_finite_raw_data}, and the resulting multi-parameter variance is plotted in Fig.~\ref{fig:metrological_gain_grid_bsb}. The variance outperforms the SQL for average phonon numbers $\langle \hat{n} \rangle \gtrapprox 1$. Similar to the results of Fig.~2, the largest metrological gain is obtained at $\langle \hat{n} \rangle = 3.22$, and measurements with the BsB protocol give a larger gain of $\SI{6.0(5)}{dB}$.

In general, we stipulate that the use of finite-energy BsB operators for multi-parameter sensing with grid states may be advantageous due to the preservation of the grid state's energy and the improved metrological gain. However, a lack of a straightforward analytical probability distribution due to the non-commuting operations within a BsB operator prevents straightforward integration with the phase estimation algorithm of section~\ref{sec:qpe}. 

\subsection{Metrological gain of number--phase states with airy envelope}

\begin{figure}
    \centering
    \includegraphics[]{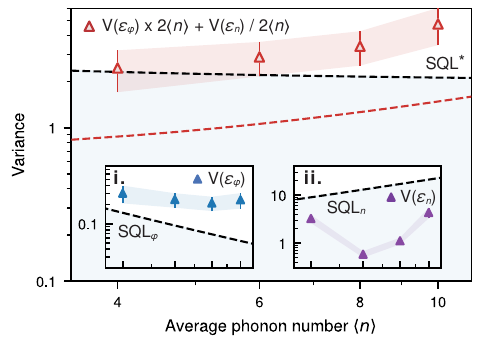}
    \caption{
    \textbf{Metrological gain of number--phase states with an airy envelope.}
    The number--phase states have a spacing $N=4$ and offset $\lambda = 2$. The number--phase state is made physical by damping the Fock coefficients with an envelope calculated from the airy function. The variances of number and phase are obtained from the classical Fisher information extracted from the reconstructed probability distributions. Insets (i., ii.) plot the individual variances obtained from the covariance matrix. The standard quantum limits of number and phase are obtained from heterodyne measurements with a coherent state, giving $\mathrm{SQL}_n = 2 \langle \hat{n}\rangle + 1$ and $\mathrm{SQL}_\phi = (2 \langle \hat{n}\rangle)^{-1} + 3 (8 \langle \hat{n}\rangle^2)^{-1}$. The combined variances are a summation of $\mathrm{V}(\epsilon_n)$ and $\mathrm{V}(\epsilon_\phi)$ after rescaling by $(2\langle \hat{n} \rangle)^{-1}$ and $2 \langle \hat{n} \rangle$, respectively, such that number and phase SQLs have a variance of 1 and the simultaneous SQL (SQL*) is equal to 2 at large $\langle \hat{n} \rangle$.
    Error bars correspond to one standard deviation calculated from quantum projection noise.
    }
    \label{fig:metrological_gain_airy}
\end{figure}

Finite-energy number--phase states are obtained by applying a damping envelope to their Fock distribution. In section~\ref{sec:num_phase_theory}, two such envelopes are presented: a sine envelope, which constrains the distribution within a cutoff, and an airy envelope, which constrains the average phonon number and does not have a cutoff. The metrological gain of sine states is explored in Fig.~4 of the main text, and we here present experimental results for the metrological gain of airy states.

The reconstructed probability distributions of airy states are plotted in Figure~\ref{fig:np_probs_raw_data}, and the resulting number and phase variances are plotted in Fig.~\ref{fig:metrological_gain_airy}. The variance of number performs similarly to the sine states, giving a large gain over the SQL for the entire range of $\langle \hat{n} \rangle$. However, the variance of phase is larger than for the sine states and remains above the SQL. The resulting combined variance matches the simultaneous SQL at $\langle \hat{n}\rangle = 4$, and worsens as $\langle \hat{n}\rangle$ increases. Overall, the airy states do not perform better than the sine states, and the increased combined variance is largely due to an increased phase variance. We stipulate that the poor performance of airy states is due to a larger tail in the distribution at large Fock states, which suffers from decoherence during state preparation and measurements.

\section{Quantum phase estimation with Bayesian inference}
\label{sec:qpe}

\subsection{Adaptive and non-adaptive protocols}

We now discuss in detail the phase estimation schemes used in the main text to perform multi-parameter displacement sensing with grid states. We first consider a scenario where $\hat{S}_x$ and $\hat{S}_p$ are applied once per measurement repetition, as shown in Fig.~1e with $N_S=1$. The entire circuit is repeated over $M$ measurement repetitions, giving $2M$ measurement outcomes. A total of $M$ outcomes are associated with $\hat{S}_x$, giving $\mathbf{m}_x = (\mathrm{m}_{x,0}, ..., \mathrm{m}_{x,M-1})$, with a vector of adjustable phases $\boldsymbol{\theta}_x = (\theta_{x,0}, ..., \theta_{x,M-1})$. Another $M$ outcomes are associated with $\hat{S}_p$, giving $\mathbf{m}_p = (\mathrm{m}_{p,0}, ..., \mathrm{m}_{p,M-1})$, with a vector of phases $\boldsymbol{\theta}_p = (\theta_{p,0}, ..., \theta_{p,M-1})$. The phase estimation procedure finds estimates $\tilde{\epsilon}_x$ and $\tilde{\epsilon}_p$ of the true parameters $\epsilon_x$ and $\epsilon_p$ from Bayesian inference by maximizing the posterior probability distributions, $ \tilde{\epsilon}_x = \mathrm{max}_{\epsilon_x} \left(\mathrm{P}_x(\epsilon_x | \mathbf{m}_x, \boldsymbol{\theta}_x,\eta_x) \right)$ and $\tilde{\epsilon}_p = \mathrm{max}_{\epsilon_p} \left(\mathrm{P}_p(\epsilon_p | \mathbf{m}_p, \boldsymbol{\theta}_p,\eta_p) \right)$. The posterior distributions are calculated using Bayes' theorem, 
\begin{subequations}
\label{eq:posterior_prob_dist}
    \begin{align}
        & \mathrm{P}_x(\epsilon_x | \mathbf{m}_x, \boldsymbol{\theta}_x,\eta_x) \propto \mathrm{P}_0(\epsilon_x) \prod_{j=0}^{M-1} \mathrm{P}_x(\mathrm{m}_{x,j}| \epsilon_x, \theta_{x,j}, \eta_{x}), \\ 
        & \mathrm{P}_p(\epsilon_p | \mathbf{m}_p, \boldsymbol{\theta}_p,\eta_p) \propto \mathrm{P}_0(\epsilon_p) \prod_{j=0}^{M-1} \mathrm{P}_p(\mathrm{m}_{p,j}| \epsilon_p, \theta_{p,j}, \eta_{p}),
    \end{align}
\end{subequations}
where the initial distributions are $\mathrm{P}_0(\epsilon_x) = \mathrm{P}_0(\epsilon_p) = 1/\sqrt{2\pi}$. The independent probability distributions $\mathrm{P}_x(\mathrm{m}_{x,j}| \epsilon_x, \theta_{x,j}, \eta_{x})$ and $\mathrm{P}_p(\mathrm{m}_{p,j}| \epsilon_p, \theta_{p,j}, \eta_{p})$ from a single measurement outcome are given in section~\ref{sec:grid_state_prob_distribution}. There, it was also found that a measurement of $\hat{S}_x$ will have backaction on a subsequent measurement of $\hat{S}_p$ within a single iteration of the QPE circuit. This backaction can be modeled as a bit flip of the measurement outcome, which is incorporated in the above estimation procedure by changing $\mathrm{m}_{p,j} \rightarrow 1 - \mathrm{m}_{p, j}$.

In the experimental implementation, the phase estimation algorithm is executed in real-time. Calculations of the posterior distribution and optimizations of phases for adaptive QPE should therefore be fast compared to the preparation and measurement of the probe. To this end, we modify the above phase estimation procedure using a protocol adapted from Refs.~\cite{Berry2000, Huang2017}. The probability distributions are conveniently expressed as Fourier series, such that all calculations can be efficiently done by manipulating a list of Fourier coefficients. As the following derivations are identical for both distributions of $\mathrm{P}_x$ and $\mathrm{P}_p$, we drop the subscripts $x$ and $p$ for simplicity. We further omit the subscripts $x$ and $p$ from $\epsilon_{x}$ and $\epsilon_{p}$, however the phase estimation of $\epsilon_x$ and $\epsilon_p$ can be straightforwardly obtained with the replacement $\epsilon \rightarrow \epsilon_x$ and $\epsilon \rightarrow \epsilon_p$. 

The posterior probability distributions of Eq.~\ref{eq:posterior_prob_dist} after $j$ measurement repetitions can be decomposed in a Fourier series as
\begin{align}\label{eq:fourier}
    & \mathrm{P}(\epsilon | \mathbf{m}_{0:j-1}, \boldsymbol{\theta}_{0:j-1}, \eta) = \frac{1}{2\pi} \sum_{k = -j}^j a_{k} e^{i k l_s \epsilon}, 
\end{align}
such that $\mathrm{P}$ is entirely describable by a vector of Fourier coefficients $\boldsymbol{a} \in \mathbb{C}^{2j+1}$ with elements $\boldsymbol{a} = (a_{-j}, a_{-j+1}, \cdots, a_{j-1}, a_{j})$. Updating the posterior distribution given a new measurement outcome then involves updating the elements of $\boldsymbol{a}$, whose dimension will also increase to $\mathrm{dim}(\boldsymbol{a}) \rightarrow \mathrm{dim}(\boldsymbol{a}) + 2$. As an example, we consider a new measurement outcome $\mathrm{m}_j$ from the $(j+1)$th measurement repetition, whose probability distribution is given by the Fourier series
\begin{align}
    & \mathrm{P}(\mathrm{m}_{j}|\epsilon, \theta_{j}, \eta) = \frac{1}{2\pi} \sum_{k=-1}^{1}\tilde{a}_{k} e^{i k l_s  \epsilon},
\end{align}
with Fourier coefficients $\tilde{\boldsymbol{a}} = (\tilde{a}_{-1}, \tilde{a}_{0}, \tilde{a}_{1})$. These coefficients are determined from the measurement outcome $\mathrm{m}_{j}$: $\mathrm{m}_{ j} = 0$ gives $\tilde{\boldsymbol{a}} = \frac{\pi}{2} ( \eta e^{- i \theta_{ j}}, 2, \eta e^{ i \theta_{j}})$ while $\mathrm{m}_{j} = 1$ gives $\tilde{\boldsymbol{a}} = \frac{\pi}{2} ( - \eta e^{- i \theta_{j}}, 2, - \eta e^{i \theta_{j}} )$. The posterior distribution, $\mathrm{P}(\mathbf{m}_{0:j}|\epsilon, \boldsymbol{\theta}_{0:j}, \eta) = \mathrm{P}(\mathrm{m}_{j}|\epsilon, \theta_{j}, \eta)\mathrm{P}(\epsilon | \mathbf{m}_{0:j-1}, \boldsymbol{\theta}_{0:j-1}, \eta)$, is describable by a Fourier series with coefficients $\boldsymbol{a}' \in \mathbb{C}^{\mathrm{dim}(\boldsymbol{a}) + 2}$. To find the elements of $\boldsymbol{a}'$, we first compute three vectors, $\boldsymbol{v}^{(1)},\boldsymbol{v}^{(2)}, \boldsymbol{v}^{(3)} \in \mathbb{C}^{\mathrm{dim}(\boldsymbol{a})}$,
\begin{subequations}
    \begin{align}
        & \boldsymbol{v}^{(1)} = \tilde{a}_{-1} \boldsymbol{a},  \\
        & \boldsymbol{v}^{(2)} = \tilde{a}_{0} \boldsymbol{a},  \\
        & \boldsymbol{v}^{(3)} = \tilde{a}_{1} \boldsymbol{a}.
    \end{align}
\end{subequations}
We then compute the vectors $\tilde{\boldsymbol{v}}^{(1)}, \tilde{\boldsymbol{v}}^{(2)}, \tilde{\boldsymbol{v}}^{(3)} \in \mathbb{C}^{\mathrm{dim}(\boldsymbol{v}) + 2}$ by padding $\boldsymbol{v}^{(1)}$, $\boldsymbol{v}^{(2)}$, and $\boldsymbol{v}^{(3)}$ with zeros, such that
\begin{subequations}
    \begin{align}
        & \tilde{\boldsymbol{v}}^{(1)} = (0, 0, v^{(1)}_{-j}, v^{(1)}_{-j+1}, \cdots, v^{(1)}_{j-1}, v^{(1)}_{j}), \\
        & \tilde{\boldsymbol{v}}^{(2)} = (0, v^{(2)}_{-j}, v^{(2)}_{-j+1}, \cdots, v^{(2)}_{j-1}, v^{(2)}_{j}, 0), \\
        & \tilde{\boldsymbol{v}}^{(3)} = (v^{(3)}_{-j}, v^{(3)}_{-j+1}, \cdots, v^{(3)}_{j-1}, v^{(3)}_{j}, 0, 0).
    \end{align}
\end{subequations}
The Fourier coefficients of the updated posterior distribution are finally found from
\begin{equation}
    \boldsymbol{a}' = \boldsymbol{v}^{(1)} + \boldsymbol{v}^{(2)} + \boldsymbol{v}^{(3)}.
\end{equation}

Repeating the above procedure for each of the $M$ measurement repetitions gives a final list of Fourier coefficients $\boldsymbol{a}' \in \mathbb{C}^{2M+1}$. An estimate $\tilde{\epsilon}$ of the true parameter $\epsilon$ can be obtained from an element of $\boldsymbol{a}'$, by noting that each coefficient gives an expectation value of the Fourier series, $a'_{-k} = \langle e^{i k l_s \epsilon}\rangle $. The estimate is then obtained from $\tilde{\epsilon} = \mathrm{arg}(a'_{-1}) / l_s = \mathrm{arg}(\langle e^{ i l_s \epsilon}\rangle ) / l_s$.

The angles $\boldsymbol{\theta}$ of the ancilla rotation can be varied between measurement repetitions to increase the information gained in subsequent measurements. In non-adaptive phase estimation, the angles are predetermined and we set $\theta_{j} = j \pi / M$ for an experiment with $M$ measurement repetitions. In adaptive phase estimation, $\theta_{j}$ is separately optimised before the $(j+1)$th measurement repetition to maximise the sharpness of the posterior probability distribution,
\begin{equation}
    s(\theta_j) = |\langle e^{i \phi_\epsilon} \rangle|,
\end{equation}
with $s(\theta_j) \in [0, 1]$ and where $\phi_\epsilon = l_s \epsilon$ is the phase that is estimated from the displacement parameter $\epsilon$. The sharpness function can also be expressed from the posterior probability distribution of Eq.~\ref{eq:posterior_prob_dist},
\begin{align}
    \label{eq:sharpness_fourier}
    s(\theta_j) = \frac{1}{2\pi} \sum_{\mathrm{m}_j=0,1} \left|\int^{\pi}_{-\pi} d\phi_\epsilon e^{i\phi_\epsilon} \mathrm{P}(\mathrm{m}_j| \epsilon, \theta_j) \mathrm{P}(\epsilon | \mathbf{m}_{0:j-1}, \boldsymbol{\theta}_{0:j-1}, \eta)\right|.
\end{align}
The optimal angle for the $(j+1)$th measurement iteration is then chosen as 
\begin{equation}
    \theta_{j, \mathrm{opt}} = \max_{\theta_j \in [0, \pi]} s(\theta_j).
\end{equation}
The sharpness function is efficiently calculated by noting that the integral of Eq.~\ref{eq:sharpness_fourier} evaluates to the coefficient $a'_{-1} \in \boldsymbol{a}'$ of the Posterior distribution. Therefore, the sharpness function is computed by summing the absolute values of the coefficients $a'_{-1}$ of the posterior distributions $\mathrm{P}(\epsilon| \mathbf{m}_{0:j-1}, \boldsymbol{\theta}_{0:j-1}, \theta_j, \mathrm{m}_j=0)$ and $\mathrm{P}(\epsilon| \mathbf{m}_{0:j-1}, \boldsymbol{\theta}_{0:j-1}, \theta_j, \mathrm{m}_j=1)$.

\subsection{Experimental results}

The results of Fig.~3 of the main text plot the variances obtained from QPE with Bayesian inference using the protocol outlined above. We randomly sample displacement parameters $(\epsilon_{x}, \epsilon_p) \in [-l_s/2, l_s/2]$ with a uniform distribution and, for a given number of measurement repetitions $M$, repeat the above estimation procedure to obtain $20$ estimates $(\tilde{\epsilon}_x, \tilde{\epsilon}_p)$. The visibility parameters are set to $\eta_x = \eta_p = 1$ for simplicity, as they have no effect on the calculation of the estimates $(\tilde{\epsilon}_x, \tilde{\epsilon}_p)$. We then calculate the Holevo variance, $\mathrm{V}_\mathrm{H}(\epsilon_{x/p}) = (|\langle e^{i l_s \tilde{\epsilon}_{x/p}} \rangle|)^{-2} - 1$. This process is repeated for 24 randomly sampled pairs $(\epsilon_x, \epsilon_p)$, and we report the variance averaged over these samples.

We use bootstrapping in the analysis of non-adaptive QPE to reduce the number of experimental measurements. For each randomly sampled pair $(\epsilon_x, \epsilon_p)$, the experiment performs the non-adaptive protocol with $M=128$ measurement iterations, and each $j$th iteration is repeated 20 times, giving 20 measurement outcomes for $\mathrm{m}_x$ and 20 measurement outcomes for $\mathrm{m}_p$. In the analysis, lists of measurement outcomes $\mathbf{m}_x$ and $\mathbf{m}_p$ are obtained by randomly selecting an outcome from each $j$th iteration, giving 100 bootstrap samples for each pair $(\epsilon_x, \epsilon_p)$ for a given $M$. We further obtain variances for $M<128$ by sub-sampling from the dataset. For example, results for $M=64$ are obtained by sub-sampling from every second group of measurement outcomes from the dataset of $M=128$ measurement iterations.

In the results of the main text, we further perform non-adaptive QPE with $M=32$ measurement repetitions while increasing the number of QPE sub-routine repetitions, $N_S$. Each $j$th measurement iteration yields $N_S$ measurement outcomes associated with $\hat{S}_x$ and $N_S$ measurement outcomes associated with $\hat{S}_p$. Estimates for $(\tilde{\epsilon}_x, \tilde{\epsilon}_p)$ are obtained by following a similar procedure as outlined above. For each $j$th measurement repetition, the posterior distributions associated with $\epsilon_x$ and $\epsilon_p$ are updated $N_S$ times for each $n$th measurement outcome. The controllable angles are set to $\theta_{x,j} = \theta_{p,j} = j \pi / M$ for all $N_S$ measurements during the $j$th iteration. We determine the visibility parameters of the $n$th iteration by modeling them as an exponential decay. We experimentally calibrate the visibility parameter at $n=0$ by measuring a point of the characteristic function, giving $\eta_{x} = \mathrm{Re}[\bigchi(i\sqrt{\pi})] = 0.72$, and we approximate $\eta_p \approx \eta_x$. The visibility parameters at the $n$th iteration are modeled as $\eta_{x,n} = \eta_{x,0} e^{- 2n \zeta }$ and $\eta_{p,n} = \eta_{p,0} e^{- (2n+1) \zeta }$, where $\zeta = T_\mathrm{meas} / T_{x/p} = 0.33$ is determined from the ratio of the measurement duration to the lifetime of $\hat{S}_x$ or $\hat{S}_p$ (see section~\ref{sec:conditional_x_p_stabilisers}). The variances are calculated from the Holevo variance averaged over 15 randomly sampled points $(\epsilon_x, \epsilon_p) \in [-l_s/2, l_s/2]$. We perform the same bootstrapping method detailed above and average each point over 100 bootstrap samples. Varying $N_S = \{1, 2, 3\}$ gives variances $\mathrm{V}_\mathrm{H}(\epsilon_x) = \{4.2, 3.2, 3.0\}\times 10^{-2}$ and $\mathrm{V}_\mathrm{H}(\epsilon_p) = \{7.6, 5.2, 5.7\}\times 10^{-2}$, and the total variance is $\mathrm{V}_\mathrm{H}(\epsilon_x) + \mathrm{V}_\mathrm{H}(\epsilon_p) = \{11.8, 8.5, 8.6\} \times 10^{-2}$. We observe that the variance decreases from $N_S=1$ to $N_S=2$ by $\SI{1.4}{dB}$, but does not decrease further at $N_S=3$.

We compare the experimentally measured gain from increasing the number of QPE sub-routines to the theoretical model of section~\ref{sec:theoretical_metro_gain}. Plugging the above expressions for the visiblity parameters into the expression of Eq.~\ref{eq:var_multi_N_th} gives
\begin{align}
    \mathrm{V}(\epsilon_x) + \mathrm{V}(\epsilon_p) \geq & \frac{1}{l_s^2 \eta_0^2} \left(\frac{1}{\sum_{n=0}^{N_S-1}  e^{- 4 n \zeta}} + \frac{1}{ \sum_{n=0}^{N_S-1}  e^{- 2 (2n+1) \zeta}} \right), 
\end{align}
where we have set $\eta_0 = \eta_{x,0} = \eta_{p,0}$. The metrological gain from performing measurements with $N_S>1$ repetitions compared to $N_S=1$ is obtained from the ratio of their variances. In a noise-free case where $T_1 \gg T_\mathrm{meas}$ and $\zeta = 0$, the gain is given by $N_S$. For the above experimental setting where $\zeta = 0.33$, the expected gain from performing $N_S=2$ repetitions over $N_S=1$ is $\SI{1.80}{dB}$, which is in good agreement with experimental results. We also estimate an additional gain of $\SI{0.70}{dB}$ from performing $N_S=3$ repetitions over $N_S=2$, in contrast to the experiment, which observed no discernible gain. This discrepancy is likely due to approximations made in the above model, such as neglecting the effects of backaction on sequential measurements (see section~\ref{sec:grid_state_prob_distribution}). 

\section{Force sensing with grid states}
\label{sec:force_sensing}

In the main text and the previous sections, grid states have been discussed in the context of multi-parameter displacement sensing, where they can simultaneously estimate displacements in both position and momentum. 
Here, we show how these displacement parameters can be related to estimating a force induced by an electric field. The action of an external E-field with frequency at or near the trapped ion's secular vibrational frequency is a displacement in the phase space spanned by position and momentum. The phase space displacement's amplitude is related to the E-field amplitude, and the displacement's phase is related to the relative phase of the electric field and the ion motion. Using grid states, one can estimate the force's amplitude from individual estimates of position and momentum shifts with a metrological gain that is independent of the force's phase. 
This is important for applications ranging from dark matter sensing to quantum-logic-enabled photon-recoil spectroscopy~\cite{Wolf2019, Gilmore2021}, which benefit from force sensing in a phase-independent manner.

More specifically, we aim to estimate the force, $f$, resulting from a Hamiltonian,
\begin{equation}
    \hat{H} = - \frac{f z_0}{2} (\hat{a} e^{- i \phi_f} + \hat{a}^\dagger e^{i \phi_f}),
\end{equation}
where $z_0 = \sqrt{\hbar/2m\omega_x} = \SI{4.7}{nm}$ is the spatial extend of the ground state wave function in the radial-$x$ direction. A system evolving under $\hat{H}$ for a duration $t_f$ results in a displacement $\hat{D}(\gamma e^{i(\phi_f + \pi/2)})$ with magnitude $\gamma = \frac{f z_0}{2 \hbar} t_f$. We assume that the phase $\phi_f$ is not known apriori but remains constant throughout the experiment. 

The uncertainty in estimating the force is related to the uncertainty of the displacement magnitude, 
\begin{equation}
    \label{eq:Delta_f_from_Delta_gamma}
    \Delta f = \frac{2 \hbar}{z_0 t_f} \Delta \gamma,
\end{equation}
and the achievable measurement uncertainty is limited by the Cram\'er-Rao bound, 
\begin{equation}
    \Delta \gamma \geq \frac{1}{\sqrt{M F_\gamma }},
\end{equation}
where $M$ is the number of measurement repetitions and $F_\gamma$ is the Fisher information associated with the displacement amplitude, $\gamma$. 

\begin{figure}
    \centering
    \includegraphics[]{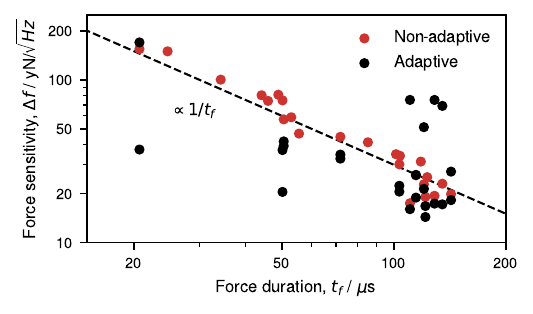}
    \caption{
    \textbf{Performance of grid states for force sensing.}
    The sensitivity to a force is measured from Bayesian QPE for increasing durations $t_f$ for which the force is applied. 
    }
    \label{fig:force_uncertainty_scaling}
\end{figure}

We estimate the uncertainty $\Delta \gamma$ achieved by the grid state from the experimental results shown in Fig.~3, which were obtained by performing multi-parameter displacement sensing with a non-adaptive QPE protocol and Bayesian inference. Estimates of the displacement amplitude, $\gamma$, are calculated from estimates of the individual displacement parameters, $\epsilon_x$ and $\epsilon_p$, with $\gamma = \sqrt{(\epsilon_x^2 +\epsilon_p^2)/2}$. We perform the same analysis of section~\ref{sec:qpe} and repeat the estimation of $\gamma$ for every pair ($\epsilon_x$, $\epsilon_p$) in the experiment. The uncertainty is then calculated from the Holevo variance, $\Delta \gamma = \sqrt{\mathrm{V}_\mathrm{H}(\gamma)}$, averaged over each pair $(\epsilon_x, \epsilon_p)$ (methods detailed in section~\ref{sec:qpe}). The resulting uncertainty with $M=128$ is $\Delta \gamma = \num{52(8)e-3}$. From this, we calculate the sensitivity, $\sigma_\gamma = \Delta\gamma /\sqrt{\mathrm{BW}}$, where $\mathrm{BW} = 1/\tau$ is the measurement bandwidth and $\tau = M t_\mathrm{exp}$ is the total integrated experimental duration. The duration of a single experiment, $t_\mathrm{exp}$, is estimated as 
\begin{equation}
    t_\mathrm{exp} = t_\mathrm{sb} + t_\mathrm{dc} + t_\mathrm{op} + t_\mathrm{ini} + 2N_S t_\mathrm{meas},
\end{equation}
where $t_\mathrm{sb} = \SI{7.48}{ms}$ is the sideband cooling duration, $t_\mathrm{dc} = \SI{6}{ms}$ is the Doppler cooling duration, $t_\mathrm{op} = \SI{50}{\mu s}$ is the optical pumping duration, $t_\mathrm{ini} = \SI{0.73}{ms}$ is the initialisation duration of the bosonic sensing state to the grid state with $\Delta = 0.41$ and $t_\mathrm{meas} \approx \SI{400}{\micro s}$ is the duration to apply and measure $\hat{S}_x$ or $\hat{S}_p$. For $N=1$ repetitions of $\hat{S}_x$ and $\hat{S}_p$, the experimental duration is $t_\mathrm{exp} = \SI{14.8}{ms}$. We neglect additional costs from post-selection and from Bayesian inference calculations. The resulting sensitivity to displacement amplitudes is $\sigma_\gamma = \SI{71(12)e-3}{\sqrt{Hz}^{-1}}$. The uncertainty $\Delta\gamma$ can also be related to an uncertainty in the ion's position with the relation $\Delta z = 2 z_0 \Delta \gamma$~\cite{Gilmore2021}, giving $\Delta_z = \SI{0.49(8)}{nm}$. We also calculate the position sensitivity, $\sigma_z = 2 z_0 \sigma_\gamma = \SI{0.67(11)}{nm/\sqrt{Hz}}$.

We finally estimate the force sensitivity from Eq.~\ref{eq:Delta_f_from_Delta_gamma}, giving $\sigma_f = (2\hbar / z_0 t_f)\sigma_\gamma$. Since each pair of displacements $(\epsilon_x, \epsilon_p)$ corresponds to a separate force duration $t_f$, we individually calculate the force sensitivity for each point (see Fig.~\ref{fig:force_uncertainty_scaling}). The minimum force sensitivity is achieved for a duration $t_f = \SI{122}{\micro s}$, giving $\sigma_f = \SI{16.8}{yN/\sqrt{Hz}}$ ($\SI{1}{yN} = \SI{1e-24}{N}$). This can equivalently be expressed as a sensitivity to electric fields, $\sigma_E = \sigma_f /q = \SI{103}{\micro V m^{-1}/ \sqrt{Hz}}$. 

The above analysis is repeated for the experimental results obtained from adaptive Bayesian estimation, which outperformed non-adaptive estimation in Fig.~3. We find an average uncertainty in the displacement amplitude of $\Delta \gamma = \num{3.7(8)e-2}$, giving a sensitivity of $\sigma_\gamma = \SI{51(11)e-3}{\sqrt{Hz}^{-1}}$. The smallest sensitivity to a force is achieved for a duration of $t_f = \SI{122}{\micro s}$, giving $\sigma_f = \SI{14.3}{yN/\sqrt{Hz}}$. The equivalent sensitivity to electric fields is $\SI{89.5}{\micro V m^{-1}/\sqrt{Hz}}$.

Our demonstrated phase-insensitive force sensitivity is comparable to the state of the art of quantumm-enhanced sensors that perform phase-sensitive force sensing. For comparison, Paul traps have achieved sensitivities of $\SI{28}{zN/\sqrt{Hz}}$~\cite{Shaniv2017} and $\SI{153}{yN / \sqrt{Hz}}$~\cite{Bonus2024} with single ions. Alternatively, Penning traps have reduced uncertainties by using crystals with a large number of ions. Ref.~\cite{Biercuk2010} achieved a sensitivity of $\SI{390}{yN/\sqrt{Hz}}$ with a crystal of 100 ions, while Ref.~\cite{Affolter2020} achieved a sensitivity of $\SI{12}{yN/\sqrt{Hz}}$ per ion with a crystal of $\sim 88$ ions. The smallest measured uncertainty to date was demonstrated in Ref.~\cite{Gilmore2021} and achieved a sensitivity of $\SI{38.4}{rN /\sqrt{Hz}}$ ($\SI{1}{rN} = \SI{1e-27}{N}$) per ion with a crystal of $\sim 150$ ions. There, the phase of the force had to be well defined a-priori to maximize the meiselogical gain. Here, we stress that we perform phase-insensitive force sensing, and our achieved sensitivities are independent of the force's phase. In comparison, Ref.~\cite{Wolf2019} performed phase-insensitive force sensing with Fock states in a single ion and achieved a sensitivity of $\SI{112}{yN/\sqrt{Hz}}$. Ref.~\cite{Shaniv2017} also performed phase-independent force sensing in a Paul trap and obtained a sensitivity of $\SI{1.7}{aN/\sqrt{Hz}}$.

Our sensitivity can be straightforwardly improved by increasing the interaction time, $t_f$, during which the force is interrogated. In our experiment, this interaction time was minimised by increasing the Rabi frequency of the SDF pulse to reduce effects from dephasing and hence reduce the variance from multi-parameter displacement sensing. However, $t_f$ is considerably shorter than the lifetime of the grid state measurement operators and can be extended.

\bibliography{bib.bib}

\newpage

\renewcommand{\figurename}{Extended Data Figure}

\begin{figure*}
    \includegraphics[]{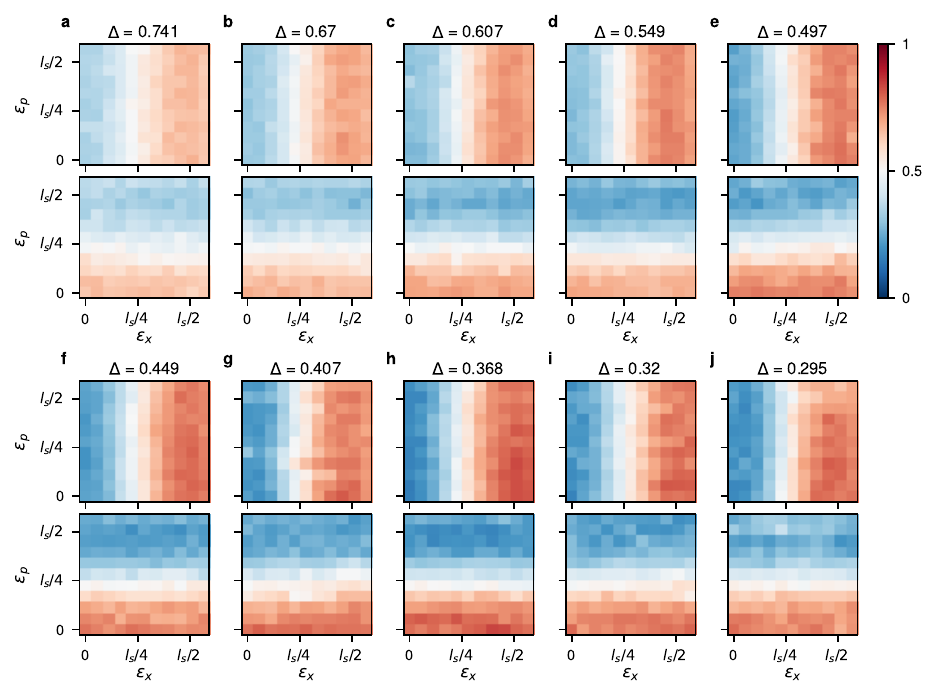}
    \caption{
    \textbf{Experimentally measured probability distributions of grid states with exact position and momentum operators}. The distributions are plotted for states with increasing energy \textbf{(a-j)}, and the subplot titles give the target squeezing parameter. Probabilities are obtained by repeating the QPE circuit of Fig.~1e for $M=10^3$ measurement repetitions for each pair $(\epsilon_x, \epsilon_p)$. Probability distributions associated with measurements of $\hat{S}_x$ (top rows) and $\hat{S}_p$ (bottom rows) are obtained from the mean of measurement outcomes $\mathrm{m}_x$ and $\mathrm{m}_p$, respectively. The displacement parameters are varied in the range $\epsilon_x, \epsilon_p \in [0, \sqrt{2}]$. $l_s = \sqrt{2\pi}$ is the modulus of the modular position--momentum variables. backaction from the first measurement of $\hat{S}_x$ can be seen from the inversion of the probability distribution of $\hat{S}_p$. 
    }
    \label{fig:grid_probs_non_finite_raw_data}
\end{figure*}

\begin{figure*}
    \includegraphics[]{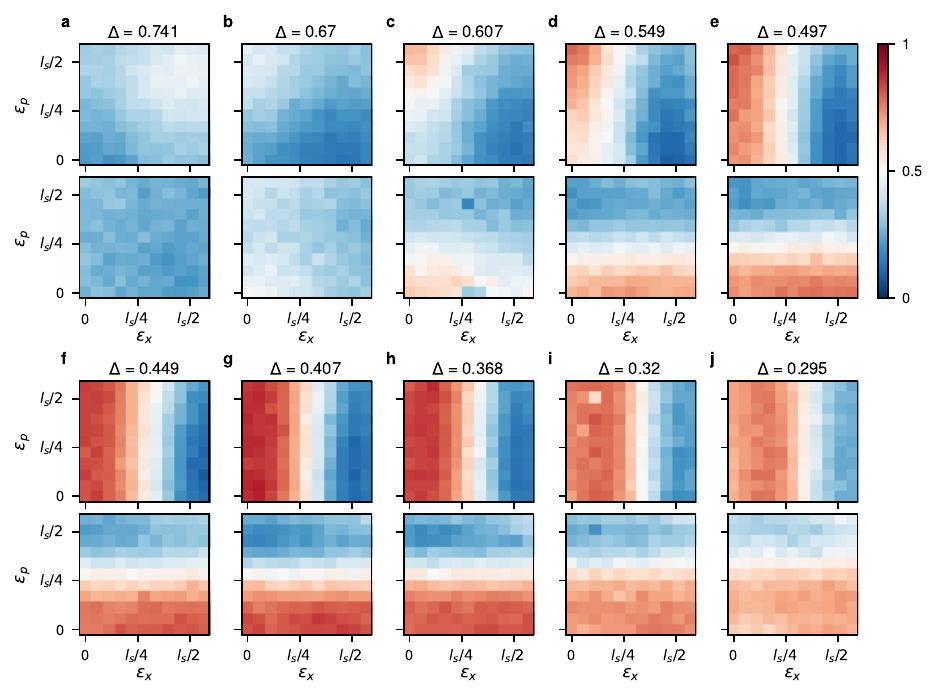}
    \caption{
        \textbf{Experimentally measured probability distributions of grid states with finite-energy position and momentum operators}. Measurements of position and momentum are performed with the finite-energy Big-small-Big protocol detailed in section~\ref{sec:metrological_gain_bsb_grid}. The distributions are plotted for states with increasing energy \textbf{(a-j)}, and the subplot titles give the target squeezing parameter. Probabilities are obtained by repeating the QPE circuit of Fig.~1e for $M=10^3$ measurement repetitions for each pair $(\epsilon_x, \epsilon_p)$. Probability distributions associated with measurements of $\hat{S}_x$ (top rows) and $\hat{S}_p$ (bottom rows) are obtained from the mean of measurement outcomes $\mathrm{m}_x$ and $\mathrm{m}_p$, respectively. The displacement parameters are varied in the range $\epsilon_x, \epsilon_p \in [0, \sqrt{2}]$. $l_s = \sqrt{2\pi}$ is the modulus of the modular position--momentum variables. 
    }
    \label{fig:grid_probs_finite_raw_data}
\end{figure*}

\begin{figure*}
    \includegraphics[]{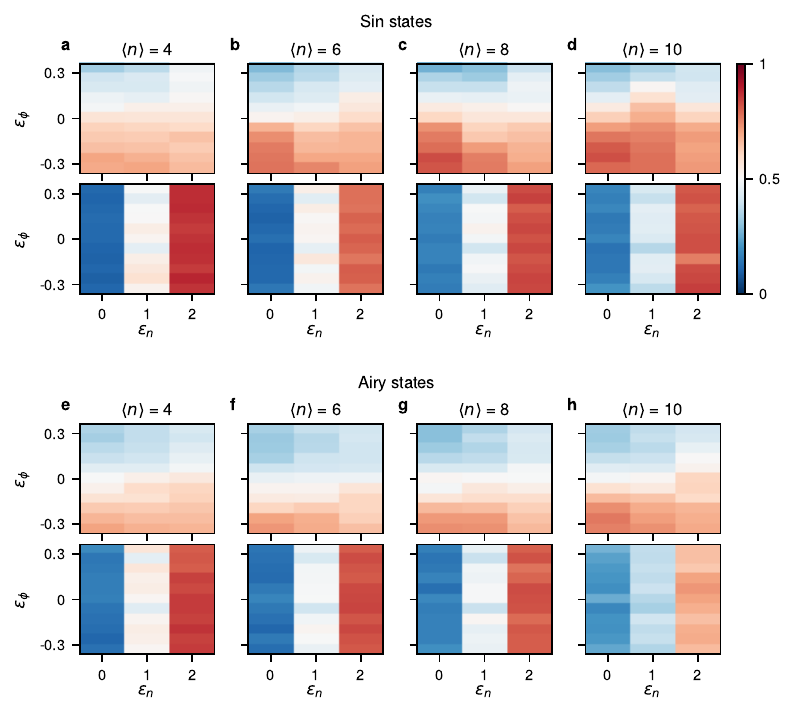}
    \caption{    
    \textbf{Experimentally measured probability distributions of number--phase states}. Distributions for sine states \textbf{(a-d)} and airy states \textbf{(e-h)} are plotted for varying target energies, where the average phonon number given by the title is varied in the range $\langle \hat{n} \rangle \in [4, 10]$. Probabilities are obtained by repeating the QPE circuit of Fig.~1e with changes outlined in the main text for $M=10^3$ measurement repetitions for each pair $(\epsilon_n, \epsilon_\phi)$. Probability distributions associated with measurements of $\hat{S}_\phi$ (top rows) and $\hat{S}_n$ (bottom rows) are obtained from the mean of measurement outcomes $\mathrm{m}_\phi$ and $\mathrm{m}_n$, respectively. The signals are varied in the range $\epsilon_\phi \in [-\pi/8, \pi/8]$ and $\epsilon_n \in  [0, 2]$. }
    \label{fig:np_probs_raw_data}
\end{figure*}

\begin{figure*}
    \includegraphics[]{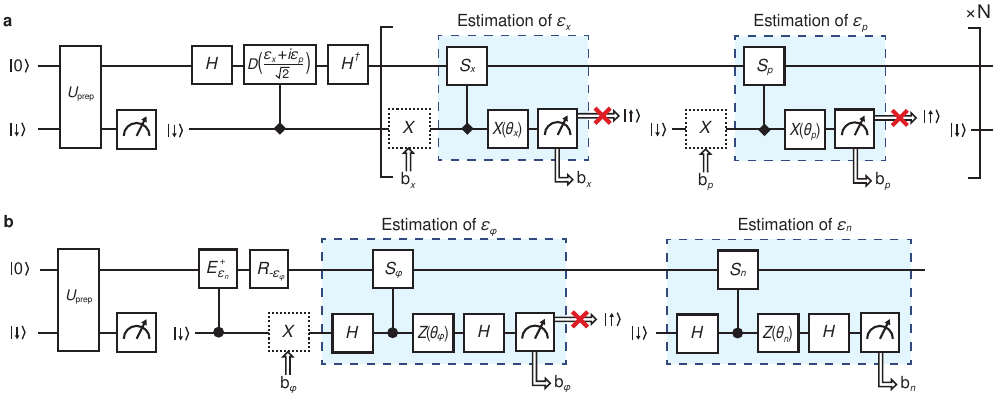}
    \caption{
    \textbf{Experimental pulse sequence for quantum phase estimation with (a) grid states and (b) number--phase states}.
    The ancilla qubit and bosonic modes are first initialised to their ground state, $\ket{\downarrow} \otimes \ket{0}$. The sensing state $\ket{\psi}$ is then prepared by applying a dynamically modulated SDF pulse, giving a unitary operation $\hat{U}_\mathrm{prep}$ which prepares the state $\ket{\downarrow} \otimes \ket{\psi}$. A mid-circuit measurement is then applied to remove residual spin-boson entanglement, and the circuit only proceeds if the measured state is $\ket{\downarrow}$ as measurements of $\ket{\uparrow}$ scatter photons which decohere the bosonic state. 
    \textbf{(a)} The grid state circuit then applies a displacement $\hat{D}((\epsilon_x + i \epsilon_p)/\sqrt{2})$. This is achieved by applying the SDF Hamiltonian in between two rotations along the $\hat{\sigma}_y$ axis which rotate the spin in and out of a $\hat{\sigma}_x$ eigenstate. QPE sub-routines for estimation of $\epsilon_x$ and $\epsilon_p$ are then performed in an alternating manner. Each QPE sub-routine applies the conditional momentum or positions operators, followed by a controllable ancilla rotation. 
    \textbf{(b)} After preparing the number--phase sensing state, the sensing signal is applied, consisting of a phonon shift by $\epsilon_n$ followed by a rotation by $-\epsilon_\phi$. The phonon shift is implemented with the conditional phase operator after initialising the ancilla in $\ket{\downarrow}$. The rotation is implemented by offseting the phases of the blue-sideband fields during subsequent interactions. The circuit then performs two QPE sub-routines which alternatingly estimate $\epsilon_\phi$ and $\epsilon_n$.
    In both \textbf{(a)} and \textbf{(b)}, the circuit only proceeds after a measurement if the state is $\ket{\downarrow}$. Measurement outcomes associated with the $\ket{\uparrow}$ state are obtained by randomly applying a rotation on the ancilla prior to the QPE sub-routine. 
    }
    \label{fig:sm_detailed_pulse_sequence}
\end{figure*}